\documentclass[iop]{emulateapj}
\newcommand{\CV}{\ion{C}{5}}
\newcommand{\CVI}{\ion{C}{6}}
\newcommand{\HI}{\ion{H}{1}}
\newcommand{\MgXI}{\ion{Mg}{11}}
\newcommand{\NV}{\ion{N}{5}}

\newcommand{\NVII}{\ion{N}{7}}
\newcommand{\NeIX}{\ion{Ne}{9}}

\newcommand{\OVII}{\ion{O}{7}}
\newcommand{\OVIII}{\ion{O}{8}}

\newcommand{\oplusseven}{O$^{+7}$}
\newcommand{\opluseight}{O$^{+8}$}

\newcommand{\Kalpha}{K$\alpha$}
\newcommand{\Kbeta}{K$\beta$}

\newcommand{\Lyalpha}{Ly$\alpha$}

\newcommand{\nH}{\ensuremath{n_{\mathrm{H}}}}

\newcommand{\NH}{\ensuremath{N_{\mathrm{H}}}}

\newcommand{\RE}{\ensuremath{R_\mathrm{E}}}

\newcommand{\nm}{\ensuremath{\mbox{\nm}}}
\newcommand{\cm}{\ensuremath{\mbox{cm}}}

\newcommand{\km}{\ensuremath{\mbox{km}}}

\newcommand{\s}{\ensuremath{\mbox{s}}}
\newcommand{\ks}{\ensuremath{\mbox{ks}}}

\newcommand{\kev}{\ensuremath{\mbox{keV}}}

\newcommand{\erg}{\ensuremath{\mbox{erg}}}

\newcommand{\sr}{\ensuremath{\mbox{sr}}}

\newcommand{\K}{\ensuremath{\mbox{K}}}

\newcommand{\ph}{\ensuremath{\mbox{photons}}}
\newcommand{\counts}{\ensuremath{\mbox{counts}}}

\newcommand{\cmsq}{\ensuremath{\cm^2}}

\newcommand{\parcminsq}{\ensuremath{\mbox{arcmin}^{-2}}}

\newcommand{\pcc}{\ensuremath{\cm^{-3}}}
\newcommand{\pcmsq}{\ensuremath{\cm^{-2}}}

\newcommand{\ps}{\ensuremath{\s^{-1}}}
\newcommand{\psr}{\ensuremath{\sr^{-1}}}

\newcommand{\flux}{\erg\ \pcmsq\ \ps}
\newcommand{\lineunit}{\ph\ \pcmsq\ \ps\ \psr}

\newcommand{\LU}{\ensuremath{\mbox{L.U.}}}
\newcommand{\kmps}{\km\ \ps}

\newcommand{\pownorm}{\ph\ \pcmsq\ \ps\ \psr\ \ensuremath{\kev^{-1}}}

\newcommand{\rassrate}{\counts\ \ps\ \parcminsq}

\newcommand{\Iovii}{\ensuremath{I_\mathrm{O\,VII}}}
\newcommand{\Eovii}{\ensuremath{E_\mathrm{O\,VII}}}
\newcommand{\Ioviii}{\ensuremath{I_\mathrm{O\,VIII}}}

\newcommand{\ace}{\textit{ACE}}
\newcommand{\asca}{\textit{ASCA}}

\newcommand{\chandra}{\textit{Chandra}}

\newcommand{\rosat}{\textit{ROSAT}}

\newcommand{\suzaku}{\textit{Suzaku}}

\newcommand{\wind}{\textit{Wind}}
\newcommand{\xmm}{\textit{XMM-Newton}}

\newcommand{\e}{\ensuremath{\mathrm{e}}}

\newcommand{\chisq}{\ensuremath{\chi^2}}

\newcommand{\mekal}{\textsc{MeKaL}}
\newcommand{\raymondsmith}{\citeauthor{raymond77} (\citeyear{raymond77} and updates)}
\newcommand{\eqref}[1]{equation~(\ref{#1})}

\newcommand{\esas}{\textit{XMM}-ESAS}

\newcommand{\cosec}{\ensuremath{\mathrm{cosec}\,}}
\newcommand{\fsw}{\ensuremath{f_\mathrm{sw}}}
\newcommand{\Ftotal}{\ensuremath{F_\mathrm{total}^{2-5}}}
\newcommand{\Fexgal}{\ensuremath{F_\mathrm{exgal}^{2-5}}}

\newcommand{\Fsource}{\ensuremath{F_\mathrm{X}^{0.5-2.0}}}
\newcommand{\uave}{\ensuremath{\bar{u}}}
\newcommand{\gammaSWCX}{\ensuremath{\gamma_\mathrm{SWCX}}}
\newcommand{\gammahalo}{\ensuremath{\gamma_\mathrm{halo}}}
\newcommand{\Igeo}{\ensuremath{I_\mathrm{geo}}}
\newcommand{\nsw}{\ensuremath{n_\mathrm{sw}}}
\newcommand{\usw}{\ensuremath{u_\mathrm{sw}}}
\newcommand{\sigmasys}{\ensuremath{\sigma_\mathrm{sys}}}
\newcommand{\sigmastat}{\ensuremath{\sigma_\mathrm{stat}}}
\newcommand{\REPL}{\ensuremath{\mathrm{R}4_\mathrm{EPL}}}
\newcommand{\Robs}{\ensuremath{\mathrm{R}4_\mathrm{obs}}}
\newcommand{\Roxygen}{\ensuremath{\mathrm{R}4_\mathrm{oxygen}}}
\newcommand{\Rhalo}{\ensuremath{\mathrm{R}4_\mathrm{halo}}}
\newcommand{\RSWCX}{\ensuremath{\mathrm{R}4_\mathrm{SWCX}}}

\newcommand{\vth}{\ensuremath{v_\mathrm{th}}}

\newcommand{\marker}[1]{#1}

\shorttitle{AN \textit{XMM-NEWTON} SURVEY OF THE SOFT X-RAY BACKGROUND. II.}
\shortauthors{HENLEY AND SHELTON}

\begin{document}

\title{An \textit{XMM-Newton} Survey of the Soft X-ray Background. II.
       An All-Sky Catalog of Diffuse \OVII\ and \OVIII\ Emission Intensities}
\author{David B. Henley and Robin L. Shelton}
\affil{Department of Physics and Astronomy, University of Georgia, Athens, GA 30602, USA}
\email{dbh@physast.uga.edu}

\begin{abstract}
We present an all-sky catalog of diffuse \OVII\ and \OVIII\ line intensities, extracted from
archival \xmm\ observations. This catalog supersedes our previous catalog, which covered the sky
between $l=120\degr$ and $l=240\degr$.
We attempted to reduce the contamination from near-Earth solar wind charge exchange (SWCX) emission by
excluding times of high solar wind proton flux from the data. Without this filtering we were able to
extract measurements from 1868 observations. With this filtering, nearly half of the observations
became unusable, and only 1003 observations yielded measurements.
The \OVII\ and \OVIII\ intensities are typically $\sim$2--11 and $\la$3~\lineunit\ (line unit, \LU),
respectively, although much brighter intensities were also recorded.
Our data set includes 217 directions that have been observed multiple times by \xmm. The time
variation of the intensities from such directions may be used to constrain SWCX models. The
\OVII\ and \OVIII\ intensities typically vary by $\la$5 and $\la$2~\LU\ between repeat observations,
although several intensity enhancements of $>$10~\LU\ were observed.
We compared our measurements with models of the heliospheric and geocoronal SWCX. The heliospheric
SWCX intensity is expected to vary with ecliptic latitude and solar cycle. We found that the
observed oxygen intensities generally decrease from solar maximum to solar minimum, both at high
ecliptic latitudes (which is as expected) and at low ecliptic latitudes (which is not as expected).
The geocoronal SWCX intensity is expected to depend on the solar wind proton flux incident on the
Earth and on the sightline's path through the magnetosheath.  The intensity variations seen in
directions that have been observed multiple times are in poor agreement with the predictions of a
geocoronal SWCX model.
We found that the oxygen lines account for $\sim$40--50\%\ of the 3/4~\kev\ X-ray background that is
not due to unresolved active galactic nuclei, in good agreement with a previous measurement.
However, we found that this fraction is not easily explainable by a combination of SWCX emission and
emission from hot plasma in the halo.
We also examined the correlations between the oxygen intensities and Galactic longitude and
latitude.  We found that the intensities tend to increase with longitude toward the inner Galaxy,
possibly due to an increase in the supernova rate in that direction or the presence of a halo of
accreted material centered on the Galactic Center. The variation of intensity with Galactic latitude
differs in different octants of the sky, and cannot be explained by a single simple plane-parallel
or constant-intensity halo model.
\end{abstract}


\section{INTRODUCTION}
\label{sec:Introduction}

The soft X-ray background (SXRB) is the diffuse X-ray emission observed in all directions in the
$\sim$0.1--10~\kev\ band \citep{mccammon90}. At energies $\ga$$1~\kev$, this emission is dominated
by unresolved emission from distant active galactic nuclei (AGN; e.g., \citealt{brandt05}).  Below
$\sim$$1~\kev$, Galactic line emission becomes important, and dominates at the lowest energies. Some
of this line emission comes from hot interstellar gas -- possible unabsorbed emission from
$\sim$million-degree gas in the Local Bubble (LB; \citealt{sanders77,cox87,mccammon90,snowden90};
although see also \citealt{welsh09}), and emission from $\sim$1--3 million-degree gas in the
Galactic halo attenuated by the Galaxy's
\HI\ \citep{burrows91,snowden91,snowden98,snowden00,wang98,pietz98,kuntz00,smith07a,galeazzi07,henley08a,lei09,yoshino09,gupta09b,henley10b}.

In addition to the emission from hot interstellar gas, X-ray line emission is produced within our
solar system, as a consequence of charge exchange reactions between solar wind ions and neutral
hydrogen and helium atoms. Solar wind charge exchange (SWCX) reactions can occur in the heliosphere
between solar wind ions and neutral hydrogen and helium atoms that have entered the heliosphere from
interstellar space \citep[e.g.,][]{cravens00,robertson03a,koutroumpa06}.  It can also occur in the
Earth's exosphere, beyond the magnetopause, between solar wind ions and geocoronal neutral hydrogen
atoms \citep[e.g.][]{robertson03b}. Charge exchange reactions can also occur in the atmospheres of
other planets and around comets \citep[e.g.,][]{cravens02,wargelin08}, but the resulting emission is
sufficiently localized as to not affect the \xmm\ data set. The heliospheric emission is in general
much brighter than the geocoronal emission, although the geocoronal and/or near-Earth heliospheric
SWCX emission can exhibit bright enhancements on timescales of $\sim$hours--days, often in
association with variations in the solar wind
\citep{cravens01,snowden04,fujimoto07,carter08,carter10,ezoe10,ezoe11}.  Longer-term variations in
SWCX line intensities, on timescales of $\sim$days--years, have also been observed
\citep{koutroumpa07,kuntz08a,henley08a,henley10a}. These variations may include variations in the
global heliospheric SWCX emission, which is expected to vary slowly with the solar cycle
\citep{robertson03a,koutroumpa06}, as well as variations in the geocoronal and near-Earth
heliospheric emission. In general it is difficult to disentangle these various effects.  From the
point of view of someone interested in the hot Galactic gas, the SWCX emission is a time-varying
contaminant of the SXRB emission.

The SXRB has been surveyed several times with rocket- and satellite-borne proportional counters
\citep{mccammon83,marshall84,garmire92,snowden97}. While the data from these surveys were essential
to establishing our current picture of the SXRB and the hot interstellar medium (ISM), these data
are of rather low spectral resolution ($E / \Delta E \sim 1$--3). The CCD cameras on board \chandra,
\xmm, and \suzaku\ offer the ability for higher-resolution spectroscopy (e.g., $E / \Delta E \sim
15$ at 1~\kev\ for the
\xmm\ EPIC-MOS\footnote{http://xmm.esa.int/external/xmm\_user\_support/documentation/\\ uhb/node14.html}).
With these cameras, some emission features in SXRB spectra can be resolved, allowing one to measure
SWCX emission line strengths
\citep{wargelin04,snowden04,fujimoto07,henley08a,henley10a,koutroumpa09b,koutroumpa11,carter10,ezoe11} or
the temperature of the halo plasma
\citep{smith07a,galeazzi07,henley08a,lei09,yoshino09,gupta09b,henley10b}.

Prior to this project, CCD-resolution spectra of the SXRB had been presented for only a few tens of
directions.  In order to greatly increase the number of directions for which good-quality SXRB
spectra are available, we have carried out a survey of the SXRB using archival data from the
EPIC-MOS cameras \citep{turner01} on board \xmm\ \citep{jansen01}. A key aspect of this survey is
that we do not use only blank-sky observations. Any observation in which the target source is not
too bright or too extended can potentially be used -- the target source is excised, and an SXRB
spectrum extracted from the remainder of the field. The first results from this survey were
published in \citet[hereafter Paper~I]{henley10a}. We presented SXRB \OVII\ and \OVIII\ intensities
extracted from 590 \xmm\ observations between $l=120\degr$ and $l=240\degr$ (i.e., one third of the
sky). We concentrated on these lines as they dominate the Galactic SXRB emission in the
\xmm\ bandpass \citep{mccammon02}. Their intensities showed considerable variation over the sky,
emphasizing the importance of extracting SXRB spectra for as many directions as possible.  Here, we
complete our survey and present an all-sky catalog of SXRB \OVII\ and \OVIII\ intensities,
comprising measurements from \marker{1880} \xmm\ observations. Note that, in the course of
completing our survey, we have completely reprocessed the observations that featured in Paper~I. We
also now include estimates of the systematic errors associated with our spectral analysis. The
current catalog therefore supersedes, rather than supplements, Paper~I.

The goal of this survey is to improve our understanding of the hot gas in our Galaxy. We have
already made progress in this direction by comparing spectra from Paper~I with the predictions of
various physical models \citep{henley10b}. These observations favor a Galactic fountain origin for
the hot halo gas, in which gas is driven from the disk into the halo by supernovae (SNe;
\citealt{shapiro76,bregman80,joung06}). This new, larger data set will allow us to further test halo
models, and investigate how the halo emission varies over the sky. In addition, our survey includes
\marker{217} sightlines for which there are multiple observations, separated in time by days to
years. In a given direction, any variation in the observed SXRB spectrum is due to variation in the
SWCX emission. Hence, a set of observations of the same direction can be used to constrain SWCX
models, which are essential for obtaining accurate spectra of the hot Galactic gas from SXRB
observations.

The remainder of this paper is arranged as follows. In Section~\ref{sec:DataReduction} we outline
the data processing (see Paper~I for more details). In Section~\ref{sec:OxygenIntensities} we
present our oxygen intensity measurements. We discuss our results in
Section~\ref{sec:Discussion}. In particular, in Section~\ref{subsec:SWCX} we discuss the
implications of our results for SWCX. In Section~\ref{subsec:3/4kevSXRB} we discuss the oxygen
lines' contribution to the 3/4~\kev\ SXRB, and the implications for the sources of the SXRB
emission. In Section~\ref{subsec:SpatialVariation} we discuss the spatial variation of the oxygen
intensities, and the implications of these results for the halo. We summarize our results in
Section~\ref{sec:Summary}.

\section{DATA REDUCTION}
\label{sec:DataReduction}

The data reduction procedure is described in full in Paper~I, to which the reader is referred for
more details. Here, we outline the main steps, and also point out any differences from Paper~I. The
results presented here were obtained from data reduced with the \xmm\ Science Analysis
System\footnote{http://xmm.esac.esa.int/sas/} (SAS) version 11.0.1, which includes the
\xmm\ Extended Source Analysis
Software\footnote{http://heasarc.gsfc.nasa.gov/docs/xmm/xmmhp\_xmmesas.html} (\esas;
\citealt{kuntz08a,snowden11}). However, much of the preliminary inspection of the data (e.g., for
identifying unusable CCDs and for determining source exclusion regions; see below) was carried out
on data processed with earlier versions of SAS. Although the latest version of \esas\ can process
data from the EPIC-pn camera \citep{struder01}, it only became available after the current data
processing was well under way, and so we consider only EPIC-MOS data here.

We used the \esas\ calibration files that were current as of 2011 April 01. After we had completed
most of the processing, we learned that these files had been superseded. We reprocessed a random
subset of our observations with the newer calibration files (released on 2011 May 16), and found
that the effect on our oxygen intensity measurements was not significant, compared with the
statistical and systematic uncertainties. We therefore decided not to reprocess our entire data set
with the newer calibration files.

\subsection{Observation Selection and Initial Data Processing}
\label{subsec:ObservationSelection}

As of 2010 August 04, there were 5698 publicly available \xmm\ observations with at least some MOS
exposure. We processed each observation with the SAS \texttt{emchain} script, excluding observations
that had been rejected in Paper~I due to obvious contamination from soft protons impinging upon the
detector, or due to bright or extended sources in the field of view. This script produces a
calibrated events list for each MOS exposure. We then used the \esas\ \texttt{mos-filter} script to
identify and remove times affected by soft-proton flaring. For each events list, this script creates
a histogram of count-rates from the 2.5--12~\kev\ light curve, using 60-s bins, fits a Gaussian to
the peak of this histogram, and rejects all times whose count-rates differ by more than 1.5$\sigma$
from the mean of the Gaussian. (Although there are some superficial differences here from the
description in Paper~I, these are solely due to changes in the way \esas\ is organized. These first
steps in the processing are essentially identical to those used in Paper~I.)

\begin{figure}
\plotone{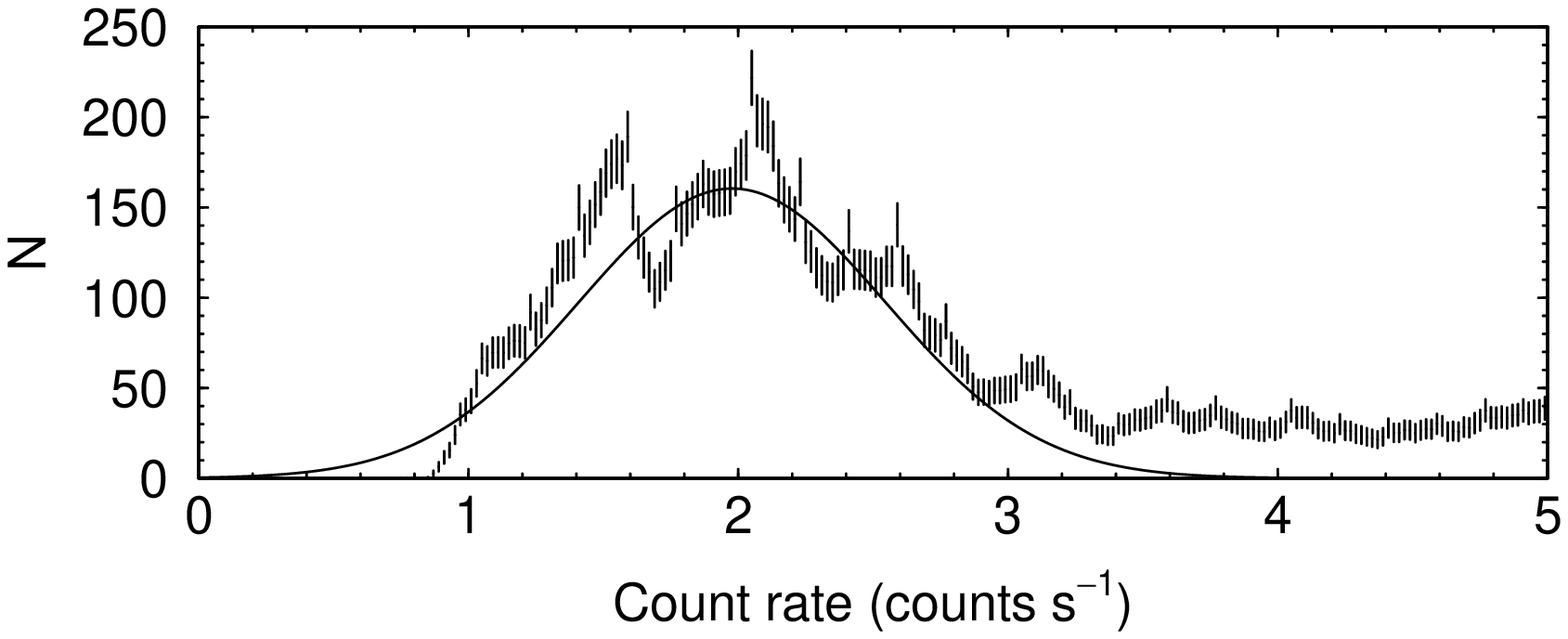}
\plotone{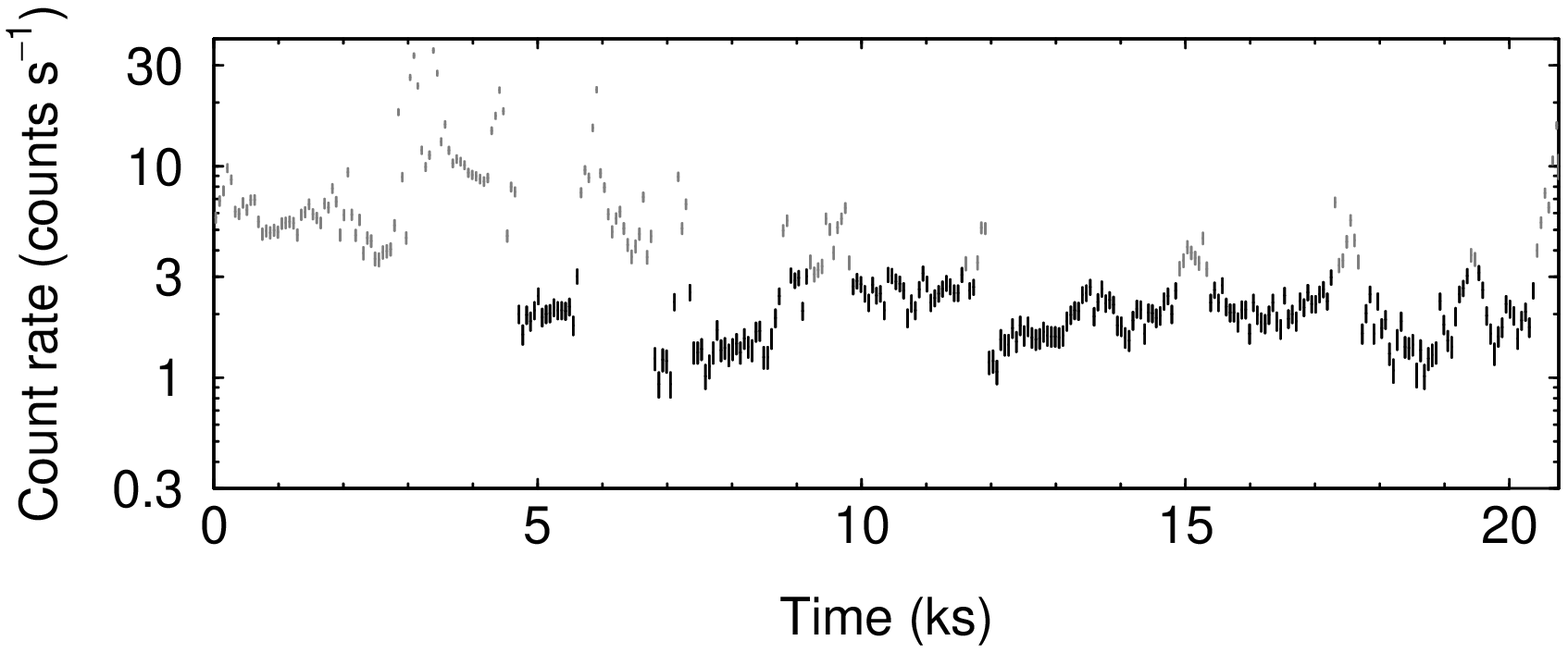}
\caption{Histogram of MOS1 count-rates (top) and MOS1 light curve (bottom) for the
  2.5--12~\kev\ band, from obs.~0051760201 (an observation for which \texttt{mos-filter} yielded
  $\ge$5~ks of good time, despite the presence of severe soft-proton contamination).  The curve in
  the upper panel is the Gaussian that was fitted to the histogram of count-rates. Times whose
  count-rates differed by more than 1.5$\sigma$ from the mean of this Gaussian were rejected; such
  times are plotted in gray in the lower panel.
  \label{fig:Lightcurve}}
\end{figure}

We discarded all exposures that did not have at least 5~ks of good time after the above filtering,
and then discarded any observation that did not have at least one good MOS1 exposure and at least
one good MOS2 exposure. We then inspected the light curves and histograms of count-rates produced by
\texttt{mos-filter}, checking for observations that were badly affected by soft protons, but for
which \texttt{mos-filter} yielded $\ge$5~ks of good time. Such observations are generally
apparent from the non-Gaussianity of the histogram of count-rates. Figure~\ref{fig:Lightcurve} shows
an example -- despite the clear soft-proton contamination, \texttt{mos-filter} yielded
$\approx$12~ks of good time for this observation. Observations like this were rejected.

For each remaining observation, we examined the cleaned images produced by
\texttt{mos-filter}. Because our ultimate goal was to extract SXRB spectra, we rejected observations
whose fields contained large diffuse sources (e.g., galaxy clusters) that filled all or most of the
field of view, or very bright sources for which the emission in the wings of the point spread
function (PSF) would dominate over the SXRB emission. We also rejected observations that exhibited
bright arcs in the field of view, due to stray-light X-rays from a bright source outside the field of
view. The above-described rejections reduced the original set of 5698 observations to \marker{2611},
a 54\%\ attrition rate (somewhat higher than the $\sim$45\%\ rate in Paper~I).

\subsection{Removal of Bright or Extended Sources, and of Unusable CCDs}
\label{subsec:BrightSourceRemoval}

The visual inspection of the cleaned images served two further purposes. We identified bright and/or
extended sources that would not be adequately excised by the automated point source removal
procedure described in Section~\ref{subsec:AutomatedSourceRemoval}, below. Such sources were removed
using circular exclusion regions. These regions were generally positioned by eye, although if the
source to be excluded was the observation target, we used the pointing direction from the events
list header to center the exclusion region. The radii of these regions were also chosen by eye,
although for some sources we used the radial surface brightness profile to aid us. As noted in
Paper~I, we erred on the side of choosing larger exclusion radii, at the expense of reducing the
number of counts in the SXRB spectra.

The final purpose of the visual inspection of the images was to identify CCDs that should be
ignored in the subsequent processing. We ignored CCDs that were in window mode, CCDs that were
missing from the images (e.g., the MOS1-6 CCD after its failure in 2005 March), and CCDs that
clearly exhibited the anomalous state identified by \citet{kuntz08a} -- such CCDs appear unusually
bright in a soft-band (0.2--0.9~\kev) X-ray image.

\subsection{Automated Source Removal}
\label{subsec:AutomatedSourceRemoval}

In Paper~I, we ran the SAS \texttt{edetect\_chain} script on each observation to automatically
identify point sources for removal. Here, we instead used data from the Second \xmm\ Serendipitous
Source Catalogue (2XMM; \citealt{watson09}), specifically the 2XMMi~DR3 data
release.\footnote{http://xmmssc-www.star.le.ac.uk/Catalogue/2XMMi-DR3/} The advantage of using 2XMM
data for the automated source removal is that it means that exactly the same point sources are
removed from a set of observations of the same direction, which is not guaranteed if the source
detection is run on an observation-by-observation basis.

For each observation, we removed all the sources listed in the catalog within the field of view with
0.5--2.0~\kev\ fluxes $\Fsource \ge 5 \times 10^{-14}~\flux$, determined by combining the fluxes for
2XMM bands 2 and 3. This is the same flux level to which sources were removed in Paper~I; this flux
threshold was originally chosen because it is the approximate level to which \citet{chen97} removed
sources when they determined the spectrum of the extragalactic background that we used in the
Paper~I spectral analysis. However, as described in Section~\ref{subsec:OxygenMethod}, below, we
used a slightly different model for modeling the extragalactic background in this paper.

We used circles of radius 50\arcsec\ to excise the above-detected sources. Such source exclusion
regions enclose $\approx$90\%\ of each source's flux (calculated from the best-fit King profiles to
the MOS telescopes' PSFs; see the \xmm\ Calibration Access and Data
Handbook\footnote{http://xmm.vilspa.esa.es/external/xmm\_sw\_cal/calib/\\documentation/CALHB/node30.html}).
Note that, in Paper~I, we used the SAS \texttt{region} task to define regions that enclosed 90\%\ of
each source's flux. However, this task uses a more centrally peaked model of the PSF than the King
profile, and thus appears to underestimate the required source exclusion radii
(30\arcsec--40\arcsec, versus $\approx$50\arcsec).

\subsection{Reducing SWCX Contamination}
\label{subsec:ProtonFluxFiltering}

In an attempt to reduce the contamination from SWCX produced near the Earth, we removed the portions
of the \xmm\ data taken when the solar wind proton flux exceeds $2 \times 10^8$~\pcmsq\ \ps\ (see
Paper~I).  As in Paper~I, we used data from OMNIWeb,\footnote{http://omniweb.gsfc.nasa.gov/} which
combines in situ solar wind measurements from several satellites. The data that we used are mainly
from the \textit{Advanced Composition Explorer} (\ace) and \wind. The data from OMNIWeb for these
satellites have been time-shifted to the Earth. The OMNIWeb data covering the \xmm\ mission also
include data from near-Earth satellites that are not time-shifted, but this lack of time-shifting
should not adversely affect our results (see Paper~I).

As noted in Paper~I, this filtering by solar wind proton flux resulted in, for some observations,
the amount of usable observation time falling below our 5~ks threshold (see
Section~\ref{subsec:ObservationSelection}). After carrying out this filtering, the number of usable
observations fell from \marker{2611} to \marker{1435}. Below, we present the results obtained both
with and without this filtering.

Although this proton flux filtering is the only step that we take during our data processing that is
aimed at reducing SWCX contamination, we also take steps during our post-processing analysis aimed
at culling contaminated observations. For this, we take advantage of the results of
\citet{carter11}, who describe a method for identifying \xmm\ observations that are contaminated by
SWCX emission (see also \citealt{carter08}).  An observation is identified as SWCX-contaminated if
it exhibits variation in a narrow energy band spanning the \OVII\ and \OVIII\ lines that is
uncorrelated with variation in a higher-energy continuum-dominated band. This method is sensitive in
particular to geocoronal SWCX emission that varies significantly during the course of an
observation.

Rather than trying to emulate the \citeauthor{carter11} method, we will make use of Table~A.1 in
\citet{carter11}, which identifies 103 \xmm\ observations affected by geocoronal SWCX. When we
present our oxygen intensity measurements in Section~\ref{subsec:OxygenResults}, below, we will
compare the intensities extracted from the observations flagged by \citet{carter11} with those
extracted from our survey as a whole.  Later, in Section~\ref{sec:Discussion}, we will discuss
various aspects of our results. In certain cases, it will be desirable to reduce the effects of SWCX
contamination on our results. In such cases, we will exclude the observations identified by
\citet{carter11} as SWCX-contaminated, and will make it clear whether or not we are doing so.

\subsection{Spectra, Response Files, and Particle Background}
\label{subsec:Spectra}

For each exposure of each observation, we extracted an SXRB spectrum from the full field of view,
minus any excluded sources or CCDs. We extracted the spectra using the \esas\ \texttt{mos-spectra}
script. This script also created the redistribution matrix files (RMFs) and ancillary response files
(ARFs), using the SAS \texttt{rmfgen} and \texttt{arfgen} tasks, respectively. Note that, because we
extract the SXRB spectra from the full field of view, our survey is insensitive to structure in the
SXRB on scales smaller than $\sim$0.5\degr. Our sensitivity to SXRB structure on larger scales
depends on the location of the archival \xmm\ observations on the sky, and on our ability to remove
the time-variable SWCX contamination.

The extracted spectra consist of the true SXRB emission (including SWCX, Galactic, and extragalactic
emission), the quiescent particle background (QPB), and residual contamination from soft protons
impinging upon the detector that remains despite the cleaning of the data by
\texttt{mos-filter}. For each extracted spectrum, we subtracted the QPB spectrum, which we
calculated using the \esas\ \texttt{mos\_back} program.  The QPB spectra were constructed from a
database of filter-wheel-closed data, scaled using data from the regions of the MOS detector outside
the field of view (see \citealt{kuntz08a} for details). We dealt with the residual soft-proton
contamination by including an extra model component in the spectral analysis (see
Section~\ref{subsec:OxygenMethod}, below).

As noted in Section~\ref{subsec:BrightSourceRemoval}, we ignored CCDs that clearly exhibited the
anomalous state identified by \citet{kuntz08a}.  However, it was not always possible to clearly
identify anomalous CCDs from a visual inspection of the X-ray images. Originally, we used plots of
hardness ratio against count-rate for the unexposed corner pixels (which are produced by the
above-described processing) to identify previously unidentified anomalous CCDs (see
Paper~I). However, the latest version of \texttt{mos\_back} (included in SAS version 11.0.1) fails
if it identifies a CCD in an anomalous state (for some observations, \texttt{mos\_back} identified
anomalous CCDs that we had failed to identify in Paper~I). In either case, if a CCD was identified
as being in an anomalous state, we ignored that CCD, and re-ran the spectral extraction and QPB
calculation for the observation in question.

\section{OXYGEN LINE INTENSITIES}
\label{sec:OxygenIntensities}

In this section we present our SXRB oxygen intensity measurements. We describe our methods for
measuring the oxygen line intensities and for estimating the systematic errors in
Sections~\ref{subsec:OxygenMethod} and \ref{subsec:SystematicErrors}, respectively. The main set of
results is presented in Section~\ref{subsec:OxygenResults}, while in
Section~\ref{subsec:MultipleObs} we present results for directions that have multiple usable
\xmm\ observations. In the subsequent subsections we discuss any differences in our new results from
those in Paper~I, the possibility of contamination from bright sources and from soft protons, the
effect of the proton flux filtering described in Section~\ref{subsec:ProtonFluxFiltering}, and the
sky coverage of our survey.

\subsection{Measurement Method}
\label{subsec:OxygenMethod}

We measured the oxygen line intensities using
XSPEC\footnote{http://heasarc.gsfc.nasa.gov/docs/xanadu/xspec/} version 12.7.0. The method that we
used is almost identical to that used in Paper~I. The differences are how we modeled the
extragalactic background, and how we modeled the residual soft-proton contamination (see below). We
now also include estimates of the systematic errors associated with some of our spectral model
assumptions (see Section~\ref{subsec:SystematicErrors}, below).

For each observation, we fitted a multicomponent spectral model simultaneously to the
0.4--10.0~\kev\ QPB-subtracted spectra extracted from the usable exposures (in most cases, there was
one MOS1 spectrum and one MOS2 spectrum). This model included two $\delta$-functions to model the
\OVII\ and \OVIII\ \Kalpha\ emission. The \OVII\ line energy was a free parameter, while the
\OVIII\ energy was fixed at 0.6536~\kev\ (from APEC; \citealt{smith01a}). This method measures the
total observed oxygen line intensities, including foreground SWCX and/or LB emission, and attenuated
halo emission, in \lineunit\ (hereafter referred to as line units, \LU). The remaining Galactic line
emission was modeled with an absorbed APEC thermal plasma model\footnote{Note that this component of
  our model also includes some thermal continuum emission, but this is typically much fainter than
  the line emission.} \citep{smith01a} with the oxygen \Kalpha\ emission disabled (see
\citealt{lei09}). We also tried other spectral codes, to estimate the size of the systematic error
associated with our choice of code (see Section~\ref{subsec:SystematicErrors}, below). The
extragalactic background was modeled with an absorbed power-law with a photon index of 1.46
(\citealt{chen97}; see below for discussion of this component's normalization). From here on we
refer to this component as the extragalactic power-law (EPL). The APEC and EPL components were
attenuated using the XSPEC \texttt{phabs} model \citep{balucinska92,yan98}, with the absorbing
column \NH\ fixed at the \HI\ column density from the LAB \HI\ survey \citep{kalberla05} for the
direction of the \xmm\ observation being analyzed.  The model also included two Gaussians to model
the bright Al-K and Si-K instrumental lines at 1.49 and 1.74~\kev\ \citep{kuntz08a}. See Paper~I for
more details about these model components. We used \citet{anders89} abundances for the APEC and
\texttt{phabs} model components.

It is possible that, despite the cleaning described in Section~\ref{sec:DataReduction}, some
contamination from soft protons impinging upon the detector will remain in the extracted
spectra. This contamination manifests itself as excess emission over the EPL at higher energies.
Following advice in the \esas\ manual \citep{snowden11}, we modeled this residual contamination by
adding a power-law that was not folded through the instrumental response to the above-described
model. We set soft limits on the power-law index at 0.5 and 1.0, and hard limits at 0.2 and
1.3. This is in contrast to Paper~I, in which we used a broken power-law with a break at
3.2~\kev\ \citep{snowden07,kuntz08a} and no constraints on the spectral indices to model the
residual soft-proton contamination.  This change appeared to have little effect on the results. The
parameters of the soft-proton power-law were independent for each exposure.

The above-described soft-proton contamination makes it difficult to independently constrain the
normalization of the EPL, because there is some degeneracy between these two model components at
higher energies. In Paper~I, we simply fixed the normalization of the EPL at
10.5~\pownorm\ (throughout this paper, we specify the normalization of the EPL at 1~\kev). This
value was found by \citet{chen97} from a joint \rosat\ + \asca\ spectrum, after a few sources with
$\Fsource \ga 5 \times 10^{-14}~\flux$ were excised from the data. However, as discussed in Paper~I,
the X-ray source counts from \citet{moretti03} imply that the extragalactic background composed of
sources with $\Fsource < 5 \times 10^{-14}~\flux$ should have a lower normalization than this. The
integrated 0.5--2.0~\kev\ surface brightness due to sources with $\Fsource < 5 \times
10^{-14}~\flux$ is $5.46 \times 10^{-12}~\flux\ \mathrm{deg}^{-2}$ \citep{moretti03}. Assuming a
photon index of 1.46 \citep{chen97}, this surface brightness implies a normalization for the EPL of
7.9~\pownorm\ (Paper~I). In this paper, we used this lower value as the nominal value for the
normalization of the EPL.

To allow for field-to-field variation in the number of unresolved sources contributing to the
extragalactic background, we experimented with allowing the EPL normalization to be a free
parameter, with soft limits at $\pm$15\%\ and hard limits at $\pm$30\%\ of the nominal value of
7.9~\pownorm. To help break the degeneracy between the EPL and the soft-proton power-law, we
increased the upper energy limit for the spectral fitting from 5~\kev\ (Paper~I) to 10~\kev. This
should have helped because the EPL falls off more rapidly than the soft-proton power-law above
$\sim$5~\kev, owing to the former having been folded through the instrumental response while the
latter was not. However, we found that best-fit EPL normalizations had a tendency to cluster near
the soft limits, rather than being distributed approximately symmetrically about the nominal value
with a peak at the nominal value. We therefore opted to make the oxygen measurements with the EPL
normalization fixed at the nominal value. In the following subsection we describe how we estimated
the size of the systematic error associated with our assuming this normalization.

\subsection{Systematic Errors}
\label{subsec:SystematicErrors}

We estimated the systematic errors associated with some of the assumptions in our spectral model,
specifically, our use of an APEC model for the non-oxygen line emission, and our fixing the EPL
normalization at 7.9~\pownorm. We include these systematic errors alongside the statistical errors
in our tables of results, below. In Paper~I, we found that any systematic error associated with our
choice of EPL normalization was typically smaller than the statistical error on the intensity, but
we did not include estimates of the systematic errors in our tables of results. In Paper~I, we also
considered contamination by bright sources or by soft protons as possible sources of systematic
error -- we will discuss these in Section~\ref{subsec:Contamination}.

To estimate the systematic error associated with our use of an APEC model for the non-oxygen line
emission, we repeated the analysis using a \mekal\ model \citep{mewe95} or a \raymondsmith\ model in
place of the APEC model. We disabled the oxygen lines for these models by setting the oxygen
abundance in XSPEC to zero. Note that this method disables all oxygen emission from the model
plasma, whereas in our original method, described above, we disabled only the oxygen
\Kalpha\ emission. This means that, in these new measurements, the \OVIII\ line may be contaminated
by emission from \OVII\ \Kbeta\ at 0.666~\kev. However, as the intensity of the \OVII\ \Kbeta\ line
relative to the \OVIII\ \Lyalpha\ line is dependent upon the temperature, which is not always well
constrained in our models, it is possible that the \OVII\ \Kbeta\ emission is not accurately
accounted for in our original \OVIII\ \Lyalpha\ measurements either. Therefore, as well as
quantifying the differences between spectral codes, this method should give an estimate of the
uncertainty in the \OVIII\ \Lyalpha\ intensity due to the uncertainty in the
\OVII\ \Kbeta\ contamination.

We measured the oxygen intensities using the three different codes for all of our observations. For each
line in each observation, our estimate of the systematic error associated with our choice of spectral code is
\begin{eqnarray}
  \sigma_\mathrm{sys,code} = \max \big\{ |I(\mbox{APEC}) &-& I(\mbox{\mekal})|,           \nonumber \\
                                       |I(\mbox{APEC}) &-& I(\mbox{Raymond-Smith})| \big\},
\end{eqnarray}
where $I(\mbox{APEC})$ is the intensity measured using the APEC code, etc.

To estimate the systematic error associated with our fixing the normalization of the EPL, we first
used a Monte Carlo simulation to estimate the expected field-to-field variation in the number of
sources contributing to the extragalactic background, and thus the variation in the flux and
normalization of the EPL. Using source counts from \citet{moretti03} and assuming a power-law photon
index of 1.46 \citep{chen97}, we estimated that the EPL normalization should vary by
$\pm$0.8~\pownorm\ ($1\sigma$) from its nominal value of 7.9~\pownorm\ among our \xmm\ fields.

We therefore repeated the oxygen measurements for each observation with the EPL
normalization fixed at 7.1 and 8.7~\pownorm. For each line in each observation, our estimate of the
systematic error associated with our fixing the EPL normalization is
\begin{equation}
  \sigma_\mathrm{sys,EPL} = \max \big\{ |I(7.9) - I(7.1)|, |I(7.9) - I(8.7)| \big\},
\end{equation}
where $I(7.9)$ is the intensity measured with the EPL normalization fixed at 7.9~\pownorm, etc. We
combined $\sigma_\mathrm{sys,code}$ and $\sigma_\mathrm{sys,EPL}$ in quadrature to get our final
estimate of the systematic error, \sigmasys.

\subsection{Results}
\label{subsec:OxygenResults}

{
  \setlength{\tabcolsep}{2pt}
  \tabletypesize{\tiny}
  \begin{deluxetable*}{llrrccccccccc}
\tabletypesize{\tiny}
\tablewidth{0pt}
\tablecaption{\OVII\ and \OVIII\ Line Intensities (Without Proton Flux Filtering)\label{tab:OxygenIntensities1}}
\tablehead{
\colhead{Obs.~ID}                    & \colhead{Start}                      & \colhead{$l$}                        & \colhead{$b$}                        & \colhead{$t^\mathrm{exp}_1$}         & \colhead{$\Omega_1$}                 & \colhead{$t^\mathrm{exp}_2$}         & \colhead{$\Omega_2$}                 & \colhead{\Iovii}                     & \colhead{\Eovii}                     & \colhead{\Ioviii}                    & \colhead{$\langle\fsw\rangle$}       & \colhead{$\frac{\Ftotal}{\Fexgal}$}  \\
\colhead{}                           & \colhead{Date}                       & \colhead{(deg)}                      & \colhead{(deg)}                      & \colhead{(ks)}                       & \colhead{(arcmin$^2$)}               & \colhead{(ks)}                       & \colhead{(arcmin$^2$)}               & \colhead{(\LU)}                      & \colhead{(keV)}                      & \colhead{(\LU)}                      & \colhead{($10^8~\pcmsq~\ps$)}        & \colhead{}                           \\
\colhead{(1)}                        & \colhead{(2)}                        & \colhead{(3)}                        & \colhead{(4)}                        & \colhead{(5)}                        & \colhead{(6)}                        & \colhead{(7)}                        & \colhead{(8)}                        & \colhead{(9)}                        & \colhead{(10)}                       & \colhead{(11)}                       & \colhead{(12)}                       & \colhead{(13)}                       
}
\startdata
0146390201                           & 2003-03-29                           & 0.005                                & $-17.799$                            & 17.1                                 & 537                                  & 17.4                                 & 474                                  & $9.72 \pm 1.03 \pm 1.77$             & 0.5575                               & $1.13 \pm 0.49 \pm 0.60$             & 8.55                                 & 2.43                                 \\
0050940401                           & 2002-03-14                           & 0.015                                & $-12.011$                            & 8.4                                  & 583                                  & 9.2                                  & 589                                  & $35.17^{+1.55}_{-2.29} \pm 2.52$     & 0.5725                               & $15.68^{+1.16}_{-1.35} \pm 6.78$     & 1.14                                 & 2.64                                 \\
0402781001                           & 2007-01-20                           & 0.271                                & $+48.208$                            & 16.1                                 & 421                                  & 16.0                                 & 583                                  & $9.34^{+1.03}_{-1.10} \pm 1.72$      & 0.5725                               & $3.61^{+0.65}_{-0.67} \pm 1.39$      & 1.27                                 & 2.41                                 \\
0021540501                           & 2001-08-26                           & 0.426                                & $+48.796$                            & \nodata                              & \nodata                              & \nodata                              & \nodata                              & \nodata                              & \nodata                              & \nodata                              & \nodata                              & \nodata                              \\
0413780101                           & 2007-02-24                           & 0.604                                & $+11.378$                            & 23.0                                 & 453                                  & 23.4                                 & 542                                  & $3.29^{+0.96}_{-1.00} \pm 1.09$      & 0.5675                               & $2.06^{+0.75}_{-0.48} \pm 0.62$      & 2.70                                 & 1.72                                 \\
0202190301                           & 2004-05-20                           & 0.939                                & $-79.578$                            & 16.8                                 & 570                                  & 17.0                                 & 578                                  & $5.71^{+0.89}_{-0.84} \pm 0.18$      & 0.5675                               & $1.21^{+0.51}_{-0.46} \pm 0.42$      & 3.07                                 & 2.39                                 \\
0413780201                           & 2007-03-03                           & 0.960                                & $+10.788$                            & 25.4                                 & 500                                  & 25.0                                 & 590                                  & $4.77^{+0.87}_{-0.86} \pm 0.56$      & 0.5775                               & $4.30^{+0.57}_{-0.58} \pm 1.10$      & 1.91                                 & 1.68                                 \\
0413780301                           & 2007-03-07                           & 1.168                                & $+10.439$                            & 16.2                                 & 430                                  & 16.6                                 & 592                                  & $5.98^{+1.26}_{-1.53} \pm 0.42$      & 0.5725                               & $4.78^{+0.77}_{-0.86} \pm 0.84$      & 0.86                                 & 2.53                                 \\
0152750101                           & 2002-09-27                           & 1.540                                & $+7.099$                             & 38.7                                 & 580                                  & 39.0                                 & 589                                  & $8.43^{+0.40}_{-0.97} \pm 2.48$      & 0.5625                               & $6.42^{+0.43}_{-0.31} \pm 1.20$      & 3.27                                 & 2.50                                 \\
0556210501                           & 2008-07-19                           & 1.649                                & $+46.606$                            & 10.8                                 & 488                                  & 10.9                                 & 578                                  & $12.01^{+1.67}_{-1.22} \pm 0.94$     & 0.5675                               & $3.95^{+0.72}_{-0.71} \pm 1.26$      & 1.65                                 & 1.88                                 \\
0201903001                           & 2004-10-24                           & 2.718                                & $-56.160$                            & 17.4                                 & 352                                  & 17.8                                 & 296                                  & $5.97^{+1.51}_{-1.31} \pm 0.38$      & 0.5725                               & $0.99^{+0.74}_{-0.75} \pm 0.20$      & 3.51                                 & 2.65                                 \\
0205390101                           & 2005-05-01                           & 3.588                                & $-64.104$                            & 27.6                                 & 311                                  & 27.6                                 & 396                                  & $7.40^{+0.90}_{-0.96} \pm 0.49$      & 0.5675                               & $1.14^{+0.57}_{-0.51} \pm 0.72$      & 1.17                                 & 2.18                                 \\
0404910801                           & 2006-05-02                           & 3.966                                & $-59.441$                            & 19.9                                 & 332                                  & 19.4                                 & 340                                  & $6.27^{+1.87}_{-0.91} \pm 1.22$      & 0.5675                               & $0.70^{+0.88}_{-0.57} \pm 1.09$      & 3.04                                 & 1.89                                 \\
0018741701                           & 2001-05-03                           & 3.967                                & $-59.448$                            & 6.6                                  & 341                                  & 6.6                                  & 340                                  & $11.29 \pm 2.03 \pm 1.31$            & 0.5675                               & $5.49 \pm 1.23 \pm 0.55$             & 2.14                                 & 2.18                                 \\
0135980201                           & 2002-04-30                           & 4.693                                & $-64.122$                            & 25.5                                 & 551                                  & 25.8                                 & 559                                  & $11.70^{+0.50}_{-0.89} \pm 0.26$     & 0.5625                               & $2.77^{+0.42}_{-0.41} \pm 1.28$      & 2.86                                 & 1.65                                 \\
0148520101                           & 2003-07-22                           & 5.456                                & $+56.770$                            & 21.6                                 & 582                                  & 21.3                                 & 588                                  & $12.42^{+0.78}_{-0.82} \pm 1.51$     & 0.5675                               & $4.47^{+0.47}_{-0.55} \pm 2.52$      & 1.18                                 & 1.74                                 \\
0057560301                           & 2001-08-09                           & 5.457                                & $+56.766$                            & 36.6                                 & 587                                  & 36.1                                 & 514                                  & $13.99 \pm 0.63 \pm 1.11$            & 0.5675                               & $4.71 \pm 0.39 \pm 2.67$             & 1.77                                 & 1.69                                 \\
0302580501                           & 2005-06-13                           & 5.676                                & $-77.683$                            & 36.1                                 & 394                                  & 37.3                                 & 481                                  & $8.46^{+1.33}_{-0.73} \pm 0.85$      & 0.5675                               & $2.01^{+0.59}_{-0.40} \pm 1.11$      & 2.10                                 & 2.32                                 \\
\enddata
\tablecomments{This table is available in its entirety in a machine-readable form in the online journal. A portion is shown here for guidance regarding its form and content.}
\end{deluxetable*}

  \begin{deluxetable*}{llrrcccccccccc}
\tabletypesize{\tiny}
\tablewidth{0pt}
\tablecaption{\OVII\ and \OVIII\ Line Intensities (With Proton Flux Filtering)\label{tab:OxygenIntensities2}}
\tablehead{
\colhead{Obs.~ID}                    & \colhead{Start}                      & \colhead{$l$}                        & \colhead{$b$}                        & \colhead{$t^\mathrm{exp}_1$}         & \colhead{$\Omega_1$}                 & \colhead{$t^\mathrm{exp}_2$}         & \colhead{$\Omega_2$}                 & \colhead{\Iovii}                     & \colhead{\Eovii}                     & \colhead{\Ioviii}                    & \colhead{$\langle\fsw\rangle$}       & \colhead{$\frac{\Ftotal}{\Fexgal}$}  & \colhead{Affected by}                \\
\colhead{}                           & \colhead{Date}                       & \colhead{(deg)}                      & \colhead{(deg)}                      & \colhead{(ks)}                       & \colhead{(arcmin$^2$)}               & \colhead{(ks)}                       & \colhead{(arcmin$^2$)}               & \colhead{(\LU)}                      & \colhead{(keV)}                      & \colhead{(\LU)}                      & \colhead{($10^8~\pcmsq~\ps$)}        & \colhead{}                           & \colhead{filtering?}                 \\
\colhead{(1)}                        & \colhead{(2)}                        & \colhead{(3)}                        & \colhead{(4)}                        & \colhead{(5)}                        & \colhead{(6)}                        & \colhead{(7)}                        & \colhead{(8)}                        & \colhead{(9)}                        & \colhead{(10)}                       & \colhead{(11)}                       & \colhead{(12)}                       & \colhead{(13)}                       & \colhead{(14)}                       
}
\startdata
0146390201                           & 2003-03-29                           & 0.005                                & $-17.799$                            & \nodata                              & \nodata                              & \nodata                              & \nodata                              & \nodata                              & \nodata                              & \nodata                              & \nodata                              & \nodata                              & \nodata                              \\
0050940401                           & 2002-03-14                           & 0.015                                & $-12.011$                            & 8.4                                  & 583                                  & 9.2                                  & 589                                  & $35.17^{+1.55}_{-2.29} \pm 2.52$     & 0.5725                               & $15.68^{+1.16}_{-1.35} \pm 6.78$     & 1.14                                 & 2.64                                 & N                                    \\
0402781001                           & 2007-01-20                           & 0.271                                & $+48.208$                            & 16.1                                 & 421                                  & 16.0                                 & 583                                  & $9.34^{+1.03}_{-1.10} \pm 1.72$      & 0.5725                               & $3.61^{+0.65}_{-0.67} \pm 1.39$      & 1.27                                 & 2.41                                 & N                                    \\
0021540501                           & 2001-08-26                           & 0.426                                & $+48.796$                            & 11.0                                 & 294                                  & 11.1                                 & 354                                  & $17.58^{+2.02}_{-2.09} \pm 0.92$     & 0.5675                               & $6.19^{+1.18}_{-1.21} \pm 1.10$      & 1.63                                 & 2.69                                 & \nodata                              \\
0413780101                           & 2007-02-24                           & 0.604                                & $+11.378$                            & \nodata                              & \nodata                              & \nodata                              & \nodata                              & \nodata                              & \nodata                              & \nodata                              & \nodata                              & \nodata                              & \nodata                              \\
0202190301                           & 2004-05-20                           & 0.939                                & $-79.578$                            & \nodata                              & \nodata                              & \nodata                              & \nodata                              & \nodata                              & \nodata                              & \nodata                              & \nodata                              & \nodata                              & \nodata                              \\
0413780201                           & 2007-03-03                           & 0.960                                & $+10.788$                            & 15.0                                 & 500                                  & 14.8                                 & 590                                  & $4.65^{+1.08}_{-1.02} \pm 0.99$      & 0.5775                               & $4.44^{+0.75}_{-0.74} \pm 1.33$      & 1.61                                 & 1.64                                 & Y                                    \\
0413780301                           & 2007-03-07                           & 1.168                                & $+10.439$                            & 16.2                                 & 430                                  & 16.6                                 & 592                                  & $5.98^{+1.26}_{-1.53} \pm 0.42$      & 0.5725                               & $4.78^{+0.77}_{-0.86} \pm 0.84$      & 0.86                                 & 2.53                                 & N                                    \\
0152750101                           & 2002-09-27                           & 1.540                                & $+7.099$                             & 12.6                                 & 580                                  & 12.7                                 & 589                                  & $6.11^{+1.14}_{-1.17} \pm 2.80$      & 0.5625                               & $5.43^{+0.63}_{-0.64} \pm 1.62$      & 1.82                                 & 2.49                                 & Y                                    \\
0556210501                           & 2008-07-19                           & 1.649                                & $+46.606$                            & 10.8                                 & 488                                  & 10.9                                 & 578                                  & $12.01^{+1.67}_{-1.22} \pm 0.94$     & 0.5675                               & $3.95^{+0.72}_{-0.71} \pm 1.26$      & 1.65                                 & 1.88                                 & N                                    \\
0201903001                           & 2004-10-24                           & 2.718                                & $-56.160$                            & \nodata                              & \nodata                              & \nodata                              & \nodata                              & \nodata                              & \nodata                              & \nodata                              & \nodata                              & \nodata                              & \nodata                              \\
0205390101                           & 2005-05-01                           & 3.588                                & $-64.104$                            & 27.6                                 & 311                                  & 27.6                                 & 396                                  & $7.40^{+0.90}_{-0.96} \pm 0.49$      & 0.5675                               & $1.14^{+0.57}_{-0.51} \pm 0.72$      & 1.17                                 & 2.18                                 & N                                    \\
0404910801                           & 2006-05-02                           & 3.966                                & $-59.441$                            & \nodata                              & \nodata                              & \nodata                              & \nodata                              & \nodata                              & \nodata                              & \nodata                              & \nodata                              & \nodata                              & \nodata                              \\
0018741701                           & 2001-05-03                           & 3.967                                & $-59.448$                            & \nodata                              & \nodata                              & \nodata                              & \nodata                              & \nodata                              & \nodata                              & \nodata                              & \nodata                              & \nodata                              & \nodata                              \\
0135980201                           & 2002-04-30                           & 4.693                                & $-64.122$                            & \nodata                              & \nodata                              & \nodata                              & \nodata                              & \nodata                              & \nodata                              & \nodata                              & \nodata                              & \nodata                              & \nodata                              \\
0148520101                           & 2003-07-22                           & 5.456                                & $+56.770$                            & 21.6                                 & 582                                  & 21.3                                 & 588                                  & $12.42^{+0.78}_{-0.82} \pm 1.51$     & 0.5675                               & $4.47^{+0.47}_{-0.55} \pm 2.52$      & 1.18                                 & 1.74                                 & N                                    \\
0057560301                           & 2001-08-09                           & 5.457                                & $+56.766$                            & 27.1                                 & 587                                  & 26.7                                 & 514                                  & $13.69^{+0.72}_{-0.76} \pm 0.28$     & 0.5725                               & $5.07 \pm 0.46 \pm 2.36$             & 1.61                                 & 1.67                                 & Y                                    \\
0302580501                           & 2005-06-13                           & 5.676                                & $-77.683$                            & 20.0                                 & 394                                  & 20.7                                 & 481                                  & $9.70^{+1.13}_{-0.98} \pm 0.43$      & 0.5625                               & $1.60^{+0.55}_{-0.50} \pm 1.05$      & 1.43                                 & 2.51                                 & Y                                    \\
\enddata
\tablecomments{This table is available in its entirety in a machine-readable form in the online journal. A portion is shown here for guidance regarding its form and content.}
\end{deluxetable*}

}

Tables~\ref{tab:OxygenIntensities1} and \ref{tab:OxygenIntensities2} contain the oxygen intensity
measurements, sorted by increasing Galactic longitude, $l$. The results in
Table~\ref{tab:OxygenIntensities1} were obtained without the proton flux filtering described in
Section~\ref{subsec:ProtonFluxFiltering}, while those in Table~\ref{tab:OxygenIntensities2} were
obtained with this additional filtering.

The columns in Tables~\ref{tab:OxygenIntensities1} and \ref{tab:OxygenIntensities2} contain the
following data.
Column~1 contains the \xmm\ observation ID.
Column~2 contains the observation start date.
Columns~3 and 4 contain the Galactic coordinates ($l,b$) of the observation pointing direction.
Column~5 contains the usable MOS1 exposure time, and Column~6 contains the solid angle
of the MOS1 detector from which the SXRB spectrum was extracted. If an observation had more than
one usable MOS1 exposure, the values for each exposure are presented as a comma-separated list.
Columns~7 and 8 contain the corresponding data for the MOS2 camera.
Column~9 contains the \OVII\ line intensity, \Iovii, measured from the SXRB spectrum. Note that
this is the observed intensity -- it has not been subjected to any deabsorption.
The first error is the $1\sigma$ statistical error from the fitting, \sigmastat, calculated with
XSPEC's \texttt{error} command using $\chi^2_\mathrm{min} + 2.30$ for two interesting parameters
\citep{lampton76,avni76}.  For a small number of observations, XSPEC's \texttt{error} command was
unable to calculate the statistical error. For such observations, we estimated the error
by varying the intensity from its best-fit value with the \texttt{steppar} command.
The second error is the estimated systematic error, \sigmasys, described above. For a small
number of observations, the different spectral codes and extragalactic normalizations consistently
yield oxygen intensities of 0~\LU. In such cases, the systematic error is quoted as zero.
Column~10 contains the photon energy of the $\delta$-function used to model the \OVII\ emission, \Eovii,
rounded to the central energy of the 5-eV wide RMF bin in which the line lies (see Paper~I).
Column~11 contains the \OVIII\ line intensity, again with \sigmastat\ and \sigmasys\
shown separately (note that the energy of this line was fixed at 0.6536~\kev).
Column~12 contains the average solar wind proton flux at the Earth, $\langle\fsw\rangle$, during the \xmm\ observation.
These data are from OMNIWeb. If there were no good solar wind data during the observation, this field is blank.
Column~13 contains the ratio of the total model flux in the 2--5~\kev\ band, \Ftotal, to the
2--5~\kev\ flux of the EPL, \Fexgal. We introduced this ratio in Paper~I as a measure of the
contamination of the count-rate due to soft protons impinging upon the detector -- any observation
for which this ratio exceeded 2 was rejected.  As we are using a lower nominal EPL normalization in
this paper (7.9 versus 10.5~\pownorm), we have increased the rejection threshold for
$\Ftotal/\Fexgal$ by a factor of $10.5/7.9$ to 2.7.  These additional rejections reduced the number
of usable observations from \marker{2611} to \marker{1868} without proton flux filtering, and from
\marker{1435} to \marker{1003} with proton flux filtering.
Column~14 (Table~\ref{tab:OxygenIntensities2} only) indicates whether or not the observation was affected
by the proton flux filtering described in Section~\ref{subsec:ProtonFluxFiltering}.

It should be noted that the \marker{1003} observations that yielded oxygen line intensities after the
proton flux filtering are not a strict subset of the \marker{1868} observations that yielded oxygen
line intensities without this filtering. \marker{Twelve} observations failed the requirement that
$\Ftotal/\Fexgal \le 2.7$ without the proton flux filtering, and so were rejected, but passed this
requirement with the proton flux filtering. These \marker{12} observations are included in
Table~\ref{tab:OxygenIntensities1}, but the entries in Columns~5 through 13 are blank. Similarly,
the \marker{877} observations that yielded oxygen intensities without the proton flux filtering but
failed to do so with this filtering are included in Table~\ref{tab:OxygenIntensities2}, but with
blank entries in Column~5 onward (of these \marker{877} observations, \marker{855} had insufficient
exposure after the proton flux filtering, and \marker{22} failed the $\Ftotal/\Fexgal \le 2.7$
requirement after the filtering). Thus, both Tables~\ref{tab:OxygenIntensities1} and
\ref{tab:OxygenIntensities2} contain the complete set of \marker{1880} \xmm\ observations that
yielded oxygen intensities.

\begin{figure*}
\centering
\plotone{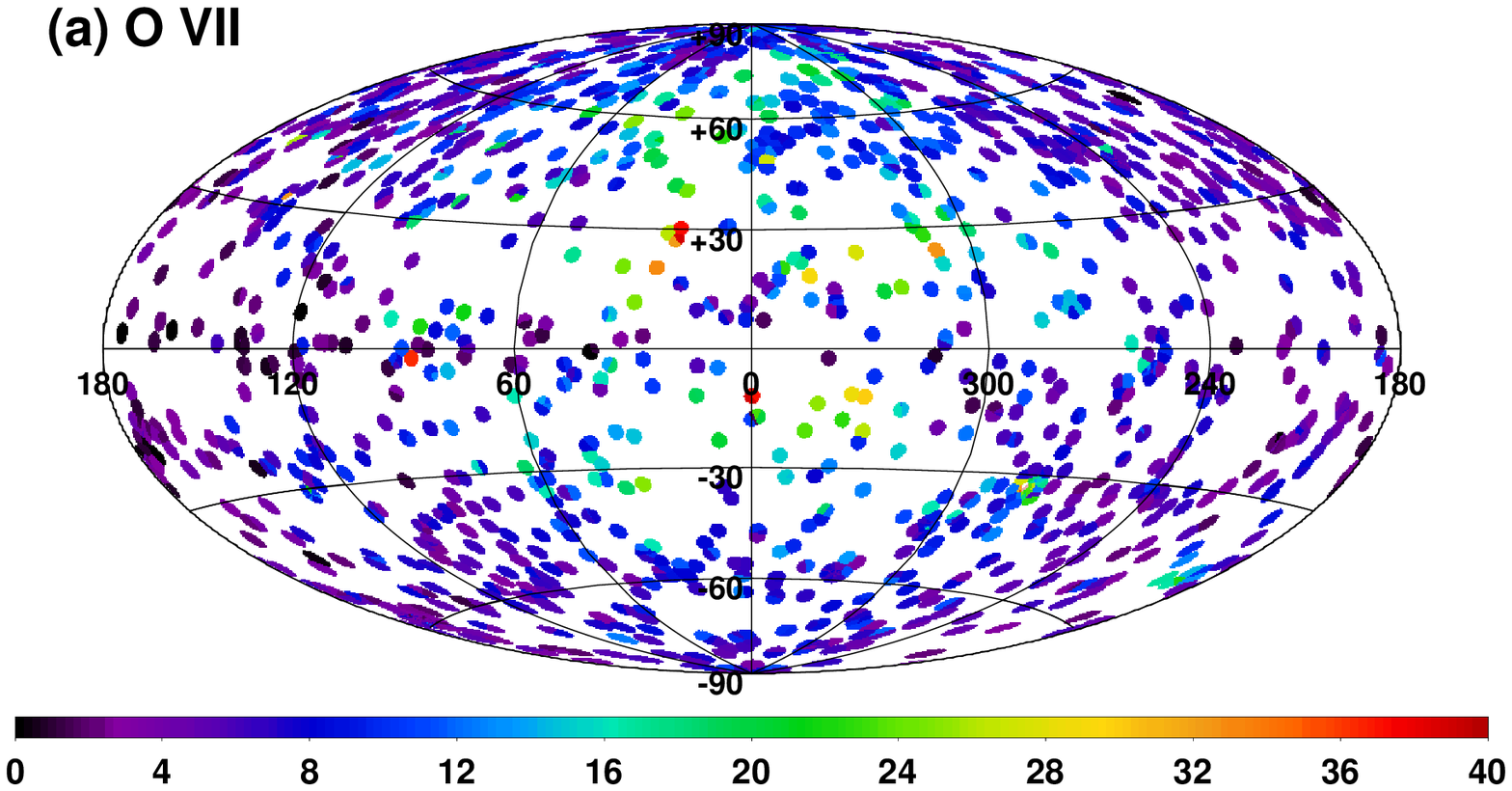}
\plotone{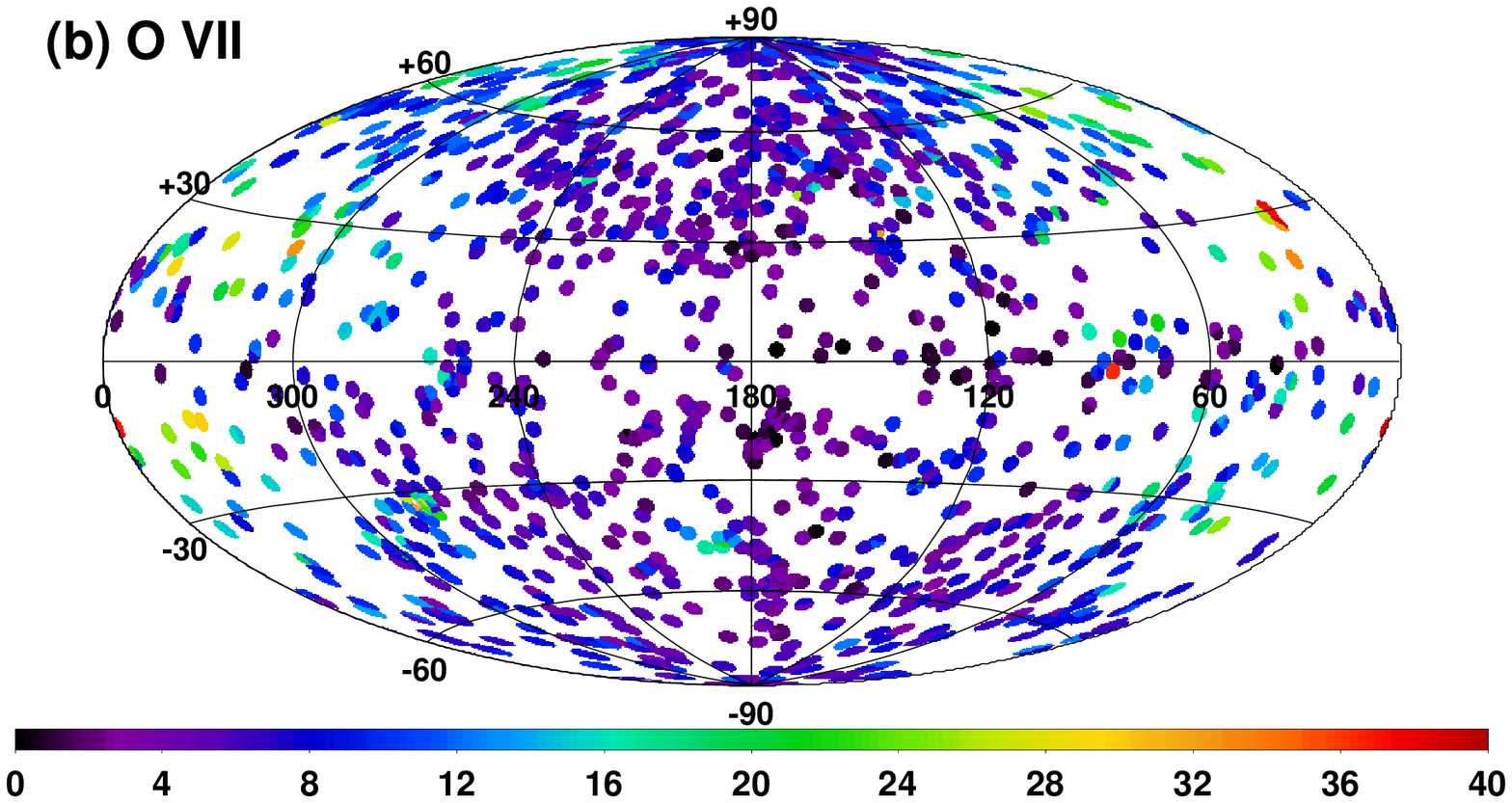}
\caption{All-sky maps of the oxygen intensities in \LU, plotted in Galactic coordinates. Panels (a)
  and (b) show the \OVII\ intensities obtained without the solar wind proton flux filtering
  described in Section~\ref{subsec:ProtonFluxFiltering}, in Hammer-Aitoff projections centered on
  the Galactic Center and the Galactic Anticenter, respectively. Panels (c) and (d) show the
  \OVIII\ intensities obtained without proton flux filtering in the same projections.  Panels (e)
  through (h) show the corresponding intensities obtained with proton flux filtering. Each intensity
  measurement is plotted as a colored circle of radius 2\degr\ centered on the \xmm\ pointing
  direction. If a point on the sky is within 2\degr\ of more than one \xmm\ observation, it is
  colored according to the nearest observation.
  \label{fig:Maps}}
\end{figure*}

\begin{figure*}
\figurenum{\ref{fig:Maps}}
\centering
\plotone{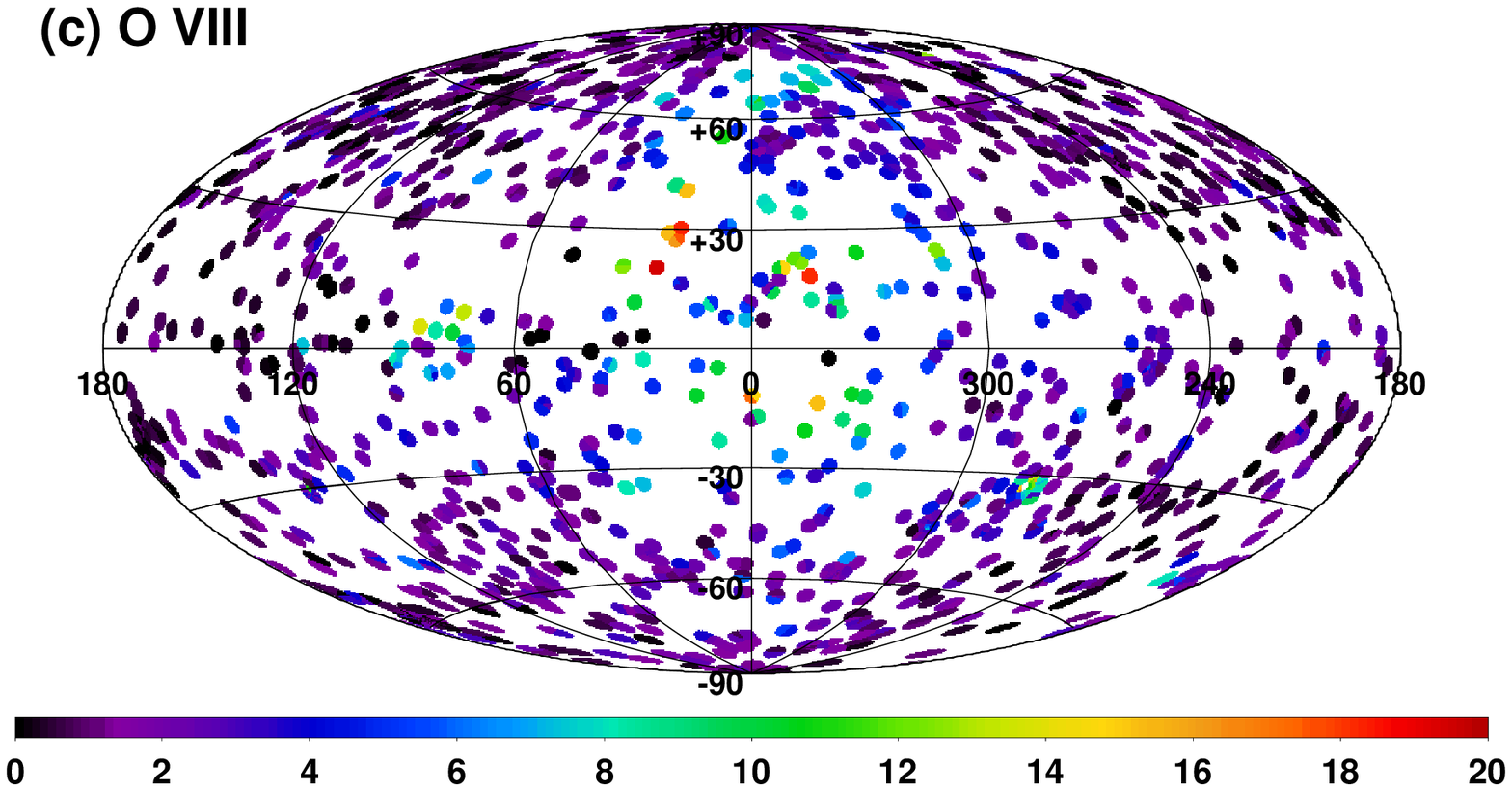}
\plotone{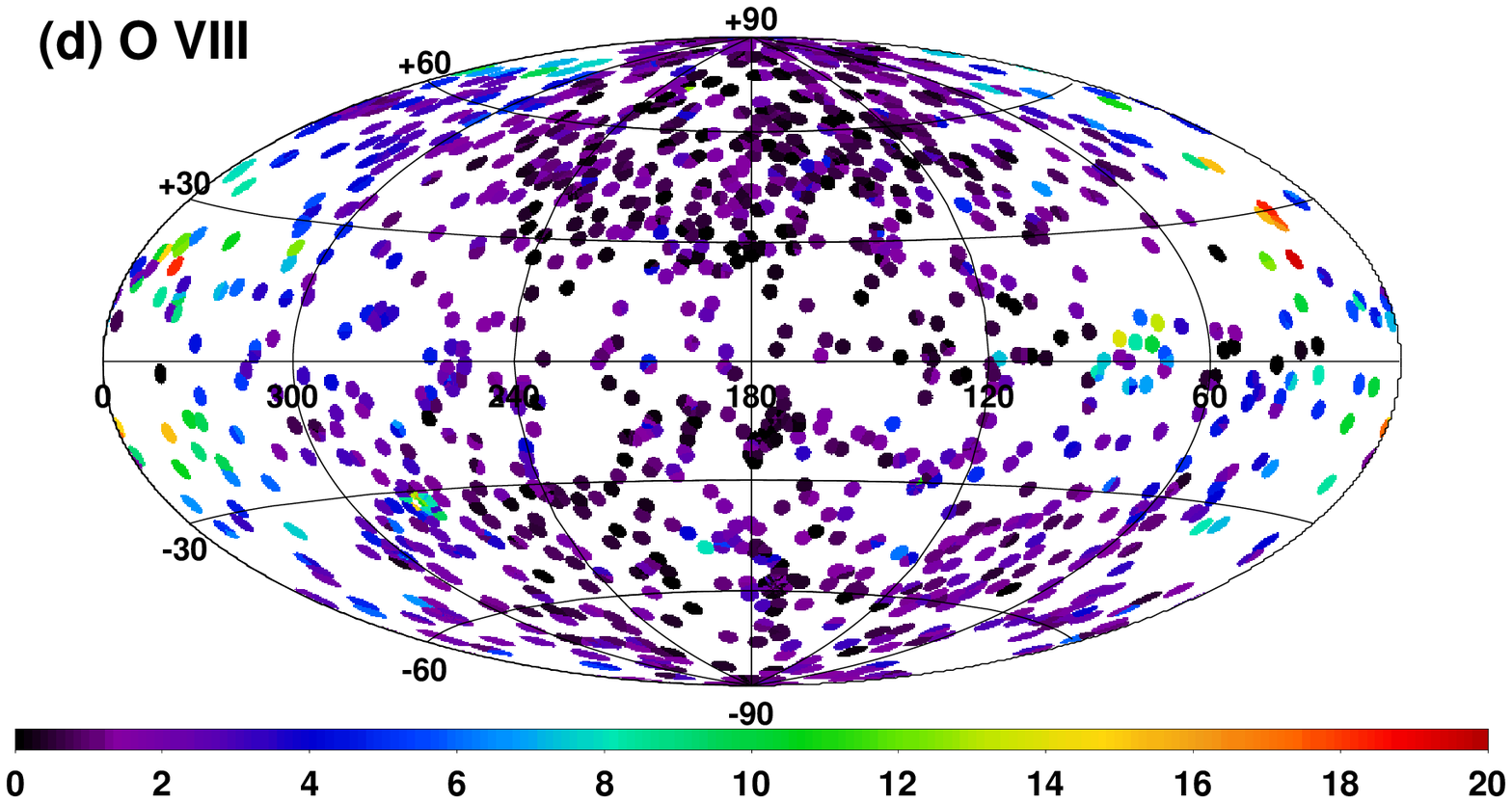}
\caption{\textit{Continued}}
\end{figure*}

\begin{figure*}
\figurenum{\ref{fig:Maps}}
\centering
\plotone{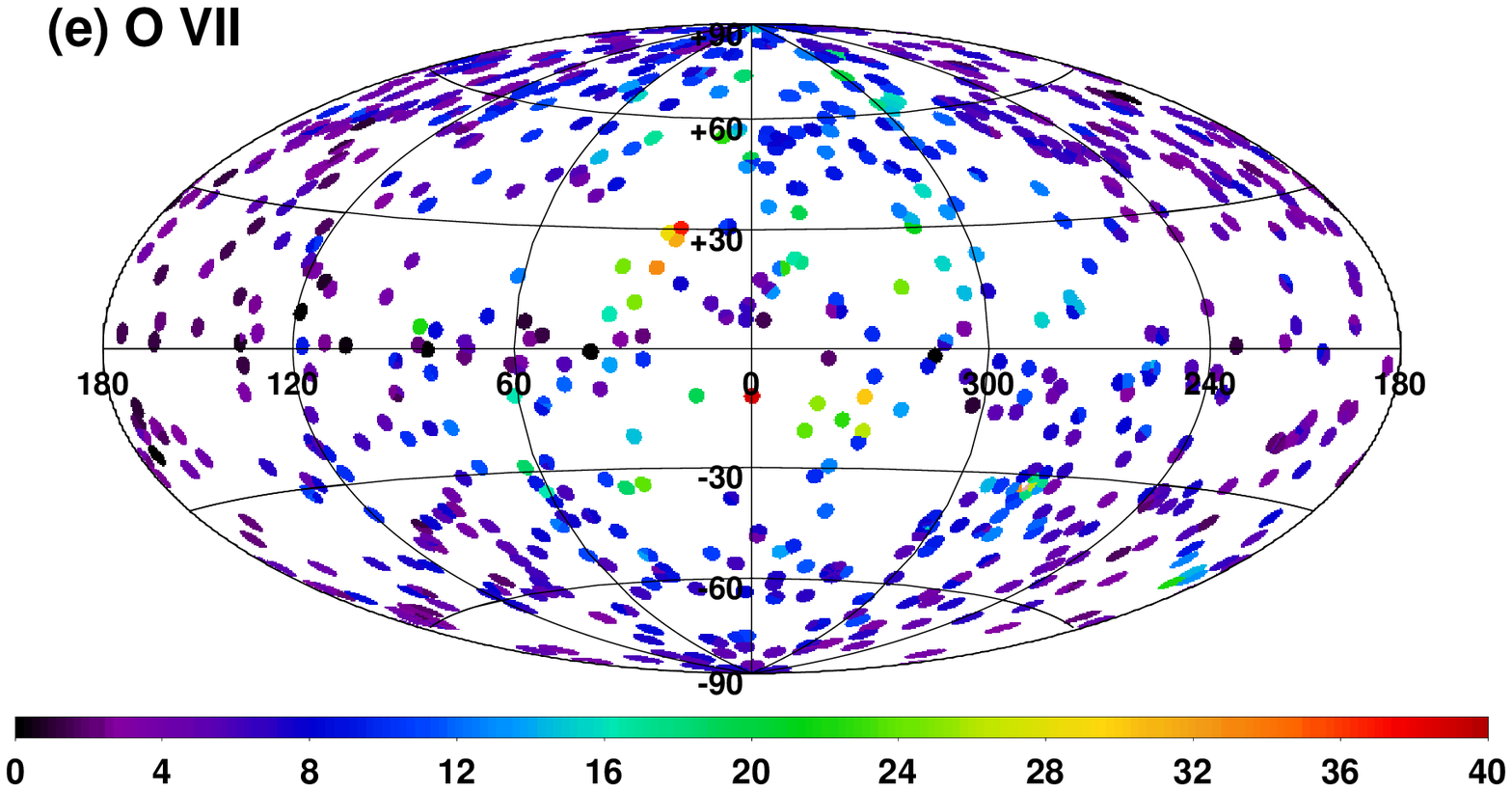}
\plotone{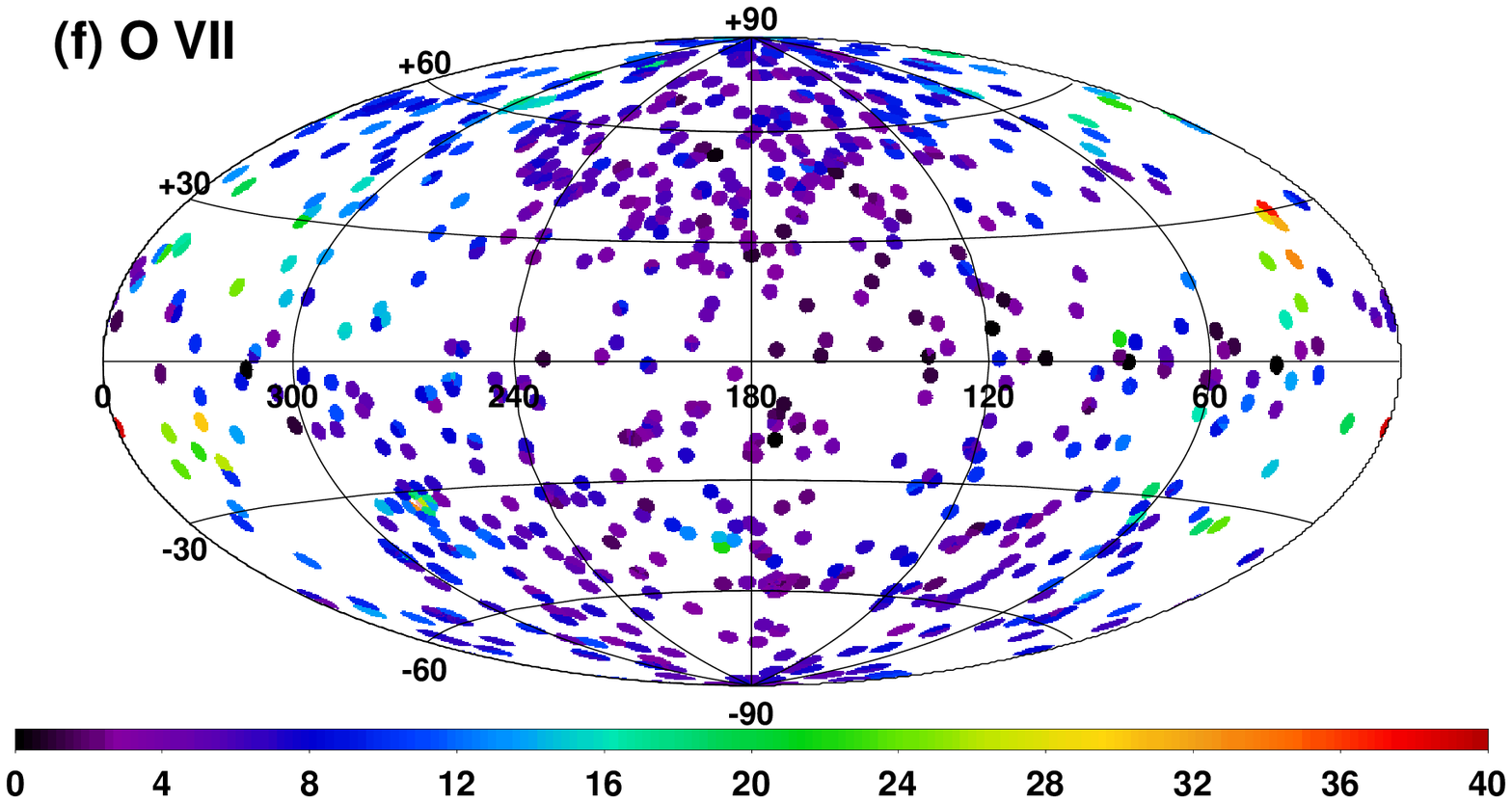}
\caption{\textit{Continued}}
\end{figure*}

\begin{figure*}
\figurenum{\ref{fig:Maps}}
\centering
\plotone{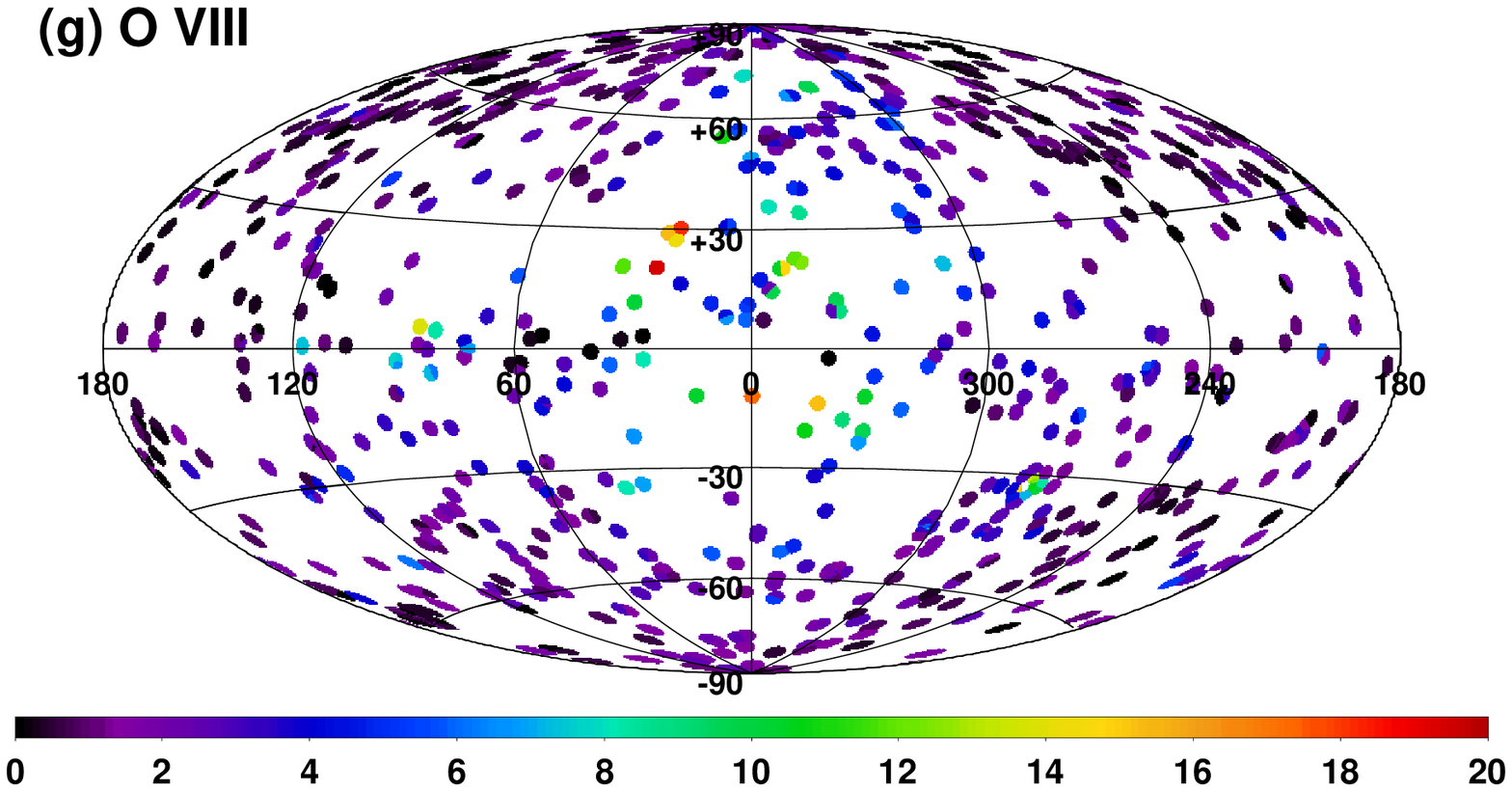}
\plotone{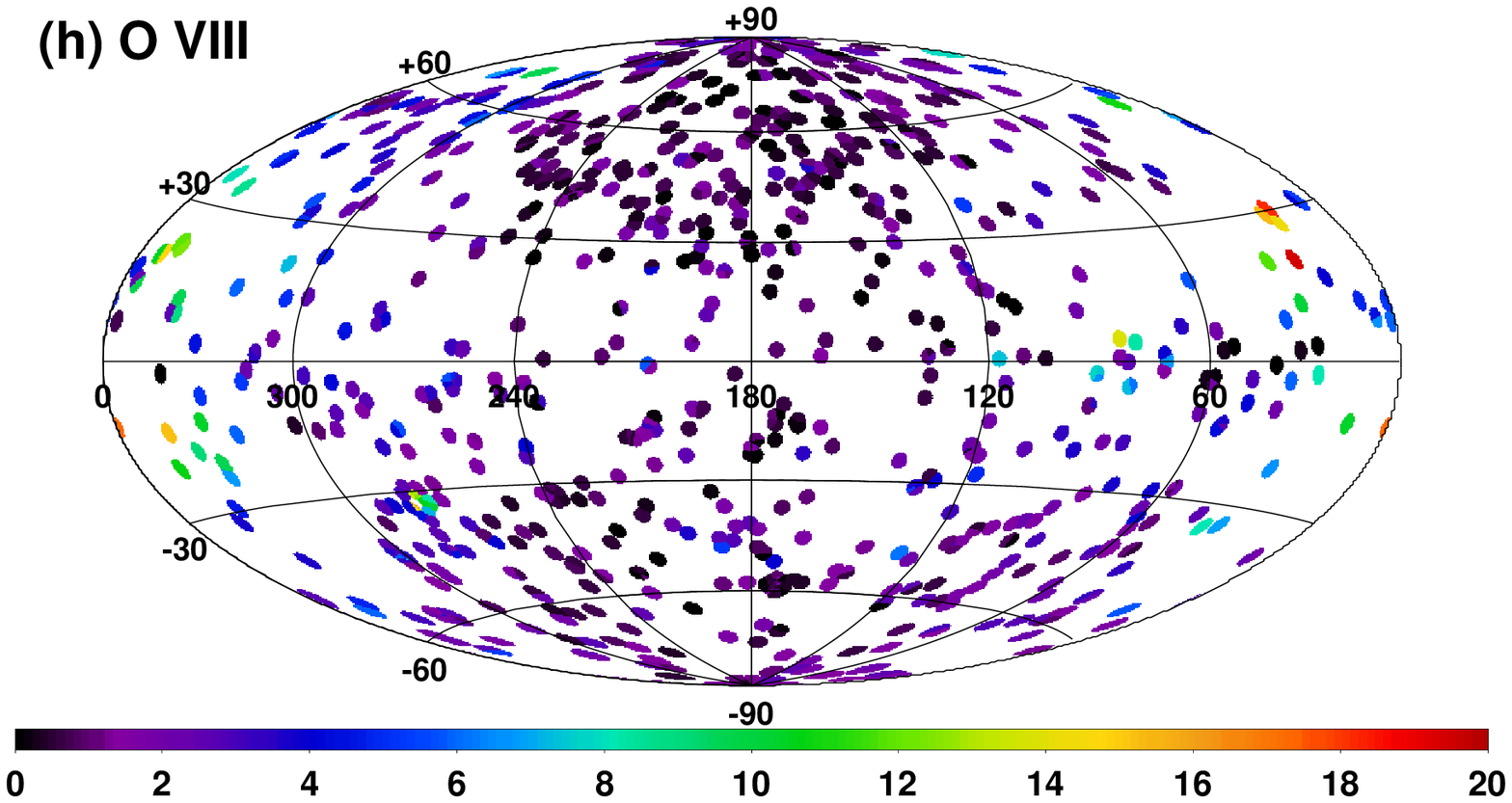}
\caption{\textit{Continued}}
\end{figure*}

Figure~\ref{fig:Maps} shows all-sky maps of the \OVII\ and \OVIII\ intensities, both without (panels
(a) through (d)) and with (panels (e) through (h)) proton flux filtering. For each line, we show
projections centered on the Galactic Center and on the Galactic Anticenter. Each colored circle in
the maps represents an SXRB line intensity measurement. Note that these circles are larger than the
\xmm\ field of view ($\mathrm{radius} = 14\arcmin$). See Section~\ref{subsec:SkyCoverage} for a
discussion of the sky coverage of our survey. We will discuss the variation in the lines'
intensities over the sky in more detail in Section~\ref{subsec:SpatialVariation}, but we note here
that the diffuse oxygen emission tends to be brighter toward the Galactic Center ($l < 90\degr$ or
$l > 270\degr$) than toward the Galactic Anticenter.

\begin{figure}
\centering
\plotone{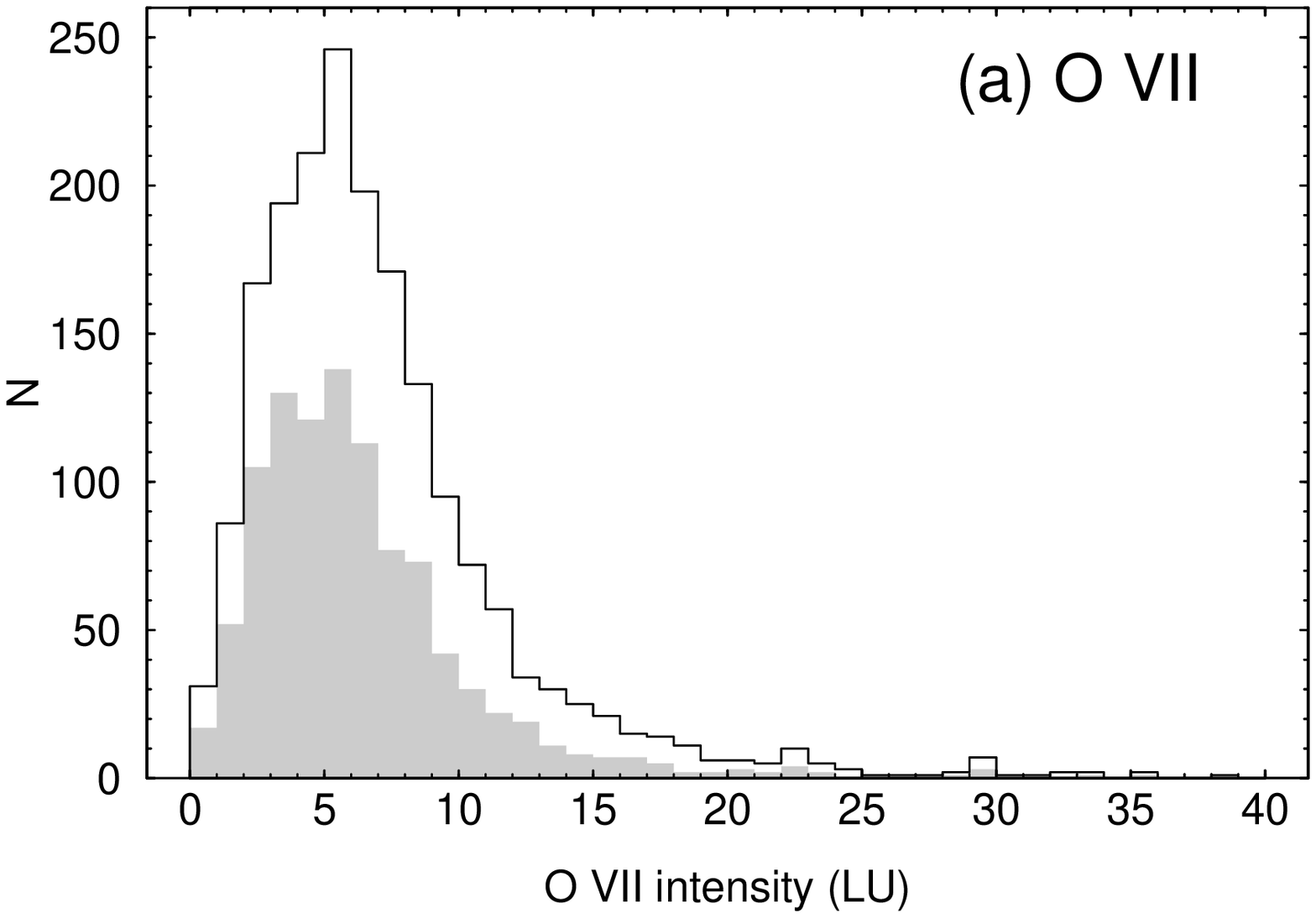}
\plotone{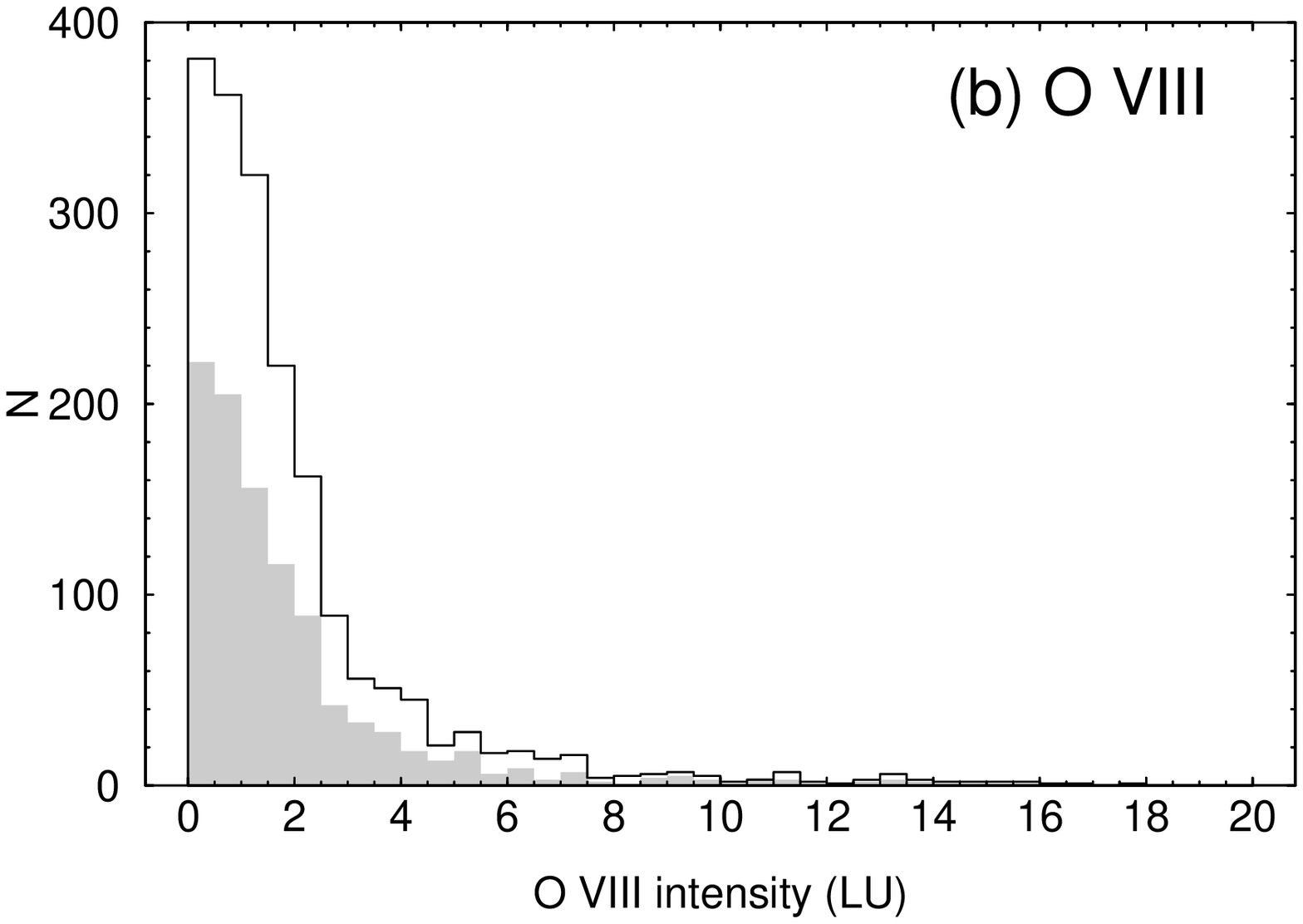}
\caption{Histograms of the (a) \OVII\ and (b) \OVIII\ intensities.  The solid black lines show the
  histograms of intensities obtained without the solar wind proton flux filtering described in
  Section~\ref{subsec:ProtonFluxFiltering}, and the gray areas show the histograms of intensities
  obtained with this filtering.
  \label{fig:IntensityHistogram}}
\end{figure}

Figure~\ref{fig:IntensityHistogram} shows histograms of the \OVII\ and \OVIII\ intensities, both
without (solid black lines) and with (gray area) the proton flux
filtering. Table~\ref{tab:Quartiles} contains the ranges and quartiles of the intensities.  The
quartiles (including the medians) of the oxygen intensities are generally significantly higher than
those in Paper~I (see Table~3 of that paper). This is because this paper includes the region around
the Galactic Center, where the intensities are generally higher (see Figure~\ref{fig:Maps}).  The
quartiles of the \OVII\ intensities obtained without the proton flux filtering are systematically
higher than the corresponding quartiles obtained with the proton flux filtering.  For \OVIII, there
are no significant differences in the intensity quartiles with and without the proton flux
filtering. We will discuss the effects of proton flux filtering in more detail in
Section~\ref{subsec:ProtonFluxFilteringEffects}.

\begin{deluxetable*}{lccccc}
\tablewidth{0pt}
\tablecaption{Ranges and Quartiles of the Oxygen Intensities\label{tab:Quartiles}}
\tablehead{
\colhead{Line}  & \colhead{Proton flux} & \colhead{Range}       & \colhead{Lower quartile} & \colhead{Median}         & \colhead{Upper quartile}      \\
                & \colhead{filtering?}  & \colhead{(\LU)}       & \colhead{(\LU)}          & \colhead{(\LU)}          & \colhead{(\LU)}
}
\startdata
\OVII           & N                     & 0.00--50.76           & 3.92 (3.78,4.15)         & 6.00 (5.90,6.14)         & 8.74 (8.54,8.97)          \\
\OVII           & Y                     & 0.00--48.81           & 3.61 (3.48,3.74)         & 5.60 (5.36,5.80)         & 7.95 (7.64,8.29)          \\
\OVIII          & N                     & 0.00--36.24           & 0.62 (0.59,0.66)         & 1.30 (1.24,1.36)         & 2.36 (2.28,2.45)          \\
\OVIII          & Y                     & 0.00--24.45           & 0.56 (0.52,0.61)         & 1.24 (1.17,1.32)         & 2.29 (2.19,2.39)          \\
\enddata
\tablecomments{The numbers in parentheses are the 90\%\ confidence intervals, calculated by bootstrapping.}
\end{deluxetable*}

We will conclude this subsection by looking at the oxygen intensities from the 103 observations
identified as being contaminated by geocoronal SWCX emission by \citet[Table~A.1]{carter11}.  Of
these observations, \marker{61} are in our final sample. Of the other \marker{42}, \marker{20} were
rejected at an early stage of the processing for a variety of reasons (e.g., the target source being
too bright), and \marker{22} failed the $\Ftotal/\Fexgal \le 2.7$ requirement.  For the remaining
\marker{61} observations, the median \OVII\ and \OVIII\ intensities (obtained without proton flux
filtering) are 9.83 and 2.19~\LU, respectively, with interquartile ranges of 7.60--13.63 and
1.07--3.18~\LU, respectively. Comparing these values with the medians and quartiles in
Table~\ref{tab:Quartiles}, we see that, unsurprisingly, the SWCX-contaminated observations
identified by \citet{carter11} yield systematically higher oxygen intensities than our survey as a
whole.

\subsection{Directions with Multiple Observations}
\label{subsec:MultipleObs}

Tables~\ref{tab:OxygenIntensities1} and \ref{tab:OxygenIntensities2} include many directions that
have been observed multiple times with \xmm\ (within a few arcmin). The times between observations
of a given direction may range from $\sim$1 day to several years. Multiple observations in the same
direction are useful for constraining models of SWCX, as only the SWCX component of the SXRB is
expected to vary on such a small timescale.

As \xmm\ observations of the same target rarely have identical pointing directions, we searched the
\marker{1868} good observations in Table~\ref{tab:OxygenIntensities1} for sets of observations whose pointing
directions are within 0.1\degr\ of each other (cf.\ the \xmm\ field of view is
$\approx$0.5\degr\ across). We found \marker{217} such sets.

{
  \setlength{\tabcolsep}{4pt}
  \tabletypesize{\tiny}
  \begin{deluxetable*}{ccccccccccc}
\tabletypesize{\tiny}
\tablewidth{0pt}
\tablecaption{Oxygen Line Intensities from Directions with Multiple Observations\label{tab:MultipleObs}}
\tablehead{
&&& \colhead{Paper I} &&& \multicolumn{2}{c}{Without proton flux filtering} && \multicolumn{2}{c}{With proton flux filtering} \\
\cline{7-8} \cline{10-11}
\colhead{Dataset}            & \colhead{$N_\mathrm{obs}$}   & \colhead{Obs.~ID}            & \colhead{Dataset}            & \colhead{$l$}                & \colhead{$b$}                & \colhead{\Iovii}             & \colhead{\Ioviii}            && \colhead{\Iovii}             & \colhead{\Ioviii}            \\
\colhead{}                   & \colhead{}                   & \colhead{}                   & \colhead{}                   & \colhead{(deg)}              & \colhead{(deg)}              & \colhead{(\LU)}              & \colhead{(\LU)}              && \colhead{(\LU)}              & \colhead{(\LU)}              \\
\colhead{(1)}                & \colhead{(2)}                & \colhead{(3)}                & \colhead{(4)}                & \colhead{(5)}                & \colhead{(6)}                & \colhead{(7)}                & \colhead{(8)}                && \colhead{(9)}                & \colhead{(10)}               
}
\startdata
1                            & 2                            & 0050940401                   & \nodata                      & 0.015                        & $-12.011$                    & $35.17^{+1.55}_{-2.29} \pm 2.52$ & $15.68^{+1.16}_{-1.35} \pm 6.78$& & $35.17^{+1.55}_{-2.29} \pm 2.52$ & $15.68^{+1.16}_{-1.35} \pm 6.78$\\
                             &                              & 0203040401                   & \nodata                      & 359.985                      & $-11.990$                    & $33.54 \pm 1.79 \pm 0.53$    & $13.35 \pm 1.08 \pm 5.71$   & & \nodata                      & \nodata                     \\
2                            & 2                            & 0018741701                   & \nodata                      & 3.967                        & $-59.448$                    & $11.29 \pm 2.03 \pm 1.31$    & $5.49 \pm 1.23 \pm 0.55$    & & \nodata                      & \nodata                     \\
                             &                              & 0404910801                   & \nodata                      & 3.966                        & $-59.441$                    & $6.27^{+1.87}_{-0.91} \pm 1.22$ & $0.70^{+0.88}_{-0.57} \pm 1.09$& & \nodata                      & \nodata                     \\
3                            & 2                            & 0057560301                   & \nodata                      & 5.457                        & $+56.766$                    & $13.99 \pm 0.63 \pm 1.11$    & $4.71 \pm 0.39 \pm 2.67$    & & $13.69^{+0.72}_{-0.76} \pm 0.28$ & $5.07 \pm 0.46 \pm 2.36$    \\
                             &                              & 0148520101                   & \nodata                      & 5.456                        & $+56.770$                    & $12.42^{+0.78}_{-0.82} \pm 1.51$ & $4.47^{+0.47}_{-0.55} \pm 2.52$& & $12.42^{+0.78}_{-0.82} \pm 1.51$ & $4.47^{+0.47}_{-0.55} \pm 2.52$\\
4                            & 2                            & 0400460301                   & \nodata                      & 6.230                        & $-38.240$                    & $6.14^{+1.08}_{-0.92} \pm 0.72$ & $2.76^{+0.72}_{-0.57} \pm 0.34$& & $7.39^{+1.39}_{-1.36} \pm 0.82$ & $2.14^{+0.83}_{-0.82} \pm 0.66$\\
                             &                              & 0400460401                   & \nodata                      & 6.229                        & $-38.240$                    & $6.29^{+1.48}_{-1.53} \pm 0.50$ & $1.70^{+0.74}_{-0.73} \pm 0.38$& & $6.29^{+1.48}_{-1.53} \pm 0.50$ & $1.70^{+0.74}_{-0.73} \pm 0.38$\\
5                            & 3                            & 0401660101                   & \nodata                      & 6.604                        & $ +7.775$                    & $7.03^{+1.16}_{-0.78} \pm 0.80$ & $4.21^{+0.61}_{-0.44} \pm 0.72$& & \nodata                      & \nodata                     \\
                             &                              & 0085582001                   & \nodata                      & 6.588                        & $ +7.785$                    & $3.84^{+1.64}_{-1.28} \pm 1.78$ & $6.01^{+1.28}_{-1.18} \pm 1.17$& & $3.84^{+1.64}_{-1.28} \pm 1.78$ & $6.01^{+1.28}_{-1.18} \pm 1.17$\\
                             &                              & 0085581401                   & \nodata                      & 6.588                        & $ +7.786$                    & $5.15^{+0.82}_{-1.55} \pm 0.99$ & $3.97^{+0.99}_{-0.95} \pm 0.97$& & $5.15^{+0.82}_{-1.55} \pm 0.99$ & $3.97^{+0.99}_{-0.95} \pm 0.97$\\
\enddata
\tablecomments{This table is available in its entirety in a machine-readable form in the online journal. A portion is shown here for guidance regarding its form and content.}
\end{deluxetable*}

}

Table~\ref{tab:MultipleObs} contain the oxygen intensity measurements for directions that have been
observed multiple times with \xmm. The sets of observations are ordered by Galactic longitude, $l$.
Each set of observations is identified by a unique number (1--\marker{217}), which is in Column~1,
while Column~2 contains the number of observations in each set. Column~3 contains the
\xmm\ observation IDs. Column~4 contains the data set number from Paper~I that each observation
belonged to, if appropriate (these data sets were numbered from 1 through 69; see Table~5 in
Paper~I). Note that \marker{data set 49} from Paper~I is not included in Table~\ref{tab:MultipleObs},
because one of the two members of this data set (obs.~0203362201) no longer passes the
$\Ftotal/\Fexgal$ test, and so is not included in this paper (see
Section~\ref{subsec:DifferencesFrom1}, below). Columns~5 and 6 contain the pointing direction in
Galactic coordinates. Columns~7 and 8 contain the \OVII\ and \OVIII\ intensities, respectively,
measured without the solar wind proton flux filtering described in
Section~\ref{subsec:ProtonFluxFiltering}. As in Tables~\ref{tab:OxygenIntensities1} and
\ref{tab:OxygenIntensities2}, the statistical and systematic errors are shown separately. Columns~9
and 10 contain the corresponding measurements obtained with solar wind proton flux filtering. If
this filtering rendered an observation unusable, these fields are left blank.

\begin{figure*}
\centering
\plottwo{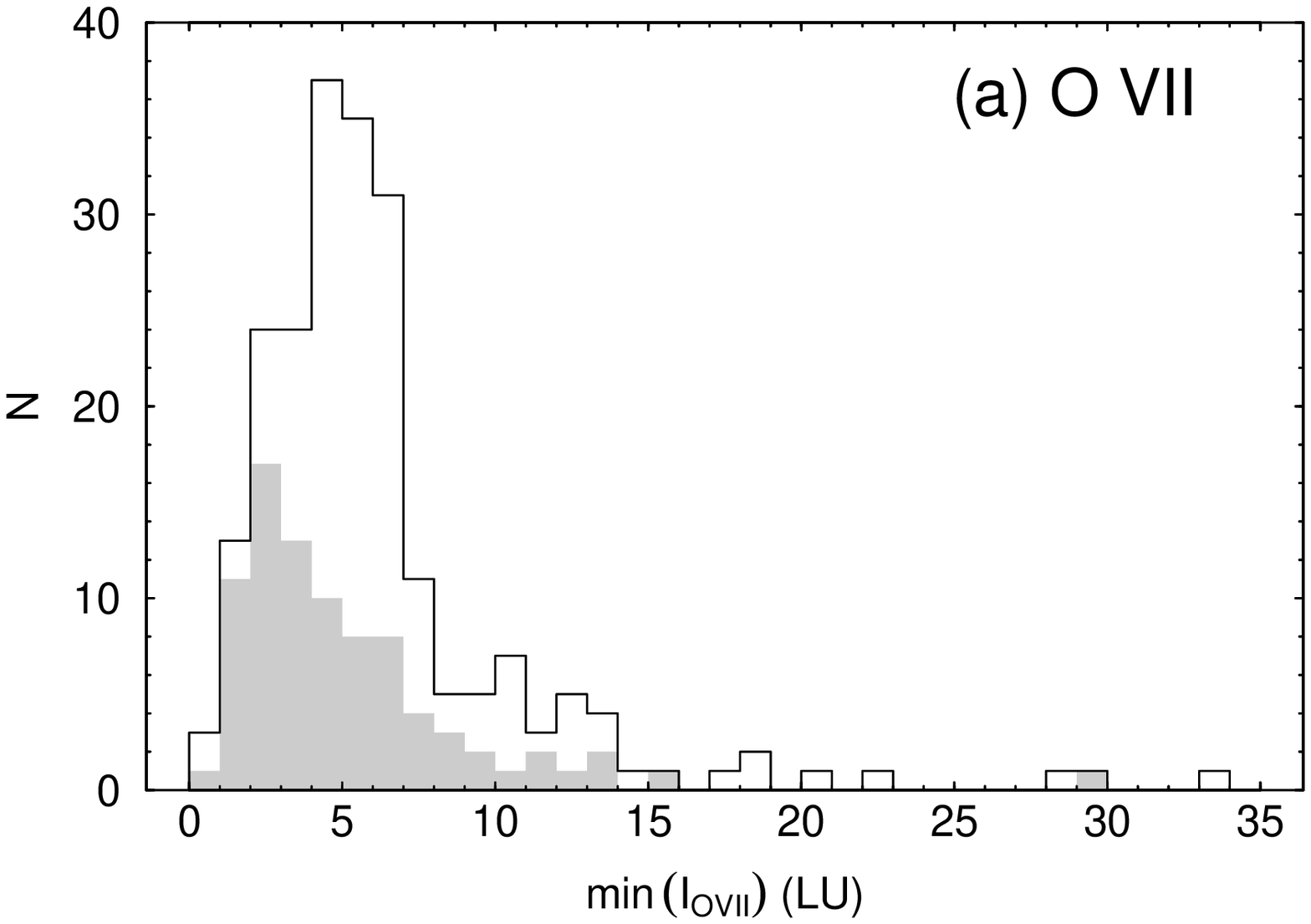}{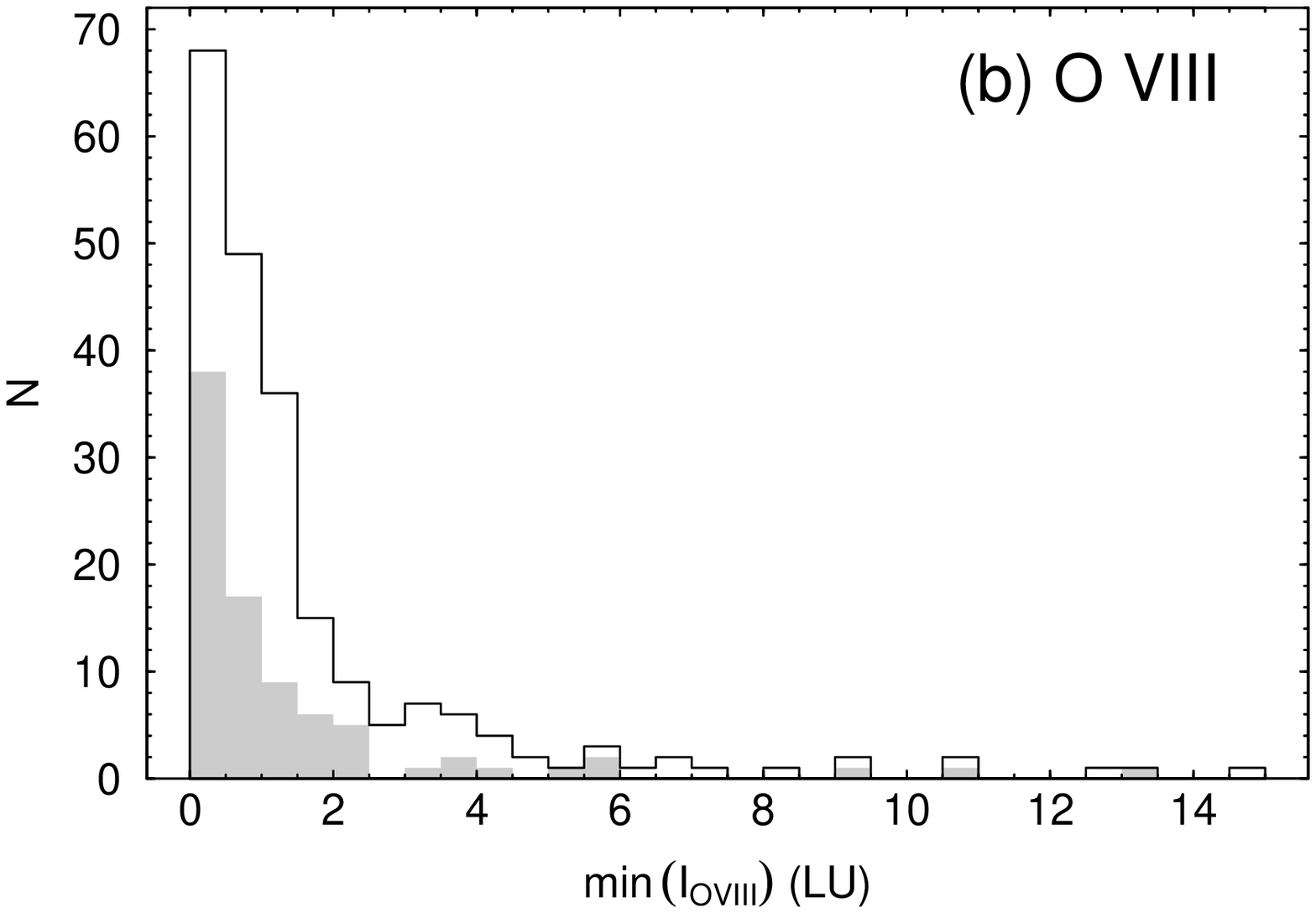}
\plottwo{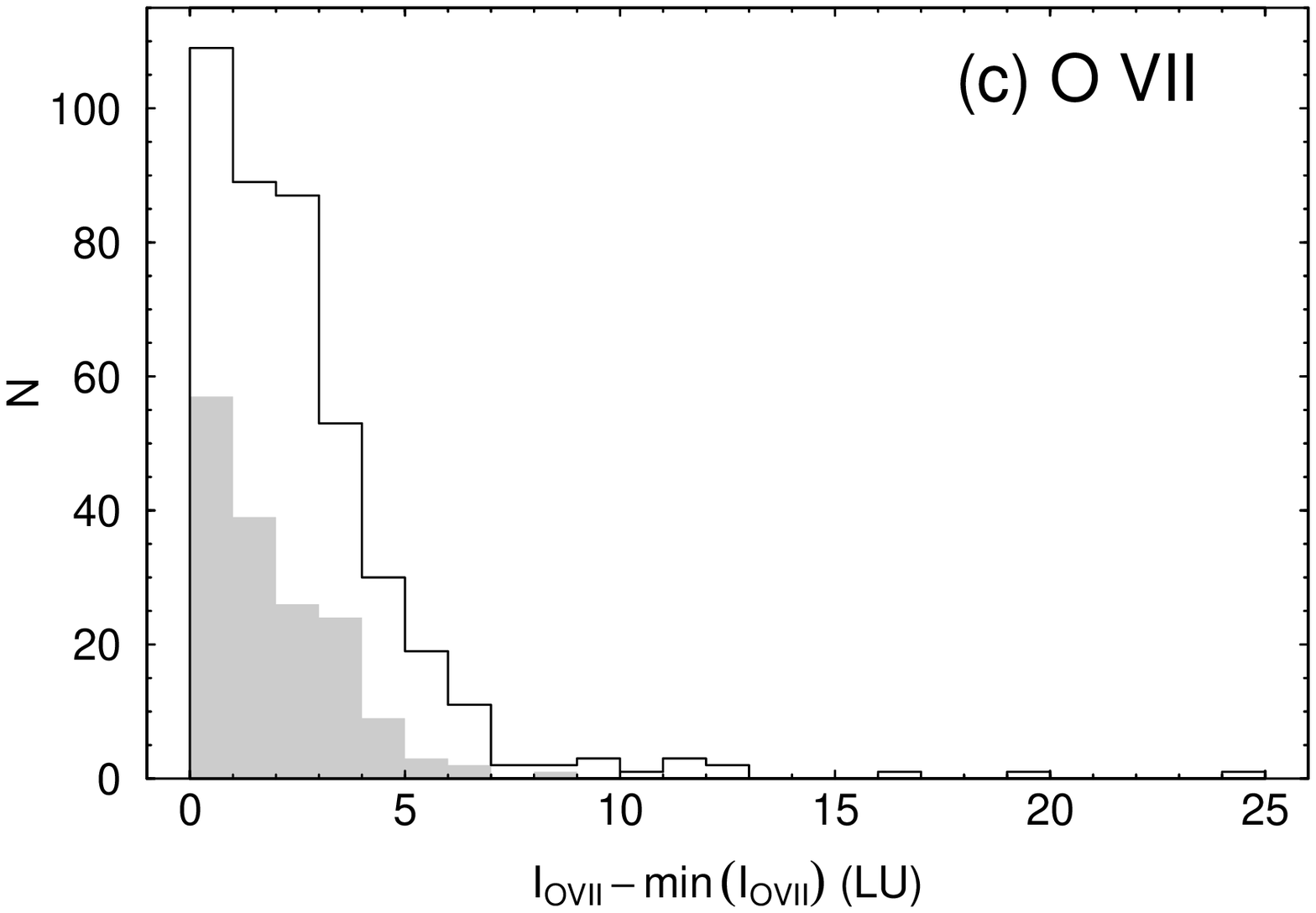}{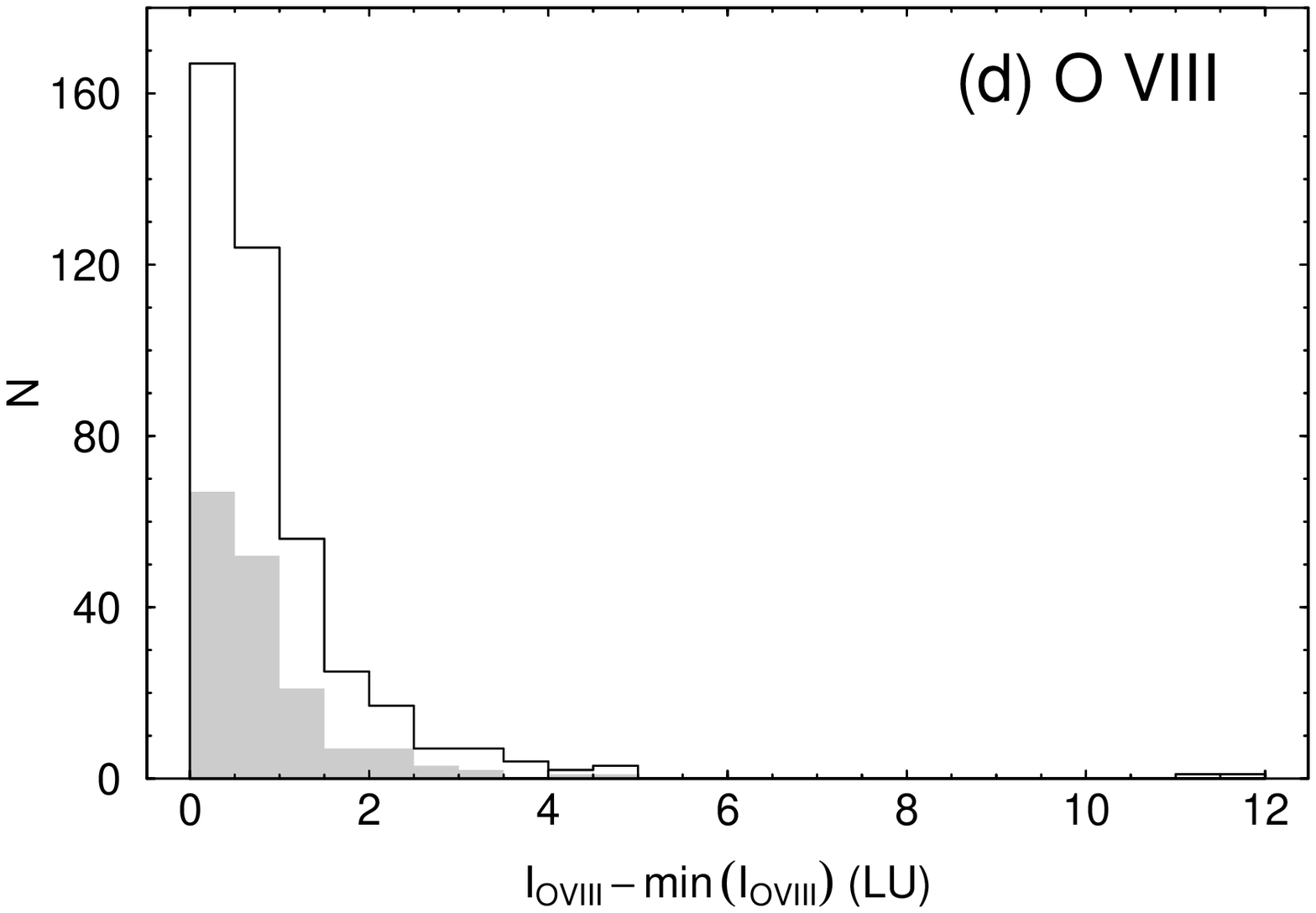}
\plottwo{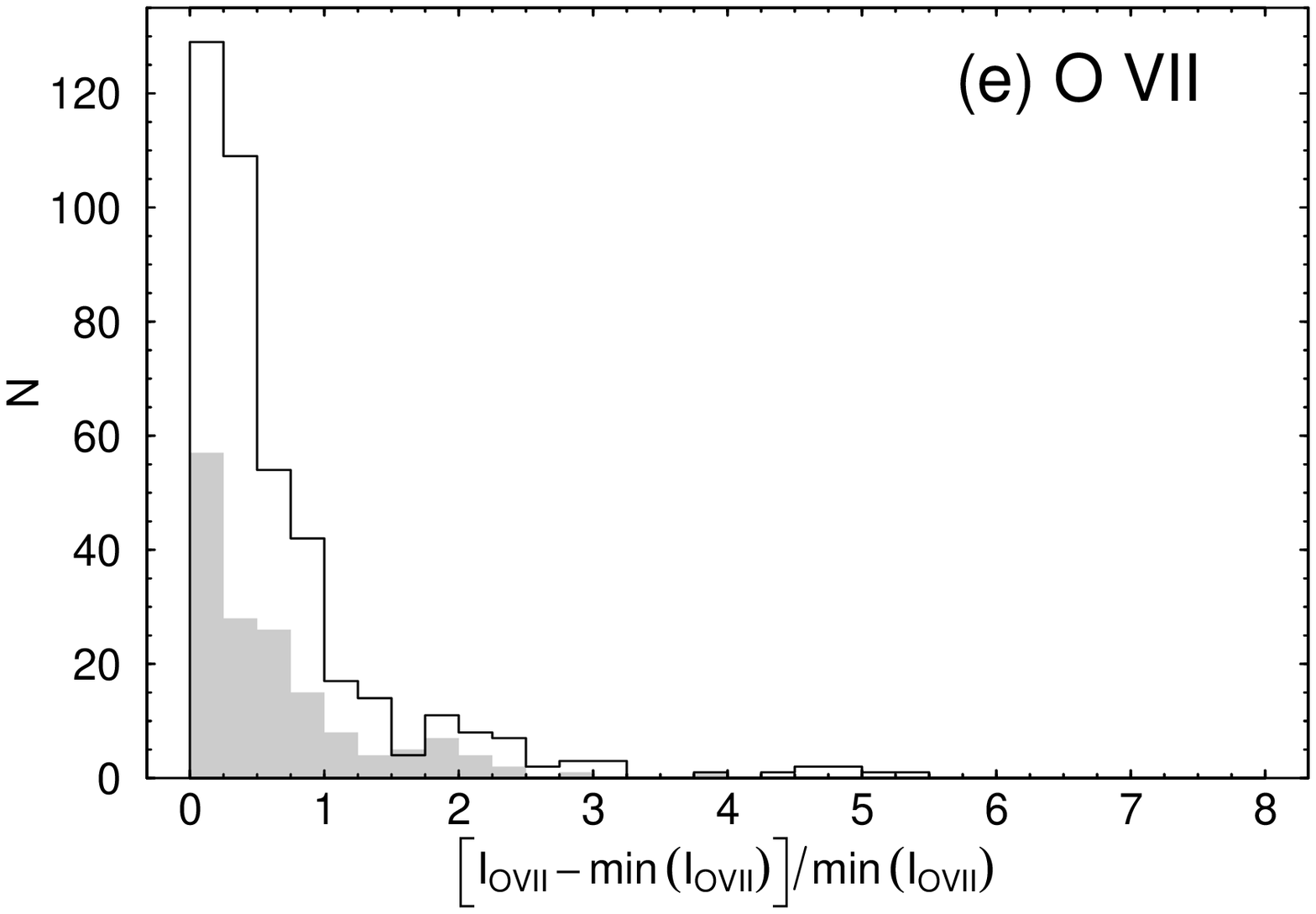}{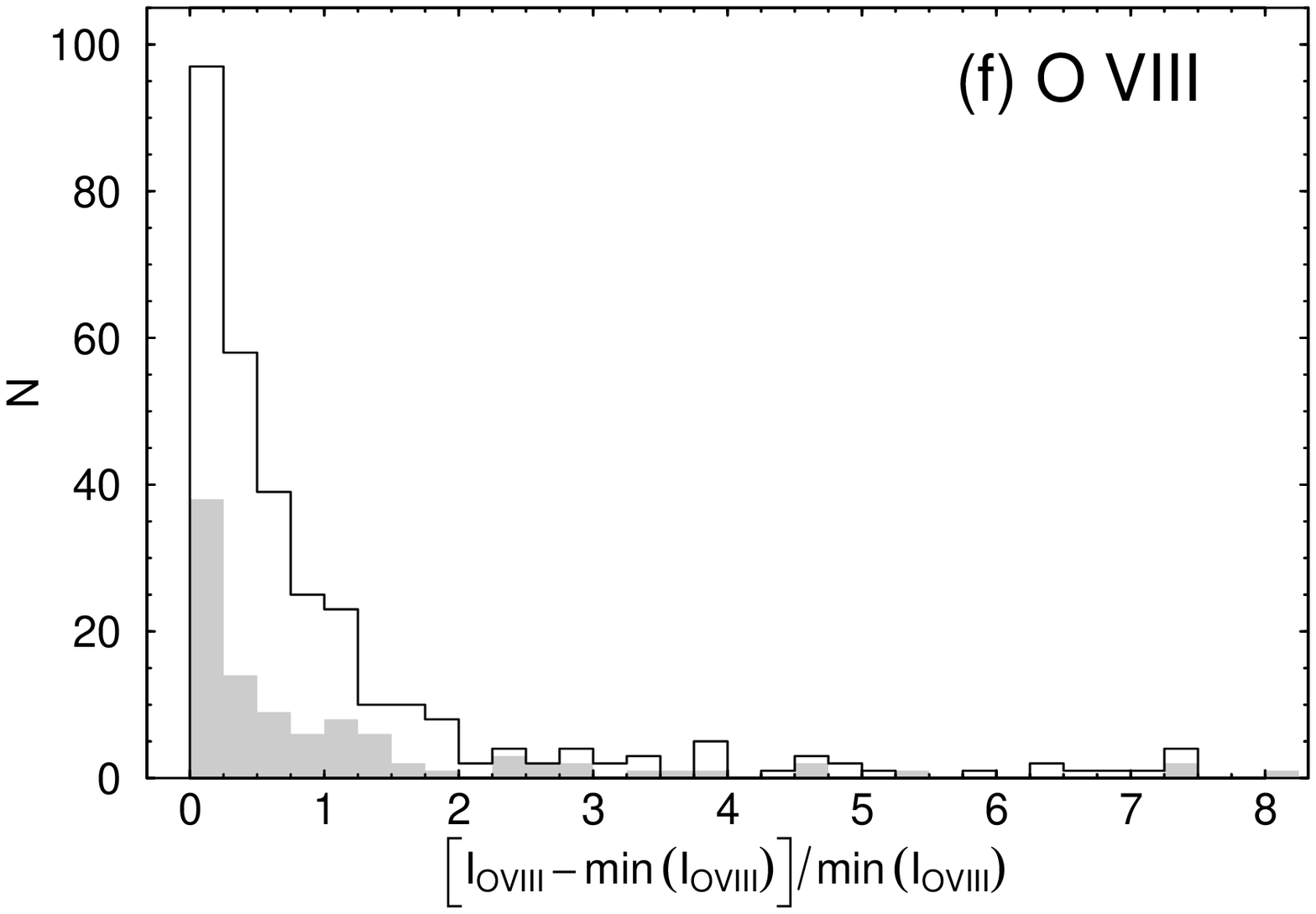}
\caption{Top row: Histograms of $\min(I)$ for (a) \OVII\ and (b) \OVIII, where $\min(I)$ is the
  minimum measured intensity in a given direction, and hence the upper limit on the intensity
  originating in the ISM.  Middle row: Histograms of $I - \min(I)$ for (c) \OVII\ and (d) \OVIII;
  here, $\min(I)$ is the minimum measured intensity in the same direction as the $I$ measurement,
  and hence $I - \min(I)$ is the lower limit on the SWCX X-ray intensity.  Bottom row: Histograms of
  $[I - \min(I)] / \min(I)$ for (e) \OVII\ and (f) \OVIII. These histograms were constructed from
  the data in Table~\ref{tab:MultipleObs}. The solid black lines show the results obtained without
  the solar wind proton flux filtering described in Section~\ref{subsec:ProtonFluxFiltering}, and
  the gray areas show the results obtained with this filtering.  For each of the \marker{217} sets
  of observations in Table~\ref{tab:MultipleObs} there is, by definition, an observation with $I -
  \min(I) = 0$; these observations are omitted from Panels~(c)--(f). In Panel~(e), \marker{three}
  observations for which the corresponding values of $\min(\Iovii)$ are 0 are omitted from both
  histograms. In Panel~(f), \marker{nine} observations with $[\Ioviii - \min(\Ioviii)] / \min(\Ioviii)
  > 8$ are omitted from both histograms, while \marker{90} (\marker{49}) observations for which the
  corresponding values of $\min(\Ioviii)$ are 0 are omitted from the histogram showing the results
  obtained without (with) proton flux filtering.
  \label{fig:I-Imin}}
\end{figure*}

For each of the \marker{217} directions in Table~\ref{tab:MultipleObs}, we can place upper limits on
the cosmic \OVII\ and \OVIII\ X-ray emission (i.e., the emission that arises beyond the solar
system, rather than the emission that is due to SWCX). These upper limits are $\min(\Iovii)$ and
$\min(\Ioviii)$, which are defined as the minimum \OVII\ and \OVIII\ intensities, respectively,
measured in a given direction. These quantities are upper limits on the cosmic oxygen intensities
because SWCX may also have contributed photons to the $\min(I)$ measurements. See the top row of
Figure~\ref{fig:I-Imin} for histograms of $\min(I)$. Then, with $\min(I)$ thus defined, $I -
\min(I)$ places a lower limit on the SWCX line intensity for a given observation (again, it is a
lower limit because SWCX may have contributed photons to the $\min(I)$ measurement).

To show the magnitude of the variability in the oxygen SWCX emission, in the middle row of
Figure~\ref{fig:I-Imin} we plot histograms of $\Iovii - \min(\Iovii)$ and $\Ioviii - \min(\Ioviii)$,
obtained with and without the solar wind proton flux filtering described in
Section~\ref{subsec:ProtonFluxFiltering}. For each set of observations, the observation with $I -
\min(I) = 0$ has been omitted. Similarly to what we found in Paper~I, the measured intensity
enhancements are typically \marker{$\la$5}~\LU\ for \OVII\ (cf.\ $\la$4~\LU\ stated in Paper~I), and
\marker{$\la$2}~\LU\ for \OVIII\ (same as stated in Paper~I). More quantitatively, the 90th
percentile of $\Iovii - \min(\Iovii)$ (excluding the faintest observation from each set) is
\marker{5.2}~\LU\ without proton flux filtering and \marker{3.9}~\LU\ with. For $\Ioviii -
\min(\Ioviii)$, the 90th percentile of $\Ioviii - \min(\Ioviii)$ is \marker{2.0}~\LU\ without proton
flux filtering and \marker{1.9}~\LU\ with.

The measured \OVII\ enhancements are generally smaller than the corresponding unenhanced intensity,
i.e., $\Iovii - \min(\Iovii)$ is typically smaller than $\min(\Iovii)$; see
Figure~\ref{fig:I-Imin}(e) (note that, although we use the term ``unenhanced'' intensity, $\min(I)$
may still include some photons due to SWCX). The median value of $[\Iovii - \min(\Iovii)] /
\min(\Iovii)$ is \marker{0.39} (\marker{0.45}) without (with) proton flux filtering, and only
\marker{19\%} (\marker{22\%}) of observations have $[\Iovii - \min(\Iovii)] / \min(\Iovii) > 1$ (the
observations that yield the $\min(\Iovii)$ measurement have been omitted from these values).  In
contrast, for \OVIII, the intensity enhancements are often somewhat brighter than the typical
unenhanced intensity; see Figure~\ref{fig:I-Imin}(f). The median value of $[\Ioviii - \min(\Ioviii)]
/ \min(\Ioviii)$ is \marker{0.82} (\marker{1.39}) without (with) proton flux filtering, and
\marker{46\%} (\marker{57\%}) of observations have $[\Ioviii - \min(\Ioviii)] / \min(\Ioviii) > 1$.
These results imply that \OVIII\ is typically brighter relative to \OVII\ in the SWCX enhancements
than in the baseline SXRB emission.

Some of the observations in Table~\ref{tab:MultipleObs} exhibit particularly bright oxygen intensity
enhancements due to SWCX. \marker{Nine} observations have $\Iovii - \min(\Iovii) > 10~\LU$; for
\marker{three} of these observations, the enhancement exceeds 15~\LU. There are \marker{two} observations
with $\Ioviii - \min(\Ioviii) > 5~\LU$ (in fact, these enhancements both exceed 10~\LU). The
observations with $\Iovii - \min(\Iovii) > 10~\LU$ or $\Ioviii - \min(\Ioviii) > 5~\LU$ are
presented in Table~\ref{tab:BrightestSWCX}, along with the observations that yielded the $\min(I)$
in each case.  Three of the bright SWCX enhancements in Table~\ref{tab:MultipleObs} were also
reported in Paper~I.  The enhancements measured in Paper~I are also shown in the table -- the new
measurements are consistent with the Paper~I measurements. Note that in Paper~I we also reported an
\OVIII\ enhancement of $\approx$8~\LU\ in obs.~0203362201 relative to obs.~0302352201. This
enhancement is not included in Table~\ref{tab:MultipleObs} because, as noted above, obs.~0203362201
is not included in this paper (see also Section~\ref{subsec:DifferencesFrom1}, below).

\begin{deluxetable*}{ccccccccc}
\tablewidth{0pt}
\tablecaption{Directions with the Brightest SWCX Emission\label{tab:BrightestSWCX}}
\tablehead{
\colhead{Dataset\tablenotemark{a}} & \colhead{Line} & \multicolumn{2}{c}{Faintest} && \multicolumn{2}{c}{Brightest} & \colhead{Difference} & \colhead{Paper I}\\
\cline{3-4} \cline{6-7}
\colhead{}                   & \colhead{}                   & \colhead{Obs.~ID}            & \colhead{$I$}                && \colhead{Obs.~ID}            & \colhead{$I$}                & \colhead{}                   & \colhead{Difference\tablenotemark{b}} \\
\colhead{}                   & \colhead{}                   & \colhead{}                   & \colhead{(\LU)}              && \colhead{}                   & \colhead{(\LU)}              & \colhead{(\LU)}              & \colhead{(\LU)}              
}
\startdata
12                           & \OVII                        & 0402080801                   & $ 5.25^{+1.88}_{-1.48}$      && 0096010101                   & $15.28^{+0.97}_{-0.97}$      & $10.03^{+1.77}_{-2.12}$      & \nodata                      \\
34                           & \OVII                        & 0206860101                   & $ 6.76^{+0.99}_{-1.01}$      && 0206860201                   & $19.55^{+2.27}_{-1.54}$      & $12.79^{+2.48}_{-1.83}$      & \nodata                      \\
38                           & \OVII                        & 0550061301                   & $ 6.45^{+1.89}_{-1.39}$      && 0041741101                   & $23.07^{+4.80}_{-4.79}$      & $16.62^{+5.00}_{-5.15}$      & \nodata                      \\
40                           & \OVII                        & 0406421401                   & $ 5.54^{+1.48}_{-0.85}$      && 0406420401                   & $17.14^{+1.71}_{-1.62}$      & $11.60^{+1.91}_{-2.20}$      & \nodata                      \\
42                           & \OVII                        & 0400360301                   & $ 4.85^{+1.82}_{-1.71}$      && 0400360801                   & $17.08^{+1.52}_{-1.45}$      & $12.23^{+2.29}_{-2.32}$      & \nodata                      \\
58                           & \OVII                        & 0305920301                   & $ 5.80^{+1.42}_{-1.41}$      && 0305920601                   & $16.83^{+1.70}_{-1.73}$      & $11.03^{+2.21}_{-2.24}$      & \nodata                      \\
83                           & \OVII                        & 0400560301                   & $ 3.83^{+0.84}_{-0.63}$      && 0059140901                   & $15.45^{+1.42}_{-1.48}$      & $11.62^{+1.56}_{-1.70}$      & $10.72^{+0.99}_{-0.96}$      \\
90                           & \OVII                        & 0112520901                   & $ 4.95^{+1.58}_{-1.71}$      && 0112520601                   & $29.28^{+1.83}_{-2.03}$      & $24.33^{+2.50}_{-2.57}$      & $26.15^{+1.52}_{-1.41}$      \\
107                          & \OVII                        & 0143150301                   & $ 3.89^{+1.25}_{-1.61}$      && 0143150601                   & $23.26^{+1.25}_{-1.38}$      & $19.37^{+2.04}_{-1.86}$      & $18.21^{+0.99}_{-1.13}$      \\
38                           & \OVIII                       & 0041740301                   & $ 3.62^{+1.87}_{-1.81}$      && 0041741101                   & $15.28^{+1.81}_{-1.80}$      & $11.66^{+2.56}_{-2.59}$      & \nodata                      \\
138                          & \OVIII                       & 0206090101                   & $ 0.02^{+0.88}_{-0.79}$      && 0206090201                   & $11.48^{+1.71}_{-1.70}$      & $11.46^{+1.89}_{-1.91}$      & \nodata                      \\
\enddata
\tablenotetext{a}{Dataset number from Table~\ref{tab:MultipleObs}}
\tablenotetext{b}{$I - \min(I)$ from Paper~I (see Table~6 of that paper).}
\end{deluxetable*}

As noted in Paper~I, the bright \OVII\ enhancements in Table~\ref{tab:MultipleObs} (i.e., those at
the bright end of the $\Iovii - \min(\Iovii)$ distribution) are of interest because they are much
larger than most measurements of \OVII\ SWCX enhancements ($\la$7~\LU;
\citealt{snowden04,fujimoto07,henley08a,gupta09b}; see also Figure~\ref{fig:I-Imin}(c)).  However,
since Paper~I, bright \OVII\ enhancements of 14--21 and 36~\LU\ have also been reported by
\citet{carter11} and \citet{ezoe11}, respectively. Note that these enhancements were found by
looking at line intensity variations within \xmm\ or \suzaku\ observations, rather than between
observations, as we have done. Note also that \citet{koutroumpa07} reported \OVII\ enhancements of
up to 10~\LU, but we found the observations in question to be badly contaminated by soft protons
(see Paper~I).

Similarly to the brightest \OVII\ enhancements, the brightest \OVIII\ enhancements (reported in
Table~\ref{tab:MultipleObs}) are much larger than those that are typically measured ($\la$2~\LU;
\citealt{henley08a,gupta09b}; see also Figure~\ref{fig:I-Imin}(d)), although \OVIII\ enhancements of
6.5 and 5.0~\LU\ were reported by \citet{snowden04} and \citet{fujimoto07}, respectively. More
recently, \OVIII\ enhancements of 26 and 12~\LU\ have been reported by \citet{carter10} and
\citet{ezoe11}, respectively, in both cases associated with coronal mass ejections.

\subsection{Differences from Paper~I}
\label{subsec:DifferencesFrom1}

The main differences in our current methodology from that in Paper~I are summarized as follows:
\begin{enumerate}
\item We used a more recent version of the SAS software (11.0.1 versus 7.0.0), which incorporates
  the \esas\ software (Section~\ref{sec:DataReduction}). The version of \texttt{mos\_back} included
  in SAS version 11.0.1 can automatically detect anomalous CCDs. For some of the observations that
  were in Paper~I, \texttt{mos\_back} detected anomalous CCDs that we had failed to identify in
  Paper~I (Section~\ref{subsec:Spectra}).
\item For the automated source removal, we used data from the \xmm\ Serendipitous Source Catalogue,
  instead of carrying out source detection ourselves on each observation. We also used larger
  circles to exclude such sources (50\arcsec\ versus $\approx$30\arcsec--40\arcsec;
  Section~\ref{subsec:AutomatedSourceRemoval}).
\item In the spectral analysis, we used an unbroken power-law to model the residual soft-proton
  contamination, as opposed to a broken power-law in Paper~I. Unlike Paper~I, we placed constraints
  on this power-law's spectral index (Section~\ref{subsec:OxygenMethod}).
\item We used a lower normalization for the EPL in our spectral model (7.9 versus 10.5~\pownorm;
  Section~\ref{subsec:OxygenMethod}). In addition, we estimated the size of systematic errors
  associated with our assuming this normalization, and with our assuming that the non-oxygen line
  emission can be modeled with an APEC model (Section~\ref{subsec:SystematicErrors}).
\item We used $\chi^2_\mathrm{min} + 2.30$ \citep{lampton76,avni76} to calculate the statistical
  errors on the oxygen intensities (although not explicitly stated, we used $\chi^2_\mathrm{min} +
  1.0$ in Paper~I, which is only valid for the error on a single interesting parameter;
  \citealt{lampton76}). These larger statistical errors, combined with the systematic errors
  mentioned above, result in confidence intervals that are typically $\sim$2 times wider than those
  in Paper~I.
\end{enumerate}
Because of the improvement in anomalous-CCD detection and the changes in the automated source
removal, there are some differences in the solid angles from which the SXRB spectra were extracted
(columns~6 and 8 in Tables~\ref{tab:OxygenIntensities1} and \ref{tab:OxygenIntensities2}). In
addition, because of differences between the OMNIWeb data used in this paper (downloaded on 2011
March 01) and that used in Paper~I (downloaded on 2009 February 12), there are, in a few cases,
significant differences in the amount of good time that remains after the proton flux filtering has
been applied (columns~5 and 7 in Table~\ref{tab:OxygenIntensities2}). For example, in
obs.~0205590301, the post-filtering good time has decreased from $\approx$41~\ks\ per camera in
Paper~I to $\approx$9~\ks. This is because the observed solar wind proton flux was close to the
threshold of $2 \times 10^8$~\pcmsq\ \ps\ during this observation. In the OMNIWeb data used in
Paper~I, the observed solar wind proton flux was generally below this threshold, whereas in the
revised proton flux data used in this paper, the observed flux was generally above this
threshold. Finally, note that for obs.~0203900101 we used two exposures per camera in
Paper~I. However, upon reinspecting this observation, we have concluded that the two shorter
exposures are likely contaminated by soft protons. We have therefore rejected these two exposures,
and just use the longer exposures (one per camera) in this paper.

Despite these differences, the new oxygen intensity measurements are generally not significantly
different from those in Paper~I. Here, we discuss the few observations for which this is not the
case.

\begin{deluxetable}{ll}
\tablewidth{0pt}
\tablecaption{Observations from Paper~I that are Missing from the Current Paper\label{tab:Paper1Missing}}
\tablehead{
\colhead{Obs.~ID} & \colhead{Reason}
}
\startdata
\cutinhead{Without Proton Flux Filtering}
0010620101 & $\Ftotal / \Fexgal > 2.7$ \\
0065820101 & $\Ftotal / \Fexgal > 2.7$ \\
0099030101 & $\Ftotal / \Fexgal > 2.7$ \\
0109270701 & $\Ftotal / \Fexgal > 2.7$ \\
0111490401 & Diffraction spikes \\
0128531501 & $\Ftotal / \Fexgal > 2.7$ \\
0134540101 & Diffraction spikes \\
0201951701 & Insufficient good time \\
0203362201 & $\Ftotal / \Fexgal > 2.7$ \\
0203610201 & $\Ftotal / \Fexgal > 2.7$ \\
0206060201 & $\Ftotal / \Fexgal > 2.7$ \\
0206360101 & $\Ftotal / \Fexgal > 2.7$ \\
0302260201 & $\Ftotal / \Fexgal > 2.7$ \\
0305290201 & $\Ftotal / \Fexgal > 2.7$ \\
0405730401 & $\Ftotal / \Fexgal > 2.7$ \\
0411980601 & $\Ftotal / \Fexgal > 2.7$ \\
\cutinhead{With Proton Flux Filtering}
0083000101 & $\Ftotal / \Fexgal > 2.7$ \\
0099030101 & $\Ftotal / \Fexgal > 2.7$ \\
0109270701 & $\Ftotal / \Fexgal > 2.7$ \\
0110950201 & Insufficient good time \\
0111100101 & $\Ftotal / \Fexgal > 2.7$ \\
0134540101 & Diffraction spikes \\
0141980601 & Insufficient good time \\
0141980701 & Insufficient good time \\
0148560501 & Insufficient good time \\
0150320201 & $\Ftotal / \Fexgal > 2.7$ \\
0158560301 & Insufficient good time \\
0162160401 & Insufficient good time \\
0200430401 & Insufficient good time \\
0200810301 & Insufficient good time \\
0201040101 & Insufficient good time \\
0201940201 & Insufficient good time \\
0203360701 & Insufficient good time \\
0203361701 & Insufficient good time \\
0203610201 & $\Ftotal / \Fexgal > 2.7$ \\
0203840101 & Insufficient good time \\
0206360101 & $\Ftotal / \Fexgal > 2.7$ \\
0305290201 & $\Ftotal / \Fexgal > 2.7$ \\
\enddata
\end{deluxetable}

The current catalog excludes \marker{16} (\marker{22}) observations that were included in the
Paper~I results obtained without (with) proton flux filtering. These observations are shown in
Table~\ref{tab:Paper1Missing}, along with the reason for their exclusion. The most common reasons
for exclusion are either that the observation now exceeds our $\Ftotal / \Fexgal$ threshold for
excluding soft-proton-contaminated observations (see Section~\ref{subsec:OxygenResults}), or that
the observation no longer has sufficient good time after being reprocessed (see
Sections~\ref{subsec:ObservationSelection} and \ref{subsec:ProtonFluxFiltering}). In two cases
(obs.~0111490401 and 0134540101), we found upon reinspection that there are diffraction spikes from
the target sources (VY~Ari and HR~1099, respectively) visible in the soft band (0.2--0.9~\kev)
images. As photons in these diffraction spikes would contaminate our SXRB measurements, we decided
to err on the side of caution and reject these observations.

\begin{deluxetable}{lcc}
\tablewidth{0pt}
\tablecaption{Observations with Oxygen Intensities Significantly Different from Those in Paper~I\label{tab:Paper1Different}}
\tablehead{
                  & \multicolumn{2}{c}{\Iovii} \\
                  & \multicolumn{2}{c}{(\LU)}  \\
\colhead{Obs.~ID} & \colhead{Paper~I} & \colhead{This paper}
}
\startdata
0108062101 &    $5.96^{+0.19}_{-0.43}$ &    $4.61^{+0.49}_{-0.48}$ \\
0147511801 &    $8.04^{+0.41}_{-0.23}$ &           $6.58 \pm 0.60$ \\
0301600101 &    $3.69^{+0.42}_{-0.20}$ &    $2.39^{+0.52}_{-0.51}$ \\
\enddata
\end{deluxetable}

There are only \marker{three} observations for which either the \OVII\ or \OVIII\ intensity
measurements differs by more than $2\sigma$ from the Paper~I value. In all \marker{three} cases, it
is the \OVII\ intensity measured without proton flux filtering that is different: the intensities
measured in this paper are lower than the Paper~I values (see Table~\ref{tab:Paper1Different}).

For obs.~0147511801 and 0301600101, the difference in \OVII\ intensity from Paper~I seems to be
mainly due to our using a smaller normalization for the EPL in this paper.  Having a smaller
normalization for the EPL results in a larger normalization for the soft-proton model. These two
model components have different spectra: the soft-proton model increases monotonically with
decreasing energy, while the EPL, being subject to Galactic absorption and the response of the X-ray
telescope, decreases with decreasing energy below $\sim$1~\kev. As a result, the net effect of
decreasing the normalization of the EPL (and thus increasing that of the soft-proton model) is to
increase the combined number of counts from these two components in the vicinity of the oxygen
lines. This in turn leads to a decrease in the number of counts attributed to the oxygen lines, and
hence a decrease in the oxygen intensity. While this effect should, in principle, affect all
observations, clearly in the vast majority of cases the effect is negligible compared with the
combined statistical and systematic error.

For obs.~0108062101, the above explanation may partly explain the difference in \OVII\ intensity.
In addition, the new version of \texttt{mos\_back} identified the MOS1.5 chip as being anomalous in
this observation, whereas we failed to identify it as such in Paper~I. Including an anomalous chip
for this observation in Paper~I may have led to an inaccurate estimate of the particle background,
and hence to an inaccurate measurement of the \OVII\ intensity.

\subsection{Contamination from Bright Sources and Soft Protons}
\label{subsec:Contamination}

In Paper~I we examined the possibilities that contamination from the wings of the PSF of bright
sources and/or contamination from soft protons were biasing our intensity measurements. We came to
the conclusion that neither was a significant source of bias. Using the same methods as in Paper~I,
we come to the same conclusion here.

To examine possible contamination from bright sources, we considered the observations in our sample
for which we had excised the target source by hand ($\sim$30\%\ of our sample).\footnote{In Paper~I,
  we considered only observations of stars, as such objects can produce bright line emission that
  could have contaminated our SXRB measurements.} For each observation, we incrementally increased
the source exclusion radius (typically in five 1\arcmin\ increments), and remeasured the oxygen
intensities each time. If photons in the wings of the PSF were biasing our measurements, we would
expect the measured intensity to decrease with increasing source exclusion radius. For each
observation, we used \chisq\ to test whether the intensities measured with larger source exclusion
radii were consistent with the original intensity measurement.

We found that for $\approx$6\%\ of the observations, the \OVII\ and/or \OVIII\ intensities measured
with larger source exclusion radii were not consistent with the original measurements at the
5\%\ level. However, upon inspecting the intensities as a function of source exclusion radius, we
found that in most cases the variation was non-monotonic, or that the intensities increased with
source exclusion radius. Neither of these forms of variation is consistent with there being
contamination from the target source. For only \marker{five} observations were there indications of
contamination from the target source.\footnote{In Paper~I, only one of the observations that we
  examined exhibited significant variation of the \OVII\ or \OVIII\ intensity with exclusion
  radius. This variation was inconsistent with there being contamination from the target source.}
For these observations, we simply increased the source exclusion radius to the point where the
intensities stopped varying with source exclusion radius (an increase of 1\arcmin\ or
2\arcmin). Having corrected these \marker{five} observations, we think that our intensity
measurements are not seriously contaminated by emission from bright sources.

We also investigated the possibility of contamination from fainter sources, i.e., the sources
removed by the automated source removal (Section~\ref{subsec:AutomatedSourceRemoval}) rather than by
hand. We reprocessed 100 of our observations, chosen at random, with the source exclusion radius
used in the automated source removal increased by 50\%\ from 50\arcsec\ to 75\arcsec. The intensities
obtained using this larger source exclusion radius were not significantly different from our original
measurements, nor was there any systematic shift in the intensities toward larger or smaller values.
We therefore conclude that our intensity measurements are not seriously contaminated by emission
from fainter, automatically removed sources.

To examine the possibility that soft-proton contamination is biasing our results, we looked for any
correlation between $I - \min(I)$ and $\Ftotal / \Fexgal$ (the latter being a measure of the amount
of soft-proton contamination). For a given direction, any variation in intensity should be solely
due to SWCX, and so, in the absence of any bias, there should be no correlation between $I -
\min(I)$ and $\Ftotal / \Fexgal$. For \OVII\ and \OVIII, the correlation coefficients (specifically,
Kendall's $\tau$; e.g., \citealt{press92}) are \marker{$-0.08$ $(-0.13,-0.04)$} and \marker{$-0.02$
  $(-0.07,+0.03)$}, respectively (the values in parentheses are the 90\%\ bootstrap confidence
intervals).  Although the correlation coefficient for $\Iovii - \min(\Iovii)$ against $\Ftotal /
\Fexgal$ is different from zero at a statistically significant level, it is still small in magnitude
($|\tau| \le 0.1$). When we use other measures of the soft-proton contamination (the normalization
of the power-law model, or its spectral index), we again find correlation coefficients that are
small in magnitude ($|\tau| \le 0.1$), and that are typically consistent with zero. Hence,
soft-proton contamination is unlikely to be seriously biasing our intensity measurements.

\subsection{Effect of Proton Flux Filtering}
\label{subsec:ProtonFluxFilteringEffects}

Here we discuss the effects of excluding the portions of the \xmm\ data taken when the solar wind
proton flux exceeded $2 \times 10^8$~\pcmsq\ \ps. This filtering was carried out in an attempt to
reduce the contamination due to SWCX emission, and is described in
Section~\ref{subsec:ProtonFluxFiltering}.  As noted in Section~\ref{subsec:OxygenResults}, the
quartiles of the \OVII\ intensities obtained without this filtering are systematically higher than
the corresponding quartiles obtained with the filtering, while for \OVIII, there are no significant
differences in the intensity quartiles with and without the proton flux filtering.  However, this
does not mean that the proton flux filtering leads to systematically lower \OVII\ intensities being
extracted from the SXRB spectra. When we look at the \marker{356} observations for which proton flux
filtering had some effect, but did not render the observation unusable (denoted by a ``Y'' in
Column~14 of Table~\ref{tab:OxygenIntensities2}), we find that the median difference in the
\OVII\ intensity without and with proton flux filtering is \marker{$-0.02~\LU$} (90\%\ bootstrap
confidence interval: \marker{$-0.10$ to 0.03~\LU}). For \OVIII\ the corresponding difference is
\marker{0.00~\LU} (\marker{$-0.01$ to 0.03~\LU}). Thus, as we found in Paper~I, proton flux
filtering does not lead to systematically lower oxygen intensities being extracted from the SXRB
spectra.

In contrast, proton flux filtering does preferentially remove the observations that yield the largest
\OVII\ intensities (i.e., during such observations the solar wind proton flux tended to exceed $2
\times 10^8$~\pcmsq\ \ps). Of the \marker{336} observations in Table~\ref{tab:OxygenIntensities1}
with $\Iovii > 10~\LU$, \marker{135} (\marker{40\%}) yield usable spectra after the proton flux
filtering, compared with \marker{856} out of \marker{1532} (\marker{56\%}) of the observations with
$\Iovii \le 10~\LU$. For $\Iovii > 15~\LU$, the fraction is \marker{45} out of \marker{118}
observations (\marker{38\%}), compared with \marker{946} out of \marker{1750} (\marker{54\%}) for
observations with $\Iovii \le 15~\LU$.  This tendency of proton flux filtering to preferentially
remove the observations that yield larger \OVII\ intensities explains why there is a significant
downward shift in the quartiles of the \OVII\ intensities when proton flux filtering is included
(Table~\ref{tab:Quartiles}). However, it should be noted that this tendency is less pronounced than
we found in Paper~I: in Paper~I, only 5 out of 31 observations (16\%) with $\Iovii > 10~\LU$, and 1
out of 9 observations (11\%) with $\Iovii > 15~\LU$ yielded usable spectra after the proton flux
filtering. This difference may be due to the fact that the oxygen intensities in the region covered
by Paper~I ($l=120\degr$--240\degr) are generally lower than those toward the Galactic Center (see
Figure~\ref{fig:Maps}), and so a bright \OVII\ line in Paper~I was more likely to be due to SWCX,
and thus more likely to be associated with a large solar wind proton flux. Despite the fact that a
larger fraction of the observations with larger \OVII\ intensities was rejected by the proton flux
filtering, in Paper~I we did not see a significant downward shift in the \OVII\ intensity quartiles
when proton flux filtering was included. Presumably this is because the observations with the
largest \OVII\ intensities were a small subset of the Paper~I sample.

For \OVIII, observations with large and small intensities yield usable spectra after the proton flux
filtering at similar rates: \marker{81} out of \marker{161} observations (\marker{50\%}) with $\Ioviii
> 5~\LU$, compared with \marker{910} out of \marker{1707} observations (\marker{53\%}) with $\Ioviii
\le 5~\LU$. In contrast, in Paper~I, only 2 out 9 observations (22\%) with $\Ioviii > 5~\LU$ yielded
usable spectra after the proton flux filtering. The fact that observations with large and small
\OVIII\ intensities yield usable spectra after the proton flux filtering at similar rates explains
the similarity of the \OVIII\ intensity quartiles in Table~\ref{tab:Quartiles} with and without
proton flux filtering.

\subsection{Survey Sky Coverage}
\label{subsec:SkyCoverage}

\begin{figure}
\centering
\plotone{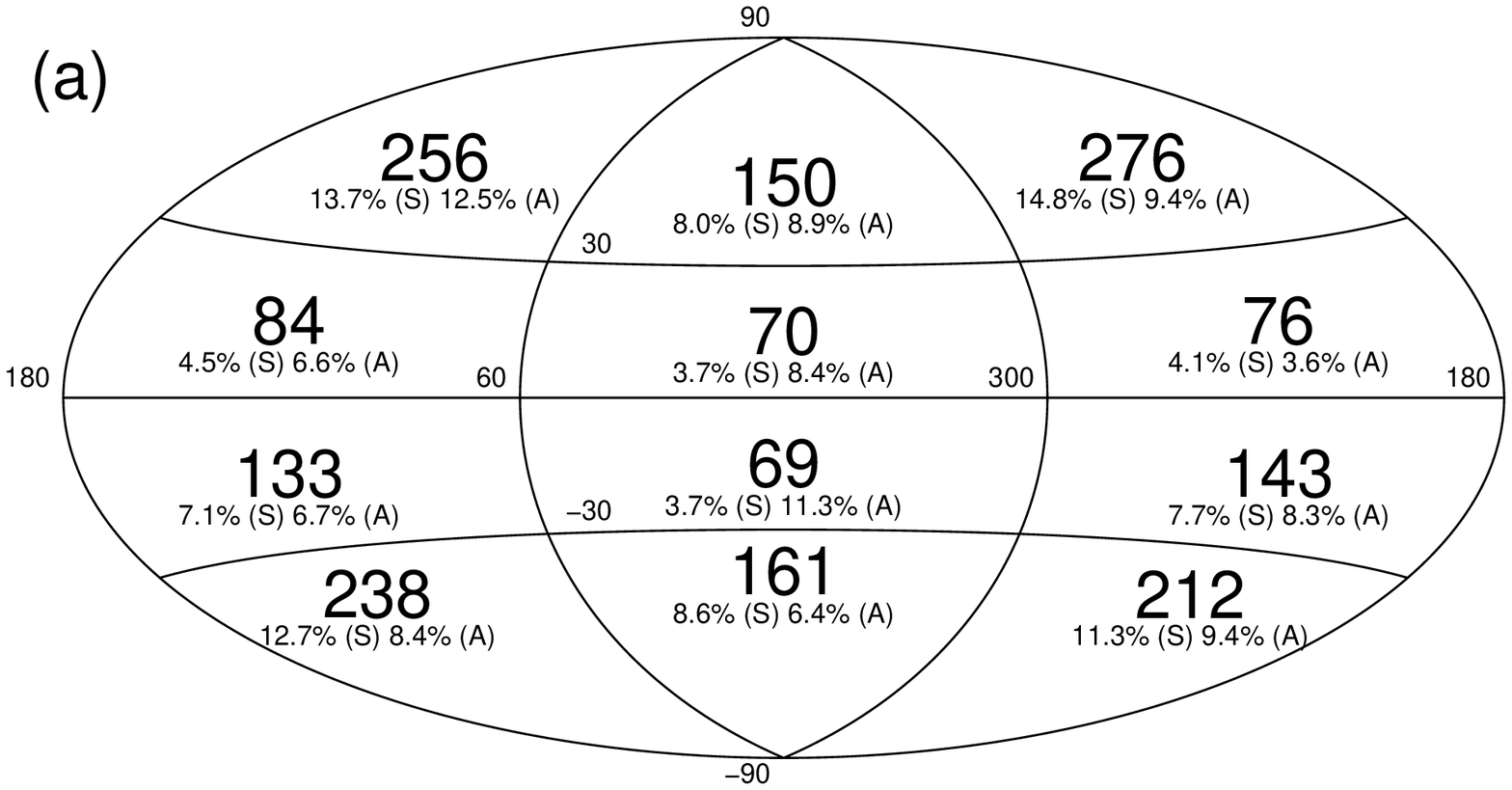}
\vspace{5mm}
\plotone{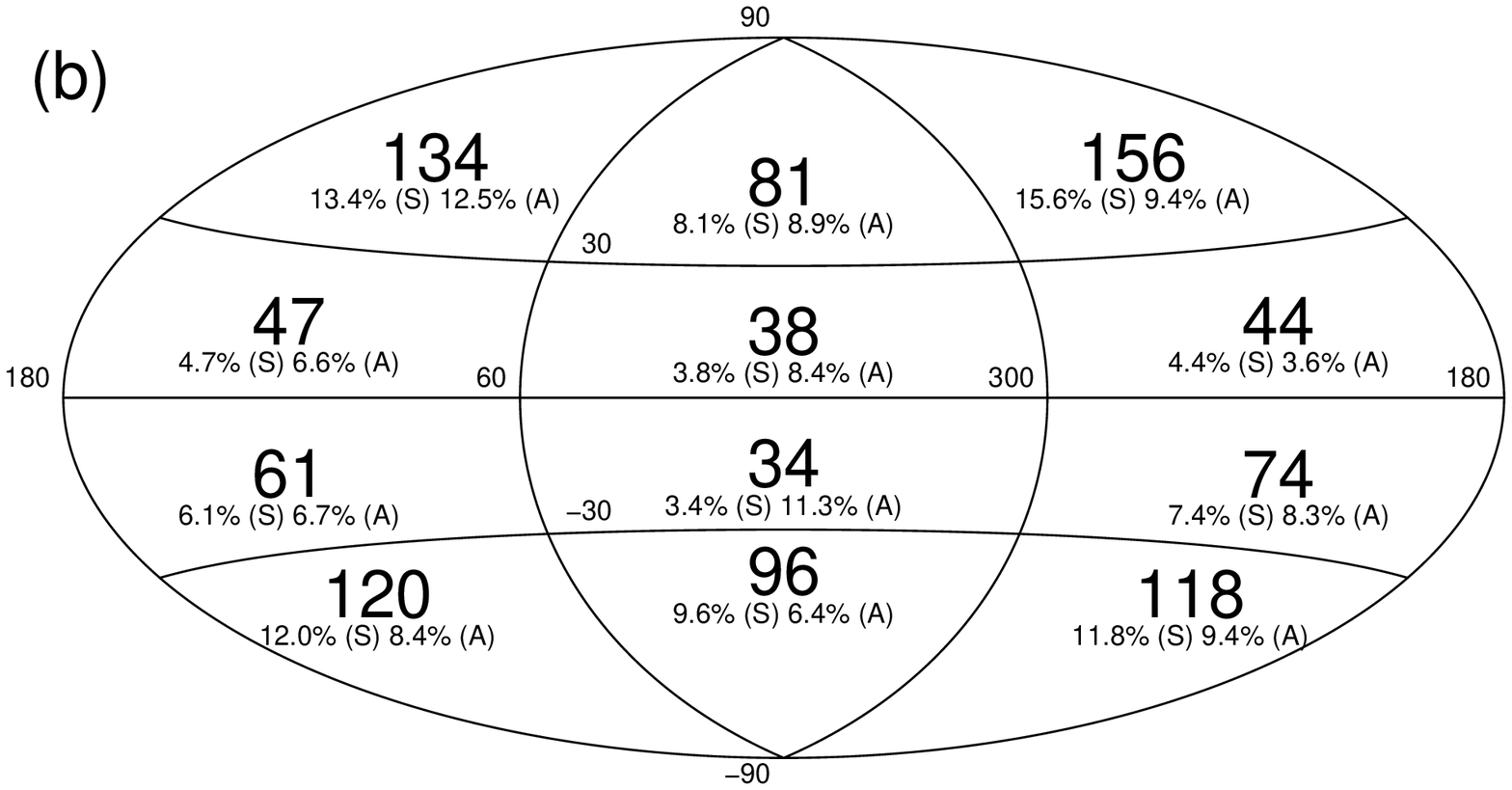}
\caption{All-sky maps in Galactic coordinates showing how our oxygen intensity measurements are
  distributed on the sky. The large numbers show the number of oxygen intensity
  measurements in each of 12 equal-area regions of the sky. The small numbers underneath show the
  percentage of observations in our survey (S), and percentage of the original set of 5698 archival
  observations (A) that fall within each region (in some cases the percentages do not sum to 100
  due to rounding). If the observations were uniformly distributed,
  each region would have 8.3\%\ of the observations. Panel (a) shows the number intensity
  measurements obtained without the solar wind proton flux filtering described in
  Section~\ref{subsec:ProtonFluxFiltering}, while panel (b) shows the numbers of intensity
  measurements obtained with this filtering.
  \label{fig:SkyCoverage}}
\end{figure}

In general our measurements are well distributed over the sky, albeit with some large scale
variation in the sky coverage (see the maps in Figure~\ref{fig:Maps}, and also
Figure~\ref{fig:SkyCoverage}, which gives the number of oxygen intensity measurements in 12
equal-area regions of the sky). There appear to be fewer observations at low Galactic latitudes
($|b| < 30\degr$) than at higher latitudes. We see from Figure~\ref{fig:SkyCoverage} that low
latitudes away from the Galactic Center ($l = 60\degr$--300\degr\ and $|b| < 30\degr$) are
relatively undersampled in our final sample, with \marker{23\%} of the observations lying in this
third of the sky. This is largely due to the relatively lower coverage of this region of the sky in
the \xmm\ archive -- 25\%\ of the archival observations are toward this third of the sky.  Higher
latitudes away from the Galactic Center are relatively oversampled in our final sample --
\marker{53\%} of the observations in our final sample are in the third of the sky with $l =
60\degr$--300\degr\ and $|b| \ge 30\degr$. This is partly because this region of the sky is
oversampled in the archive, with 40\%\ of the archival observations lying in this third of the
sky. It is not clear why a larger fraction of the observations in this region of the sky make it to
our final sample.

There also appear to be fewer observations toward the inner Galaxy ($l < 60\degr$ or $l > 300\degr$)
than away from the inner Galaxy. Higher latitudes toward the inner Galaxy ($l < 60\degr$ or $l >
300\degr$, $|b| \ge 30\degr$) may appear undersampled in Figure~\ref{fig:Maps} in comparison to the
relative oversampling of the adjacent high-latitude regions.

The region immediately around the Galactic Center ($l < 60\degr$ or $l > 300\degr$, $|b| < 30\degr$)
is undersampled in our final sample -- only \marker{7\%} of our final set of observations lie in
this sixth of the sky, compared with 20\%\ of the archival observations. This is for a number of
reasons. First, observations in this region of the sky are rejected at a higher rate by the light
curve filtering described in Section~\ref{subsec:ObservationSelection}, for reasons that are
unclear.  Second, observations in this region of the sky are rejected at a higher rate due to the
presence of bright point sources, bright diffuse emission, or bright arcs in the field of view (see
Section~\ref{subsec:ObservationSelection}). Finally, observations in this region of the sky are
rejected at a higher rate due to $\Ftotal/\Fexgal$ exceeding 2.7 (this threshold was introduced in
Section~\ref{subsec:OxygenResults}, and is intended for the identification and rejection of
observations with excessive soft-proton contamination). Approximately 3/4 of the observations in
this region of the sky that were rejected in this way are toward very low latitudes ($|b| <
3\degr$), and thus their SXRB spectra likely exhibit the hard X-ray emission from the Galactic Ridge
\citep[e.g.,][]{worrall82,revnivtsev06}. This additional hard emission would tend to increase
\Ftotal\ relative to \Fexgal. While such observations are not necessarily badly contaminated by soft
protons, we erred on the side of caution and rejected all observations with $\Ftotal/\Fexgal > 2.7$.
Note also that our spectral model (Section~\ref{subsec:OxygenMethod}) is designed for the analysis
of spectra in which counts above $\sim$2~\kev\ are from either the EPL or from residual soft-proton
contamination. Re-analyzing the spectra with a model that is better suited to modeling the Galactic
Ridge emission is beyond the scope of this survey.

\section{DISCUSSION}
\label{sec:Discussion}

\subsection{Implications for Solar Wind Charge Exchange}
\label{subsec:SWCX}

Several factors affect the intensity of the SWCX emission in a given observation of the SXRB. The
geocoronal SWCX intensity is expected to depend on the solar wind proton flux near the Earth, and on
where the sightline passes through the magnetosheath
\citep[e.g.,][]{robertson03a,robertson03b,robertson06}. The heliospheric SWCX intensity is expected
to depend on the phase of the solar cycle, the ecliptic latitude, and the location of the Earth in
its orbit \citep[e.g.,][]{robertson03a,koutroumpa06}. Variations in the solar wind ion composition,
or the passage of a coronal mass ejection across the line of sight
\citep[e.g.,][]{koutroumpa07,henley08a,carter10} can also affect the SWCX intensity.

A detailed study of all the above factors is beyond the scope of this catalog paper. However, we
will point out a few interesting features of our data, and discuss them in the context of simple
SWCX models. First, we will consider the heliospheric SWCX emission, and its variation with ecliptic
latitude and phase of the solar cycle. Then, in Section~\ref{subsubsec:GeocoronalSWCX}, we will
consider the geocoronal SWCX emission, and compare the observed variations in intensity with those
expected from a simple model of the emission.

\subsubsection{Heliospheric SWCX Emission}
\label{subsubsec:HeliosphericSWCX}

\paragraph{Model Expectations}
To determine how the heliospheric SWCX emission is expected to vary with ecliptic latitude and with
phase of the solar cycle, we refer to the heliospheric SWCX model of
\citet{koutroumpa06}. Sightlines toward low ecliptic latitudes sample the slow solar wind (and hence
approximately the same density of solar wind \oplusseven\ and \opluseight\ ions) throughout the
solar cycle. During solar minimum, the photoionization of neutral H and He is less efficient,
leading to greater densities of these atoms. Hence, the oxygen intensities at low ecliptic latitudes
are expected to be higher during solar minimum than at solar maximum \citep{koutroumpa06}.  In
contrast, sightlines toward high ecliptic latitudes only experience the slow solar wind (in which
\oplusseven\ and \opluseight\ are relatively abundant) during solar maximum; they experience the
fast solar wind (in which \oplusseven\ and \opluseight\ are less abundant) during solar minimum.
Thus, we expect the X-ray intensities at high ecliptic latitudes to be higher during solar maximum.
Furthermore, during solar maximum the intensities are not expected to be strongly dependent on
ecliptic latitude, whereas during solar minimum the intensities are expected to be higher at low
ecliptic latitudes (see Figures~1 and 11--13 in \citealt{koutroumpa06}).

\begin{figure}
\plotone{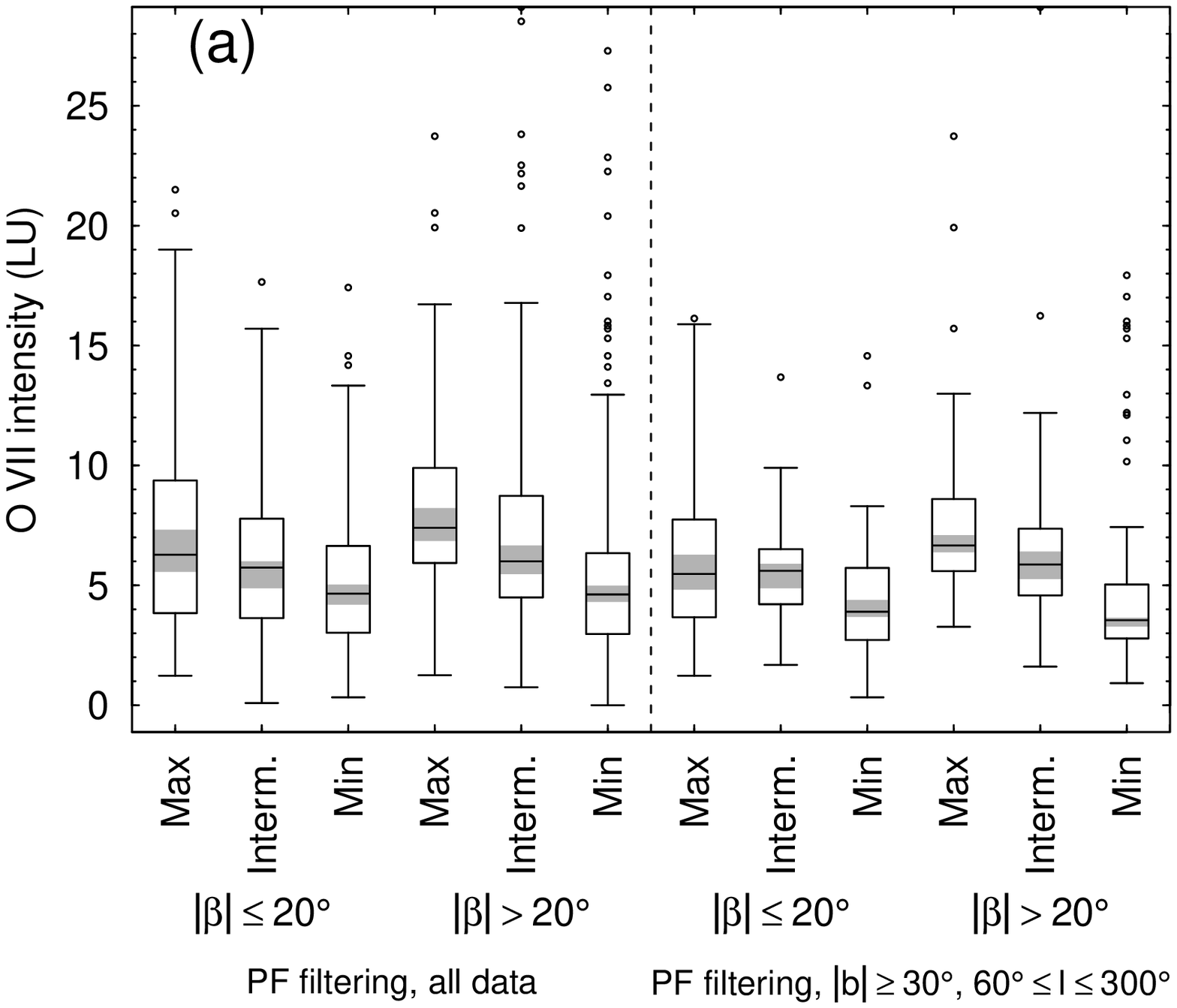}
\plotone{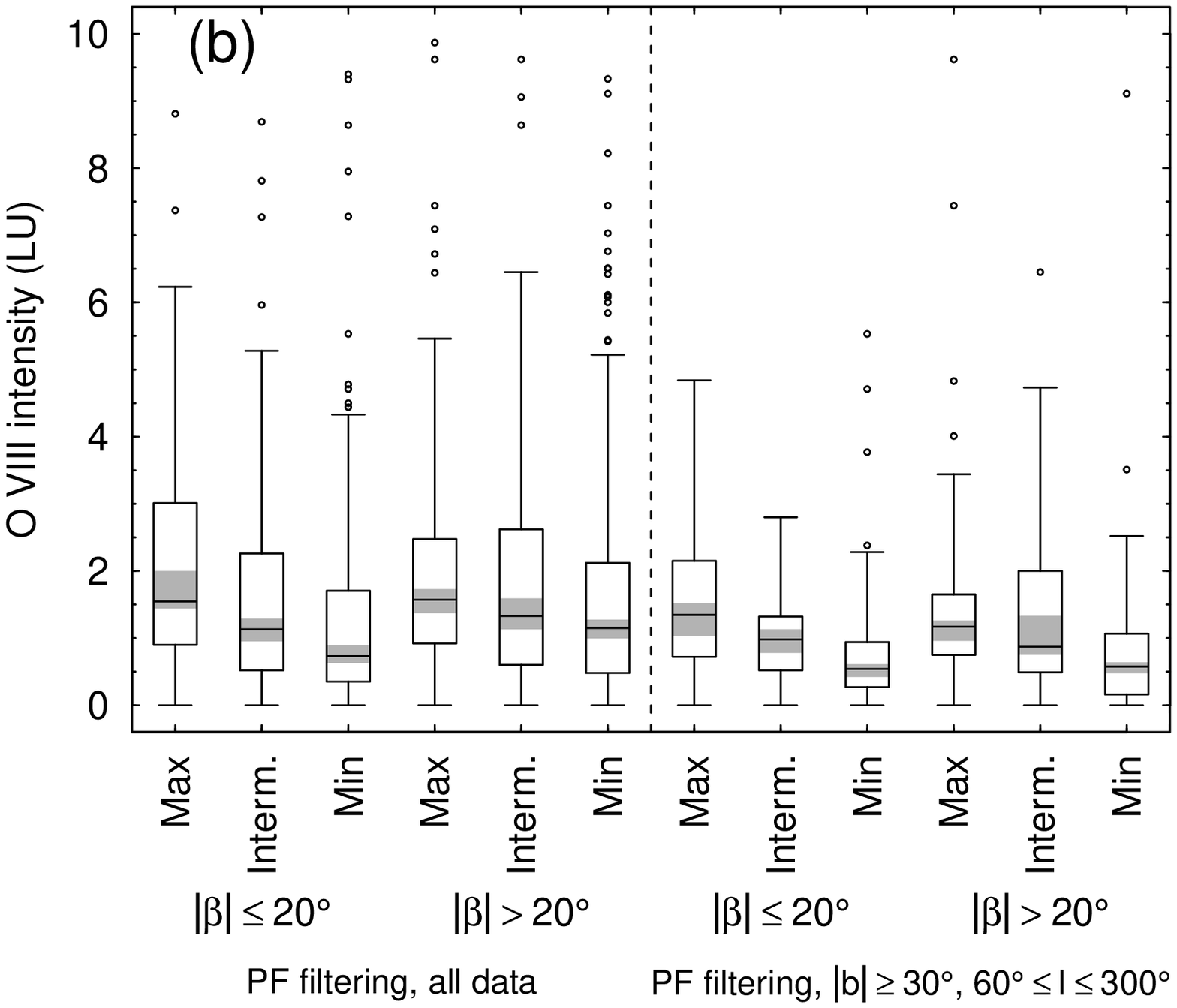}
\caption{Boxplots showing the intensity of (a) \OVII\ and (b) \OVIII\ for different ecliptic
  latitudes ($|\beta| \le 20\degr$ versus $|\beta| > 20\degr$) and for different phases in the solar
  cycle (see text for details). Each box indicates the median and the lower and upper quartiles, and
  the whiskers extend to the most extreme data point that is no more than twice the interquartile
  range from the box. Outliers are plotted individually with open circles. The gray bands indicate
  the 90\%\ confidence intervals on the medians, calculated by bootstrap resampling. The boxes in
  the left-hand region of each panel show the results for measurements obtained with the proton flux
  (PF) filtering described in Section~\ref{subsec:ProtonFluxFiltering}, with observations identified
  as SWCX-contaminated by \citet{carter11} excluded. The boxes in the right-hand regions show the
  same data, but with observations toward low Galactic latitudes ($|b| < 30 \degr$) or toward the
  inner Galaxy ($l > 60\degr$ or $l < 300\degr$) excluded.
  \label{fig:SWCXboxplot}}
\end{figure}

\paragraph{Comparison with Observations}
In order to test these ideas, we present the distributions of the \OVII\ and \OVIII\ intensities
(from the proton-flux-filtered measurements, having excluded the observations identified as
SWCX-contaminated by \citealt{carter11}). The boxplots presented in Figure~\ref{fig:SWCXboxplot} are
categorized by ecliptic latitude and by phase of the solar cycle. For example, the first box in each
panel shows the intensity distribution for observations toward low ecliptic latitudes made during
solar maximum (``Max''). Here, we define solar maximum as being before 00:00UT on 2003 Feb 01
($\mbox{MJD} = 52,671$), and solar minimum (``Min'') as being after 00:00UT on 2005 Jun 01
($\mbox{MJD} = 53,522$), with an intermediate phase (``Interm.'') between these
times.\footnote{Similarly to Paper~I, these dates were estimated from sunspot data from the National
  Geophysical Data Center (http://www.ngdc.noaa.gov/stp/SOLAR/). Note that in Paper~I we did not
  define an intermediate phase between solar maximum and minimum.} The left-hand regions of the
plots show the results for all of the measurements, while in the right-hand regions we have excluded
observations toward low Galactic latitudes or toward the inner Galaxy (to reduce the confounding
effects of the variation in intensity with Galactic latitude and longitude; see
Section~\ref{subsec:SpatialVariation}, below).

Although the distributions of the intensities are rather broad, there is a clear systematic decrease
in the intensities from solar maximum to solar minimum. While this is expected at high ecliptic
latitudes, it is the opposite of what is expected at low ecliptic latitudes (\citealt{koutroumpa06};
see \textit{Model Expectations}, above). As we have used the proton-flux-filtered data, the
unexpected trend at low ecliptic latitudes is unlikely to be due to systematic differences in the
geocoronal SWCX intensity with solar cycle. A possible explanation is that, during solar maximum,
observations at all latitudes are more likely to be affected by SWCX emission from a coronal mass
ejection moving across the line of sight than during solar minimum.

We searched for systematic differences between the intensities toward high and low ecliptic
latitudes, by checking whether or not the confidence intervals on the median intensities
overlapped. We considered only the data plotted in right-hand regions of
Figures~\ref{fig:SWCXboxplot}(a) and \ref{fig:SWCXboxplot}(b), to reduce the possibility of the
variation with Galactic coordinates affecting the results. The \OVII\ intensities measured during
solar minimum tend to be slightly higher toward low ecliptic latitudes. Although the difference is
not large, this trend is at least qualitatively as expected (\citealt{koutroumpa06}; see
\textit{Model Expectations}, above). During solar maximum, the \OVII\ intensities tend to be higher
toward high ecliptic latitudes, which is not as expected (see \textit{Model Expectations},
above). In contrast, for the \OVIII\ measurements in the right-hand region of
Figure~\ref{fig:SWCXboxplot}(b), the median intensities measured toward high and low ecliptic
latitudes are consistent with each other, for all three phases of the solar cycle (maximum,
intermediate, and minimum).

\subsubsection{Geocoronal SWCX Emission}
\label{subsubsec:GeocoronalSWCX}

We now consider the geocoronal SWCX emission. In Paper~I we showed that the solar wind proton flux
alone is not a good indicator of the degree of SWCX contamination in an SXRB spectrum: for several
directions with multiple \xmm\ observations, we found that the oxygen intensity decreased with
increasing proton flux. We also found that there was no clear tendency for sightlines that pass
close to or through the sub-solar region of the magnetosheath (that is, the region near the
Earth-Sun line) to yield systematically higher oxygen intensities, contrary to expectations
\citep{robertson03b}.

Our new, larger dataset leads us to the same conclusions. However, rather than re-creating
Figures~13 and 15 from Paper~I with our new data, here we will look at the geocoronal SWCX in a
different way, using a model that is similar to that described in \citet{robertson03b}.

\paragraph{Geocoronal SWCX Model}
The geocoronal SWCX intensity is given by
\begin{equation}
  \Igeo = \frac{1}{4\pi} \int \alpha \nsw \uave \nH dl,
  \label{eq:Igeo}
\end{equation}
where $\alpha$ is the product of the relevant charge exchange cross-section, line yield, ion
fraction, and elemental abundance, \nsw\ is the solar wind proton density, \uave\ is the mean
ion-neutral collision speed (denoted by $\langle g \rangle$ in \citealt{robertson03b}), and \nH\ is
the hydrogen density in the Earth's exosphere. The integral is carried out along the line of
sight. The mean ion-neutral collision speed is given by
\begin{equation}
  \uave \approx \left( \usw^2 + \vth^2 \right)^{1/2} = \left( \usw^2 + 3kT/m \right)^{1/2},
  \label{eq:uave}
\end{equation}
where \usw\ is the bulk solar wind speed, and \vth\ is the thermal speed of an ion of mass $m$ in solar
wind gas of temperature $T$.

If we assume that $\alpha$ is approximately constant along each line of sight, and approximately the
same for each observation, then the difference in the geocoronal SWCX intensity between two
observations of the same direction would be proportional to the difference in $\int \nsw \uave \nH
dl$ between the two observations. If we also assume for now that the heliospheric SWCX intensity is
approximately constant in a given direction, we would expect
\begin{equation}
  I - \min(I) \approx \frac{\alpha}{4\pi} \Delta \left\{ \int \nsw \uave \nH dl \right\},
  \label{eq:IminI}
\end{equation}
where $\min(I)$ is as defined in Section~\ref{subsec:MultipleObs}, and the $\Delta$ indicates the
difference in the integral between the observations that yielded $I$ and $\min(I)$.

For comparison with the observed values of $I - \min(I)$, we estimated the right-hand side of
Equation~(\ref{eq:IminI}) for each observation in Table~\ref{tab:MultipleObs}. In particular, we
calculated the time-averaged value of $\int \nsw \uave \nH dl$ during each observation, including
only the good times from our data processing, and taking into account \xmm's changing position and
the variation in the incident solar wind. For the exospheric hydrogen density, we assumed $\nH(r) =
(25~\pcc)(10 \RE / r)^3$, where $r$ is the distance from the center of the Earth, and \RE\ is the
radius of the Earth \citep{cravens01}. For the solar wind density, speed, and temperature in the
magnetosheath, we used the magnetosphere model of \citet[Figures~10 and 11]{spreiter66}, assuming
that the distance from the Earth's center to the magnetopause along the Earth-Sun line is
$10\RE$. For the pre-bowshock solar wind density, speed, and temperature, we used data from
OMNIWeb. We assumed that changes in the incident solar wind occurred instantaneously across the
domain of the \citet{spreiter66} model. As the time-resolution of the solar wind data is 1~h, and
the solar wind takes $\la 30\RE / (400~\kmps) = 8$~min to cross the model domain, this is a
reasonable approximation.  We calculated $\alpha$ for \OVII\ ($3.8 \times 10^{-19}~\cmsq$) and
\OVIII\ ($2.2 \times 10^{-19}~\cmsq$) using the slow solar wind values from Table~1 of
\citet{koutroumpa06}, assuming a line yield of 1.

\begin{figure*}
\plottwo{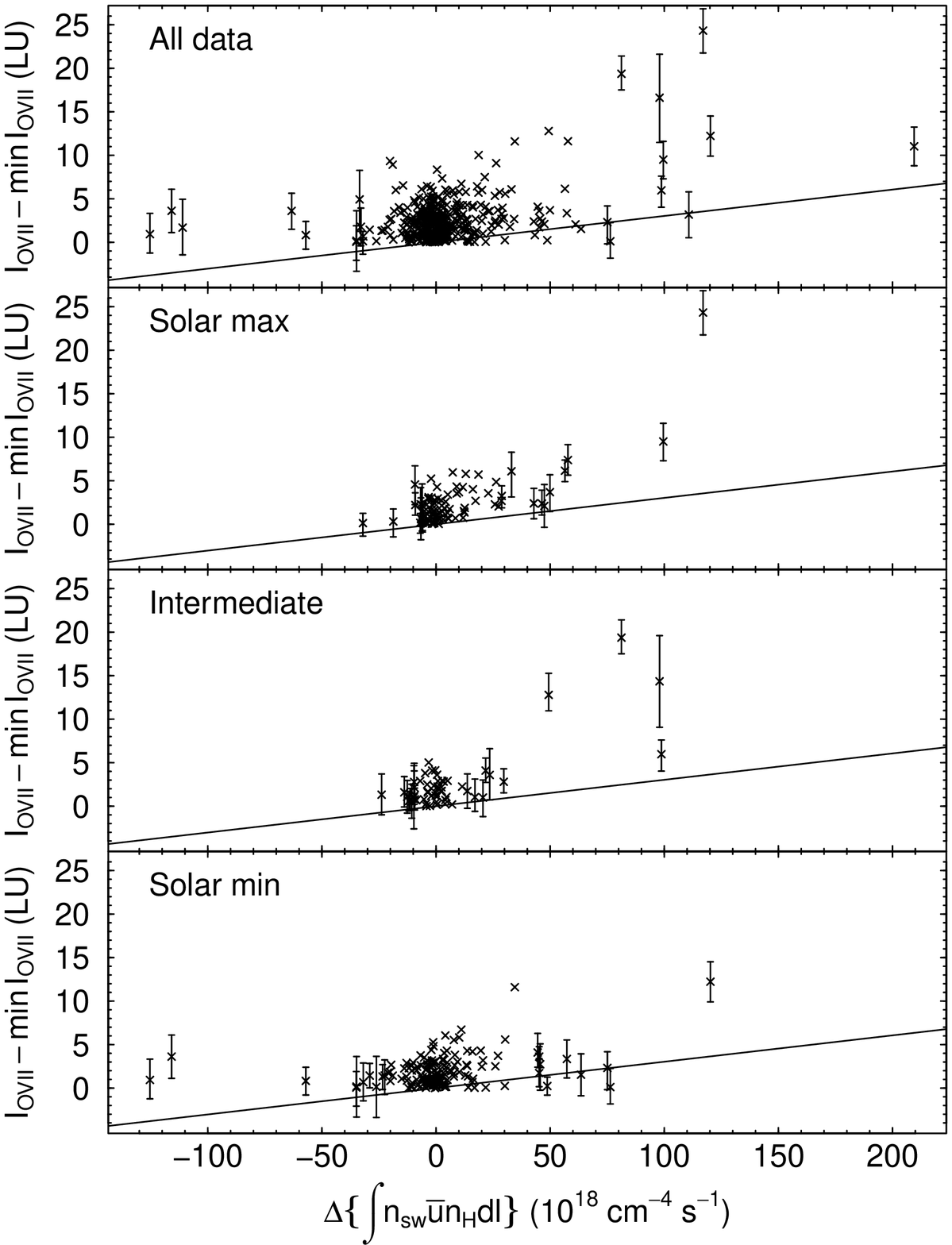}{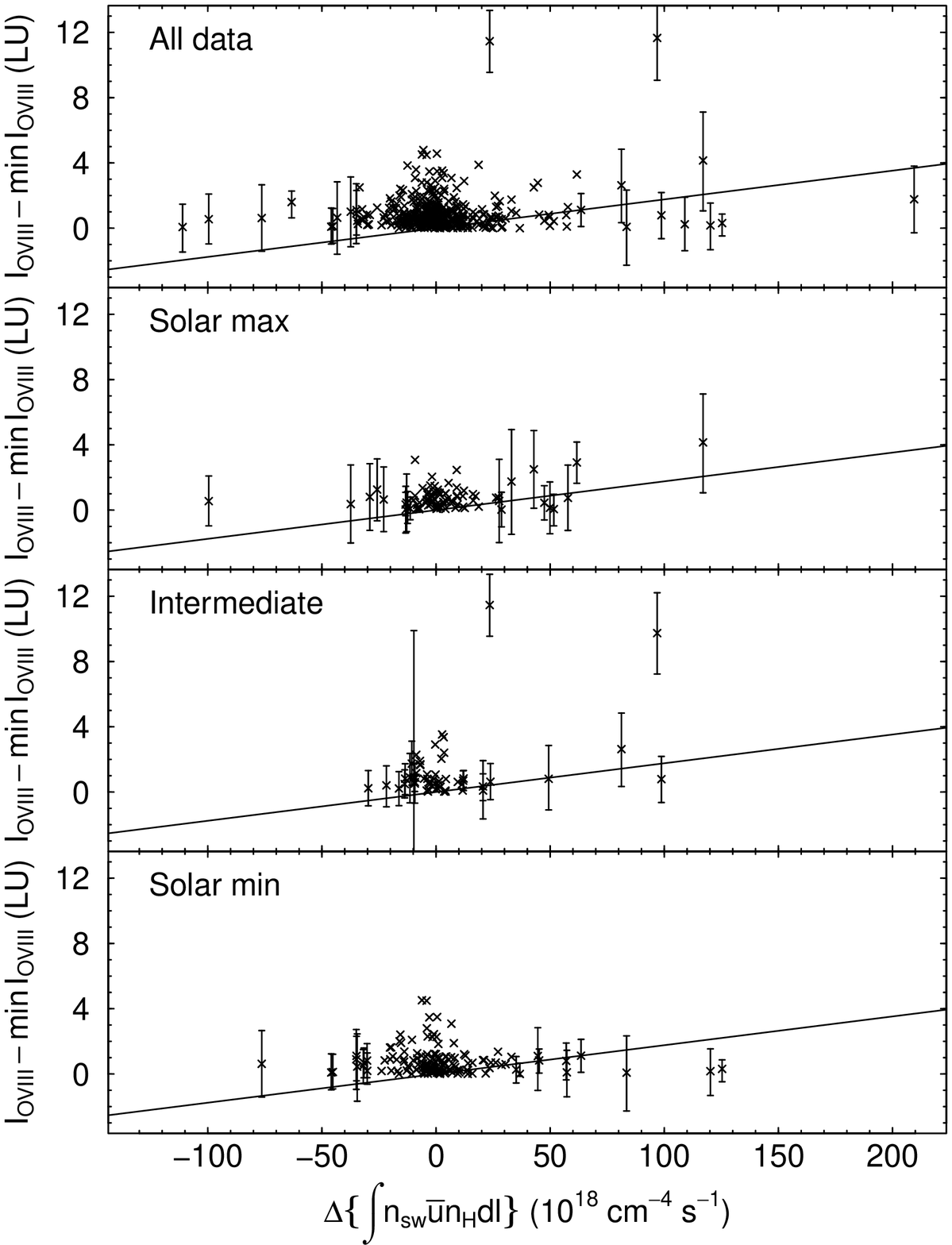}
\caption{Plots of $I - \min(I)$ against the difference in the time-averaged value of $\int \nsw
  \uave \nH dl$ between the observations that yielded $I$ and $\min(I)$. The left and right hand
  sets of plots show \OVII\ and \OVIII, respectively. The top panels show the results for all
  the observations in Table~\ref{tab:MultipleObs} (without proton flux filtering). The second
  through fourth panels show the results for different phases of the solar cycle (see text for
  details). For each set of observations of a given direction, there is, by definition, a data point
  at the origin; such points have been omitted. For clarity, only the outermost data points'
  error bars are shown. The solid lines show the relation expected from Equation~(\ref{eq:IminI}).
  \label{fig:GeocoronalIntegral}}
\end{figure*}

\paragraph{Comparison with Observations}
Figure~\ref{fig:GeocoronalIntegral} shows $\Iovii - \min(\Iovii)$ and $\Ioviii - \min(\Ioviii)$
against $\Delta \left\{ \int \nsw \uave \nH dl \right\}$.  As we wish to look at the influence of
the solar wind on the geocoronal SWCX, we use the results obtained without proton flux filtering,
and do not exclude the observations identified as SWCX-contaminated by \citet{carter11}.  To reduce
the effect that the variation of the heliospheric SWCX with solar cycle phase has on $I - \min(I)$,
and to allow us to concentrate on variations in the geocoronal SWCX, we present various solar cycle
phase-based subsets of the data in Figure~\ref{fig:GeocoronalIntegral}. The top panels show the
results for all the observations in Table~\ref{tab:MultipleObs}.  The other panels show the results
for different phases of the solar cycle (as defined earlier). Note that, in the (say) solar maximum
panels, $\min(I)$ means the minimum intensity measured in a given direction \textit{during solar
  maximum}, which is not necessarily the absolute minimum intensity measured in that direction.
Figure~\ref{fig:GeocoronalIntegral} also shows the relation expected from Equation~(\ref{eq:IminI}).

For \OVII, $\Iovii - \min(\Iovii)$ is positively correlated with $\Delta \left\{ \int \nsw \uave \nH
dl \right\}$, as expected. However, the relation is clearly not as expected from
Equation~(\ref{eq:IminI}). Some observations exhibit \OVII\ intensity enhancements much larger than
those expected from Equation~(\ref{eq:IminI}). In particular, some observations exhibit intensity
enhancements of up to $\sim$5~\LU\ while $\Delta \left\{ \int \nsw \uave \nH dl \right\} \approx 0$.
In contrast, $\Ioviii - \min(\Ioviii)$ is not significantly correlated with $\Delta \left\{ \int
\nsw \uave \nH dl \right\}$.

Although we have gone beyond a simple examination of the effect of solar wind proton flux on oxygen
intensity, our geocoronal SWCX model is still inadequate to describe the observed variations in the
oxygen intensities. There are various possible reasons for this. First, despite our dividing up the
data by solar cycle phase, it is possible that the heliospheric SWCX is varying significantly
between some our observations. Second, some observations may be affected by coronal mass ejections
traveling across the line of sight. Third, changes in the solar wind composition would cause the
observations to deviate from the relation in Equation~(\ref{eq:IminI}), as we assumed that $\alpha$,
and hence the oxygen abundance and the ion fractions, are constant between observations. Fourth, our
model does not take into account the variation in the size of the magnetosheath with the incident
ram pressure (increasing the ram pressure would tend to move the magnetosheath closer to the Earth,
resulting in denser regions of the exosphere being sampled). If we wish to be able to accurately
remove the SWCX contamination from SXRB spectra, a SWCX model that includes these various effects is
needed. Our $I - \min(I)$ measurements will provide useful constraints on such a model.

\subsection{The Oxygen Lines and the 3/4~\kev\ Soft X-ray Background}
\label{subsec:3/4kevSXRB}

In this subsection we consider the contribution of the oxygen lines to the 3/4~\kev\ SXRB. We will
compare our measurements of the fraction of the 3/4~\kev\ emission due to the oxygen lines with a
previous measurement from a higher-resolution SXRB spectrum \citep{mccammon02}. In addition, we will
compare the observed fraction due to oxygen with model predictions, and discuss the implications for
the expected sources of the SXRB emission.

\subsubsection{The 3/4~\kev\ SXRB Intensity}

The \rosat\ All-Sky Survey (RASS) was conducted in multiple bands, of which the R4 band
($\approx$0.44--1.01~\kev) is the most sensitive to the oxygen lines \citep{snowden97}. We will take
the RASS measurements in this band in the directions of the \xmm\ observations as our measure of the
3/4~\kev\ SXRB intensity, to which we will compare our measured \OVII\ and \OVIII\ intensities. We
used the \texttt{sxrbg}
tool\footnote{http://heasarc.gsfc.nasa.gov/Tools/xraybg\_help.html\#command} to extract the observed
\rosat\ count-rates, \Robs, for each \xmm\ observation direction. Note that the RASS was carried out
in 1990 and 1991 \citep{snowden95}, during a solar maximum.

The EPL's contributions to the RASS R4 count-rates, \REPL, must be removed. To this end, we assumed
an intrinsic EPL spectrum equal to $10.5 (E/1~\kev)^{-1.46}~\pownorm$ \citep{chen97}\footnote{Note
  that the normalization of this spectrum is larger than the value we used in our spectral analysis
  (7.9~\pownorm; see Section~\ref{subsec:OxygenMethod}). However, this is the extragalactic model
  that was used by \citet{snowden00} and \citet{kuntz00} in their \rosat\ analyses.} and an
absorbing column density from the LAB \HI\ survey \citep{kalberla05}. The absorbed spectrum for each
\xmm\ observation direction was folded through the \rosat\ response to yield our estimates of the
EPL's contribution to the R4 count-rate. We then subtracted this from the observed count-rate,
yielding the non-EPL portion of the R4 count-rate, $\Robs - \REPL$. This consists of SWCX emission,
emission from hot Galactic gas, and possibly emission from the warm-hot intergalactic medium (WHIM;
e.g., \citealt{cen99}).

For comparison with $\Robs - \REPL$, we calculated the R4 count-rates expected from our oxygen line
measurements, \Roxygen, by folding $\delta$ functions with the measured intensities through the
\rosat\ response.

\subsubsection{The Fraction of the 3/4~\kev\ SXRB Due to Oxygen}
\label{subsubsec:FractionDueToOxygen}

\begin{figure}
\plotone{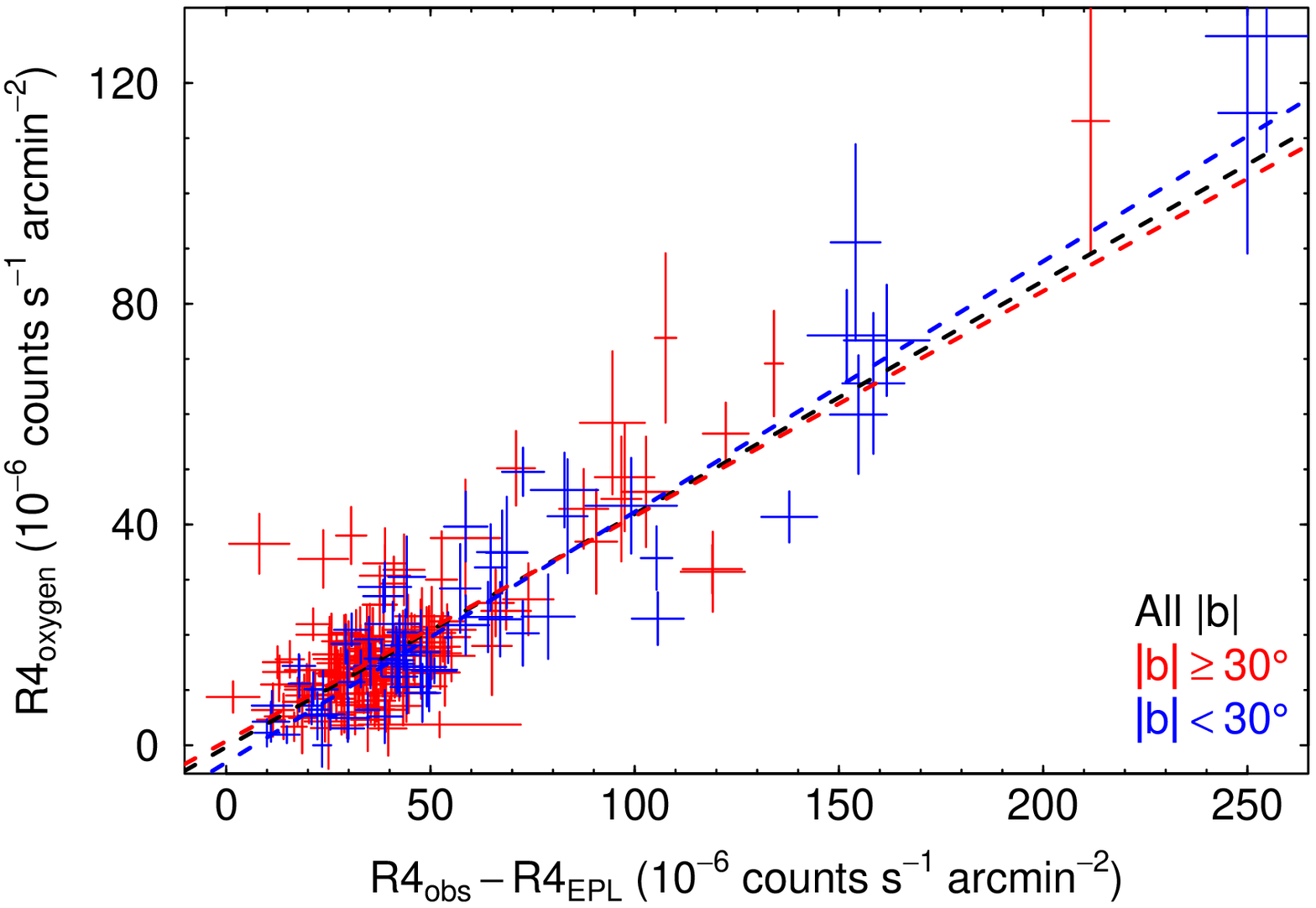}
\caption{R4 count-rate expected from our measured oxygen intensities against the non-EPL portion of
  the observed R4 count-rate, $\Robs - \REPL$. We show only the results obtained using
  $\min(\Iovii)$ and $\min(\Ioviii)$ from directions with multiple \xmm\ observations (see text for
  details). The data points are colored according to the Galactic latitude of the observation
  direction. The black, red, and blue dashed lines show the best-fit straight lines for all
  directions that have been observed multiple times, for directions with $|b| \ge 30\degr$, and for
  directions with $|b| < 30\degr$, respectively.
  \label{fig:Oxygen-vs-R4}}
\end{figure}

Figure~\ref{fig:Oxygen-vs-R4} shows \Roxygen\ plotted against $\Robs - \REPL$. In order to use
values of \Roxygen\ that are as close as possible to the interstellar values, we use only results
from directions with multiple \xmm\ observations, using the minimum \OVII\ and \OVIII\ intensities
measured in each such direction. From here on, when calculating $\min(I)$ for a given direction, we
select the smallest intensity measured in that direction, whether it was measured with proton flux
filtering or not. For example, for data set 4 in Table~\ref{tab:MultipleObs}, the smallest
\OVII\ intensity was obtained from obs.~0400460301 without proton flux filtering, while for data set
19, the smallest \OVII\ intensity was obtained from obs.~0106660401 with proton flux
filtering. Because we are using the faintest (and hence least SWCX-contaminated) observation from
each direction, here we do not exclude the observations identified as SWCX-contaminated by
\citet{carter11}.

\begin{deluxetable}{lccccccc}
\tablewidth{0pt}
\tablecaption{Straight-Line Fit Results for \Roxygen\ against $\Robs - \REPL$\label{tab:LinearRegression}}
\tablehead{
\colhead{Dataset}    & \colhead{Gradient}   & \colhead{Intercept\tablenotemark{a}} 
}
\startdata
\sidehead{$\min(I)$\tablenotemark{b}}
All $|b|$            & $0.422 \pm 0.024$    & $-0.3 \pm 1.1$       \\
$|b| \ge 30\degr$    & $0.408 \pm 0.039$    & $+0.7 \pm 1.6$       \\
$|b| < 30\degr$      & $0.454 \pm 0.032$    & $-3.1 \pm 2.0$       \\
\sidehead{All data\tablenotemark{c}, without proton flux filtering}
All $|b|$            & $0.485 \pm 0.011$    & $+1.2 \pm 0.5$       \\
$|b| \ge 30\degr$    & $0.509 \pm 0.016$    & $+1.4 \pm 0.7$       \\
$|b| < 30\degr$      & $0.477 \pm 0.014$    & $-0.4 \pm 0.8$       \\
\sidehead{All data\tablenotemark{c}, with proton flux filtering}
All $|b|$            & $0.483 \pm 0.012$    & $+0.2 \pm 0.6$       \\
$|b| \ge 30\degr$    & $0.503 \pm 0.017$    & $+0.2 \pm 0.7$       \\
$|b| < 30\degr$      & $0.479 \pm 0.020$    & $-1.1 \pm 1.1$       \\
\enddata
\tablecomments{Each value is the median $\pm$ the median absolute deviation of the relevant marginalized
    posterior probability distribution.}
\tablenotetext{a}{$10^{-6}~\rassrate$.}
\tablenotetext{b}{Results obtained using $\min(\Iovii)$ and $\min(\Ioviii)$ from directions with
    multiple \xmm\ observations (see text for details).}
\tablenotetext{c}{Excluding observations identified as SWCX contaminated by \citet{carter11}.}
\end{deluxetable}

Clearly \Roxygen\ is well correlated with $\Robs - \REPL$. We fitted straight lines to the data,
using the method described in Section~8 of \citet{hogg10} for data with errors on the $x$ and $y$
values and intrinsic scatter. We used a Metropolis-Hastings Markov Chain Monte Carlo method
\citep[e.g.,][Section~15.8]{press07} to sample the parameter space.\footnote{Specifically, we used
  the \texttt{metrop()} function from the R \texttt{mcmc} package \citep{geyer10}.} We defined the
``best-fit'' line using the medians of the gradient and the intercept from the marginalized
posterior probability distributions, with uncertainties given by the median absolute deviations of
these distributions. The best-fit lines for all directions that have been observed multiple times,
and for directions with $|b| \ge 30\degr$ and with $|b| < 30\degr$, are plotted in
Figure~\ref{fig:Oxygen-vs-R4} in different colors, and the best-fit values are shown in
Table~\ref{tab:LinearRegression}.

Although not shown in plots (which would be rather crowded), we also fit straight lines to the
larger datasets composed of all measurements, except for those from observations identified as
SWCX-contaminated by \citet{carter11}. We carried out two versions of these fits, one using the
measurements obtained without proton flux filtering, and the other using measurements obtained with this
filtering.  Table~\ref{tab:LinearRegression} also contains the best-fit values from these fits.

The gradients in Table~\ref{tab:LinearRegression} imply that $\sim$40--50\%\ of the non-EPL portion
of the R4 count-rate is due to the oxygen \Kalpha\ lines. This fraction is lower when we use the
$\min(I)$ oxygen intensities than when we use all of the oxygen intensities. This is unsurprising,
as when we use the $\min(I)$ measurements we discard the measurements that would yield larger values
of $\Roxygen/(\Robs - \REPL)$. For all the fits, the intercepts are generally consistent with zero,
implying that the fraction of the non-EPL R4 count-rate due to oxygen is independent of
\Roxygen\ and $\Robs - \REPL$. This implies that the spectrum of the non-EPL 3/4~\kev\ emission does
not vary systematically with its brightness.  Finally, we note that the fraction of the non-EPL R4
count-rate due to oxygen is similar for low and high Galactic latitudes.

For comparison, \citet{mccammon02} obtained a high-spectral-resolution microcalorimeter spectrum of
the SXRB, and found that 52\%\ of the R4 emission at high Galactic latitudes not due to AGN is due
to the oxygen \Kalpha\ lines, which they observed at their laboratory wavelengths.  The gradients in
Table~\ref{tab:LinearRegression} obtained for all observations with $|b| \ge 30\degr$ (with or
without proton flux filtering) are in excellent agreement with this fraction. In contrast, the
gradient obtained from the $\min(I)$ measurements with $|b| \ge 30\degr$ is significantly smaller
than the \citet{mccammon02} fraction.  These results most likely mean that the brightness of the
SWCX oxygen emission in \citeauthor{mccammon02}'s spectrum is similar to that in a typical
\xmm\ spectrum, whereas the \xmm\ spectra that yielded the $\min(I)$ measurements typically contain
less SWCX emission than \citeauthor{mccammon02}'s spectrum.

\subsubsection{Implications for Sources of SXRB Emission}
\label{subsubsec:ImplicationsForSources}

Here we will compare the observed fraction of the non-EPL 3/4~\kev\ SXRB due to the oxygen lines
with the fractions expected from different sources. We will show that the observed values are lower
than those expected from collisional ionization equilibrium (CIE) plasma in the halo, or from
SWCX. Hence, it is possible that the true SWCX emission is harder than the model SWCX spectrum we
have used, that the halo is out of equilibrium, and/or that oxygen is depleted in the halo.

For this comparison, it will be helpful to plot the observed fractions against \NH, as the fraction
expected from hot plasma in the halo varies with \NH\ (see below). We do this in
Figure~\ref{fig:r4frac-v-NH}. The \xmm\ observations are color-coded according to solar cycle phase,
using the definitions introduced in Section~\ref{subsubsec:HeliosphericSWCX}, above (as noted
earlier, the RASS was carried out during a solar maximum). As in Figure~\ref{fig:Oxygen-vs-R4}, we
just show the results obtained using $\min(\Iovii)$ and $\min(\Ioviii)$ from directions with
multiple \xmm\ observations.\footnote{Recall that here we are defining $\min(I)$ as the lowest
  intensity measured in a given direction, regardless of whether it was obtained with or without
  proton flux filtering.}

\begin{figure}
\plotone{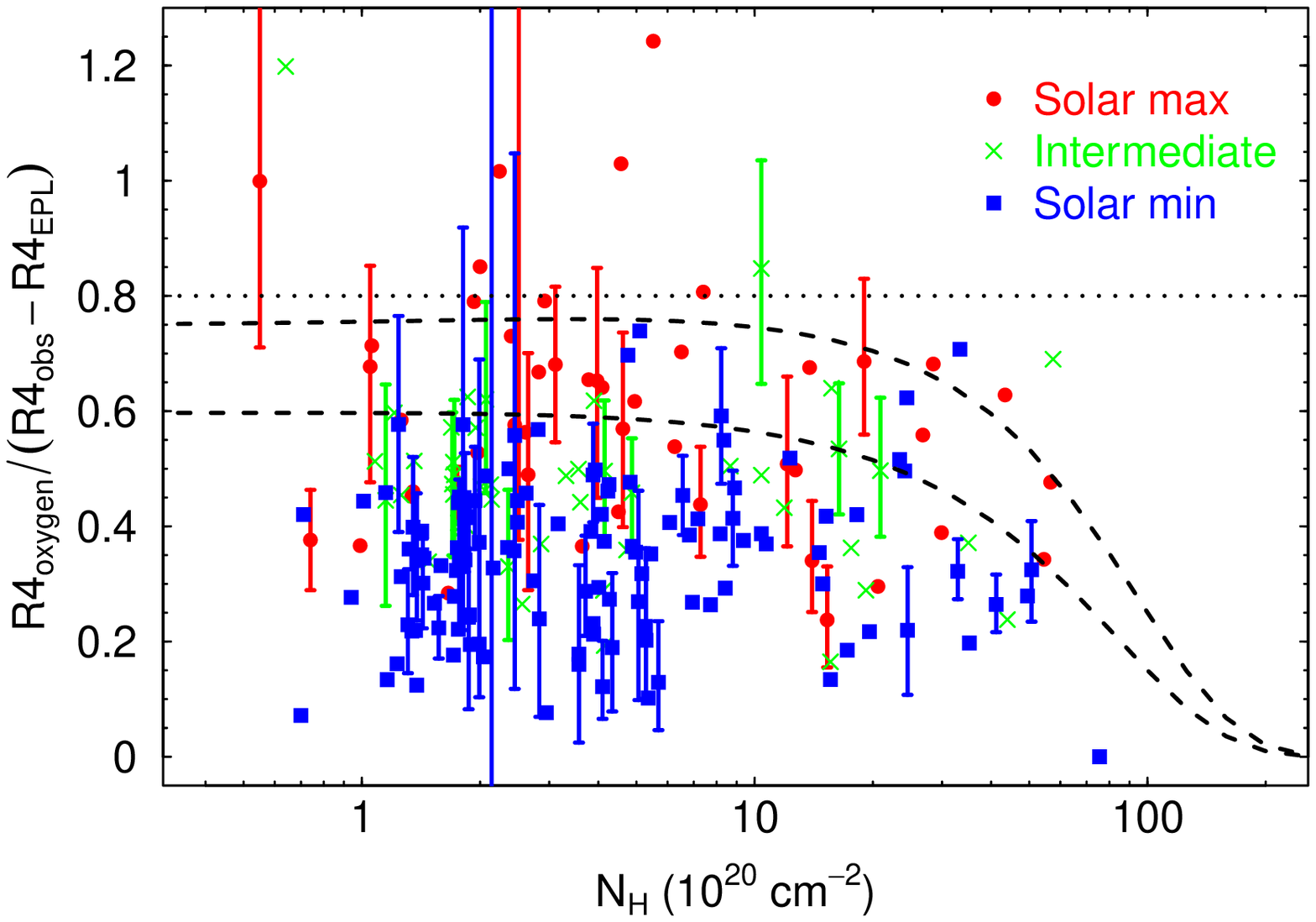}
\caption{Fraction of the non-EPL portion of the \rosat\ R4 count-rate that is due to the oxygen
  \Kalpha\ lines, as a function of the \HI\ column density, \NH\ \citep{kalberla05}. We just show
  the results obtained using $\min(\Iovii)$ and $\min(\Ioviii)$ from directions with multiple
  \xmm\ observations (see text for details). For clarity, only a subset of the data points' error
  bars are plotted. The data are color-coded according to the phases of the solar cycle during which
  the \xmm\ observations were carried out (see Section~\ref{subsubsec:HeliosphericSWCX}).  The upper
  and lower dashed curves show the expected R4 fraction for a single-temperature CIE halo plasma of
  temperature $2 \times 10^6$ and $3 \times 10^6~\K$, respectively. The horizontal dotted line shows
  the approximate R4 fraction expected for SWCX emission.
  \label{fig:r4frac-v-NH}}
\end{figure}

\paragraph{Fractions of R4 Emission Due to Oxygen Expected from Halo Plasma and from SWCX}
The dashed curves in Figure~\ref{fig:r4frac-v-NH} show how $\Roxygen/(\Robs - \REPL)$ would vary
with \NH\ if the non-EPL 3/4~\kev\ emission were dominated by emission from a 2 or $3 \times
10^6~\K$ plasma in CIE in the Galactic halo. Note that these model values decrease toward large
\NH. This is because the photoelectric absorption cross-section increases with decreasing photon
energy, and so as \NH\ increases, the lower-energy emission in the R4 band (including the oxygen
lines) is attenuated more than the higher-energy emission.

Even though we are using the $\min(I)$ measurements, SWCX may still be contributing to the
3/4~\kev\ SXRB emission. To estimate the fraction of the R4 count-rate from a typical SWCX spectrum
due to the oxygen \Kalpha\ lines, we use the spectrum shown in Figure~2 of
\citet{koutroumpa09d}. The line intensities for both solar minimum and solar maximum shown in this
plot imply that $\approx$80\%\ of the SWCX R4 count-rate is due to the oxygen \Kalpha\ lines
(horizontal dotted line in Figure~\ref{fig:r4frac-v-NH}), which is slightly larger than the fraction
expected from CIE plasma in the halo for $\NH \la \mathrm{few} \times 10^{21}~\pcmsq$. We obtain
similar fractions if we use the SWCX spectrum shown in Figure~1 of \citet{koutroumpa09a}, or if we
calculate the relative intensities of SWCX lines in the R4 band using cross-sections for \CV, \CVI,
\NV, \NVII, \OVII, \OVIII\ lines from Table~8.2 of \citet{bodewits07} and for \NeIX\ and
\MgXI\ lines from Table~1 of \citet{koutroumpa06}, with ion abundances also from Table~1 of
\citet{koutroumpa06}.\footnote{For this calculation we consider only ions interacting with H, as
  Table~8.2 of \citet{bodewits07} does not give data for ions interacting with He. However, it
  should be re-iterated that this calculation yields a similar value to those obtained from the
  \citet{koutroumpa09d,koutroumpa09a} SWCX spectra, which do include emission from ion-He
  interactions.}

The observed values of $\Roxygen/(\Robs - \REPL)$ are generally much lower than those expected from
a $\sim$$(\mbox{2--3}) \times 10^6~\K$ CIE plasma in the halo, or from SWCX emission.  Hence, a
combination of these two emission mechanisms seems to be unable to explain the observed values of
$\Roxygen/(\Robs - \REPL)$. In the remainder of this subsection we will explore possible
explanations for the low observed values of $\Roxygen/(\Robs - \REPL)$.

\paragraph{Is the SWCX Emission Fainter in the \xmm\ Data than in the \rosat\ Data?}
Here, we consider the possibility that the SWCX emission (specifically, the oxygen emission) is
typically fainter in the \xmm\ observations than it was during the RASS; i.e., the values of
\Roxygen\ derived from the \xmm\ data may be smaller than the values we would have obtained had we
been able to measure \Roxygen\ at the same time as $\Robs - \REPL$. Such an occurrence would have
the effect of lowering the $\Roxygen/(\Robs - \REPL)$ ratio. This explanation for the low observed
ratios is plausible because the \xmm\ observations that yield low values of $\Roxygen/(\Robs -
\REPL)$ were typically taken during solar minimum, while the RASS was carried out during solar
maximum, and, as we showed in Section~\ref{subsubsec:HeliosphericSWCX}, the oxygen intensities are
typically fainter during solar minimum than during solar maximum (see Figure~\ref{fig:SWCXboxplot}).
However, it should be noted that a few \xmm\ observations taken during solar maximum (i.e., the
equivalent phase as the RASS) also yield low values of $\Roxygen/(\Robs - \REPL)$. Furthermore, when
we examine this explanation more quantitatively, we find that it is unable to reproduce the observed
values of $\Roxygen/(\Robs - \REPL)$ for \xmm\ observations taken during solar minimum, as we will
now demonstrate.

We begin by defining \RSWCX\ and \Rhalo\ as the contributions to the RASS R4 count-rate from SWCX
and the halo, respectively; i.e., $\Robs - \REPL = \RSWCX + \Rhalo$ (we are assuming the
contribution from the LB is negligible). Note that \RSWCX\ is the count-rate due to SWCX during
solar maximum. If \gammaSWCX\ and \gammahalo\ are the expected fractions of \RSWCX\ and \Rhalo,
respectively, due to oxygen, and $\xi$ is the typical factor by which oxygen SWCX intensities are
reduced in going from solar maximum (epoch of the RASS) to solar minimum (blue data points in
Figure~\ref{fig:r4frac-v-NH}), then a typical value of \Roxygen\ for an \xmm\ observation taken
during solar minimum would be given by
\begin{equation}
  \Roxygen = \xi \gammaSWCX \RSWCX + \gammahalo \Rhalo.
\end{equation}
Since $\gammaSWCX \approx 0.8$ and $\gammahalo \approx 0.7$ (see Figure~\ref{fig:r4frac-v-NH}),
\begin{equation}
  \frac{\Roxygen}{\Robs - \REPL} \approx 0.8 (1 - f) \xi + 0.7 f,
  \label{eq:r4frac}
\end{equation}
where $f \equiv \Rhalo / (\Robs - \REPL)$ is the typical fraction of the EPL-subtracted R4
count-rate that is from the halo. For the R45 band, Table~2 in \citet{kuntz00} implies that 78\%\ of
the count-rate not due to the EPL or point sources is from the halo. An equivalent measurement for
the R4 band is not available, but using the temperatures and normalizations of the relevant model
components from \citet{kuntz00}, we estimate that $f \approx 0.7$.

For the observations taken during solar minimum, $\Roxygen/(\Robs - \REPL)$ is typically 0.24--0.44
(these are the lower and upper quartiles, respectively). For $f \approx 0.7$, we find that
Equation~(\ref{eq:r4frac}) can only yield $\Roxygen/(\Robs - \REPL) = 0.24$--0.44 if $\xi < 0$,
which is unphysical. Hence, while the lower oxygen SWCX intensities during solar minimum may be
partly responsible for the low observed values of $\Roxygen/(\Robs - \REPL)$, they do not fully
explain the observations. Also, as noted above, a few \xmm\ observations taken during solar maximum
also yield low values of $\Roxygen/(\Robs - \REPL)$.

\paragraph{Is the Halo Much Hotter than $\mathit{\sim}$$\mathit{3 \times 10^6~\K}$?}
For a CIE plasma with $T \ga 2 \times 10^6~\K$, $\Roxygen/(\Robs - \REPL)$ is expected to decrease
with increasing temperature (e.g., compare the curves for $T = 2$ and $3 \times 10^6~\K$ in
Figure~\ref{fig:r4frac-v-NH}).  This is because, as the temperature increases, the harder emission
in the R4 band brightens relative to the oxygen emission. However, the low observed values of
$\Roxygen/(\Robs - \REPL)$ are unlikely to mean that the halo temperature is much greater than $3
\times 10^6~\K$. Such a high temperature would exceed the Galaxy's virial temperature
\citep[e.g.,][]{bregman09b}, and the hot halo gas would escape from the Galaxy's potential well.

It should also be noted that some observations that have yielded typical halo temperatures elsewhere
yield low values of $\Roxygen/(\Robs - \REPL)$ here.  For example, 22 out of the 26
\xmm\ observations studied by \citet{henley10b} yielded halo temperatures between $1.5 \times 10^6$
and $2.5 \times 10^6~\K$, assuming CIE. Of these 22 observations, 14 yield $\Roxygen/(\Robs - \REPL)
< 0.5$.

We have excluded differences in the SWCX oxygen intensities between the RASS data and the
\xmm\ data, and a very high halo temperature as possible explanations for the low observed values of
$\Roxygen/(\Robs - \REPL)$. In the following paragraphs, we consider three further explanations that
we cannot exclude. Currently, we cannot discriminate between these proposed explanations.

\paragraph{Is the SWCX Emission Harder than Expected?}
It is possible that the true SWCX spectrum is harder than the model SWCX spectra we have used, which
would mean that $\gammaSWCX \approx 0.8$ is an overestimate. This could arise from inaccuracies in
the assumed charge exchange cross-sections or ion fractions. If there were relatively more flux from
lines with energies greater than those of the oxygen \Kalpha\ lines, it would help explain the low
observed values of $\Roxygen/(\Robs - \REPL)$.

\paragraph{Is the Halo out of Equilibrium?}
It is also possible that the true halo spectrum is harder than that expected from a
$\sim$$(\mbox{2--3}) \times 10^6~\K$ CIE plasma. We ruled out above the possibility that the halo
temperature is much greater than $3 \times 10^6~\K$. Instead, these results may imply that the halo
is overionized and recombining \citep{breitschwerdt94,breitschwerdt99}. The emission from an
overionized, recombining plasma may include relatively bright recombination continuum emission at
higher energies than the oxygen lines. Hence, the fraction of the EPL-subtracted R4 emission due to
oxygen may be lower than that expected from a few-million-degree CIE plasma, as observed. If this is
true, it would have important implications for spectral analyses of the halo, as it suggests that
CIE halo models with best-fit temperatures of $\sim$$(\mbox{2--3}) \times 10^6~\K$
\citep[e.g.,][]{galeazzi07,lei09,yoshino09,gupta09b,henley10b} may not accurately model the
higher-energy emission from the halo.

\paragraph{Is Oxygen Depleted in the Halo?}
When calculating the values of $\Roxygen/(\Robs - \REPL)$ expected from a hot halo plasma
(Figure~\ref{fig:r4frac-v-NH}), we assumed solar abundances \citep{anders89}. However, it is
possible that oxygen is depleted in the halo. \citet{yoshino09} found evidence from a subset of
their \suzaku\ observations that neon and iron are enhanced in the halo relative to oxygen. If
oxygen were depleted in the halo, $\Roxygen/(\Robs - \REPL)$ would be lower than expected for a
solar-abundance plasma, which is what we observe. If oxygen is underabundant in the halo then,
similarly to above, models that assume solar abundances
\citep[e.g.,][]{galeazzi07,lei09,gupta09b,henley10b} may not accurately model the halo's
higher-energy emission.

\subsection{Spatial Variation of the Oxygen Intensities}
\label{subsec:SpatialVariation}

\begin{figure*}
\plottwo{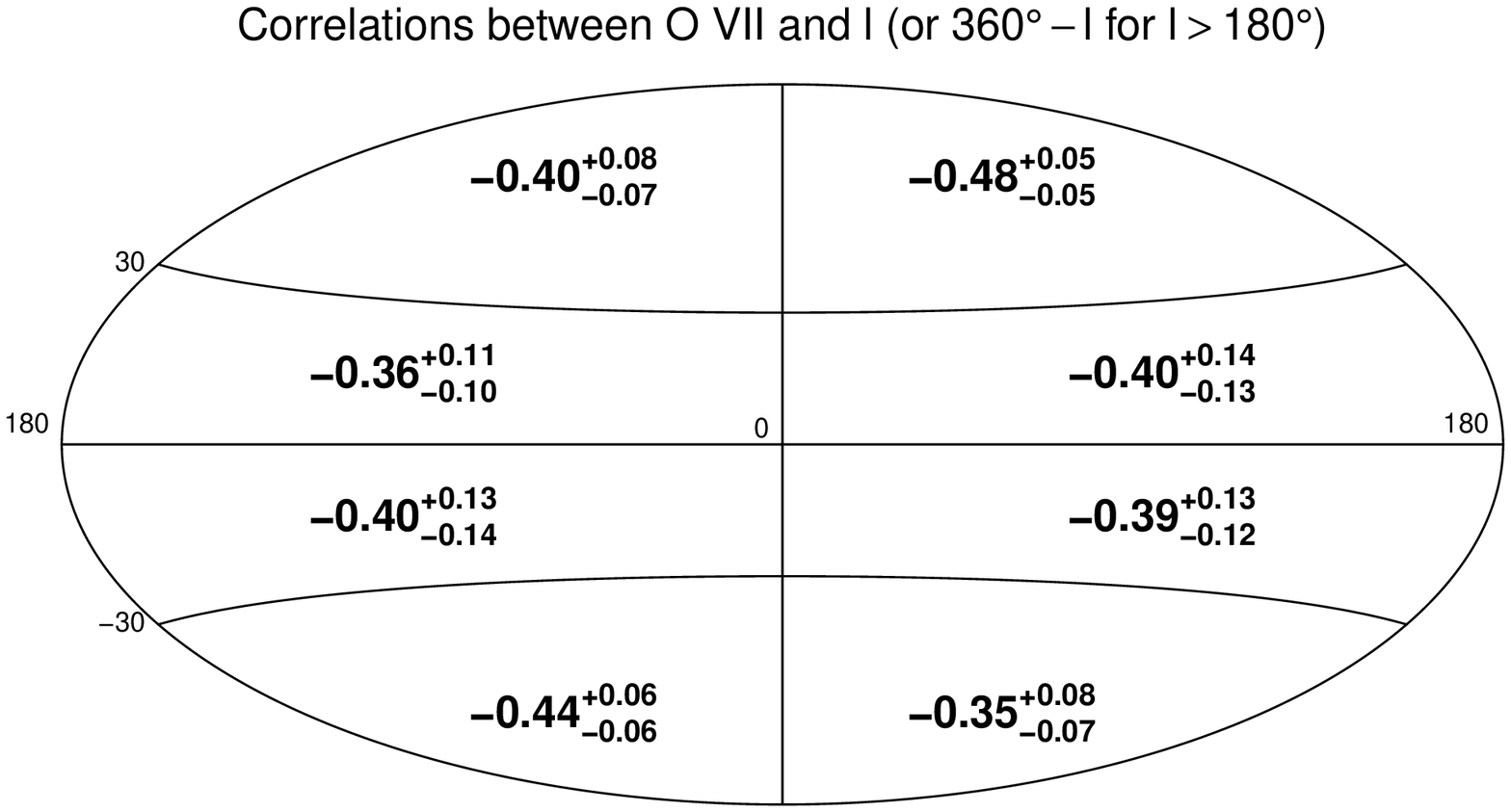}{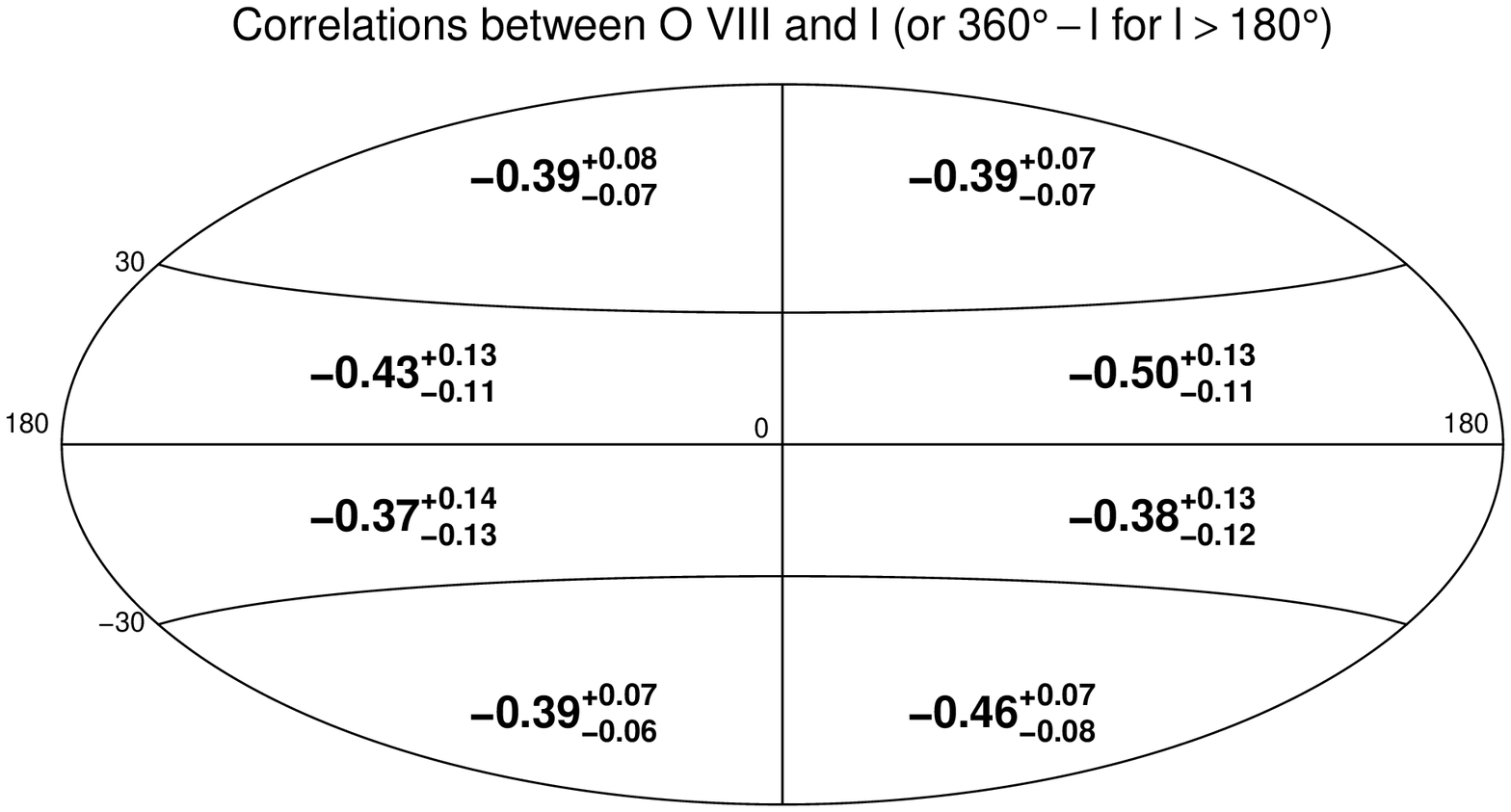}
\plottwo{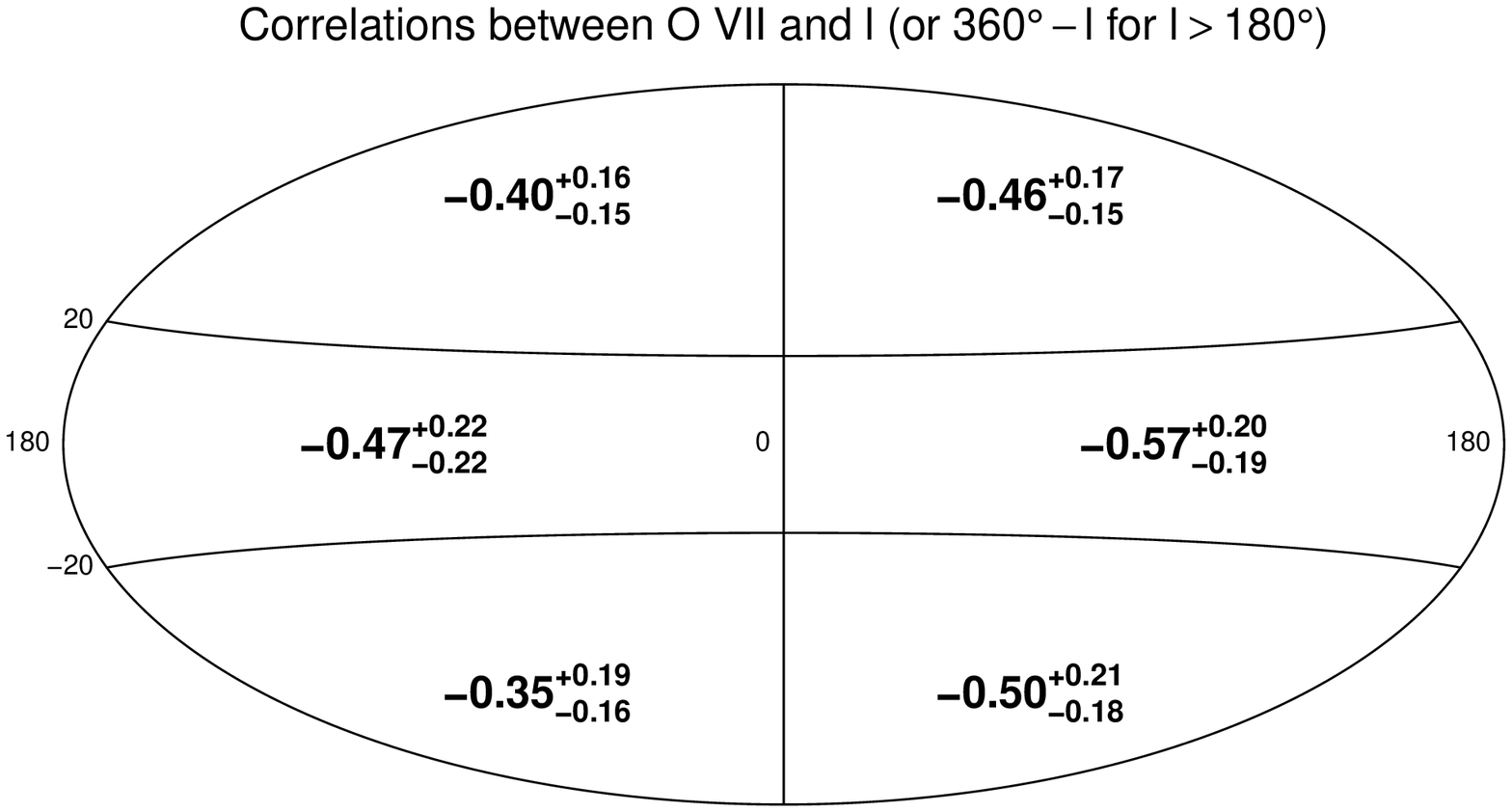}{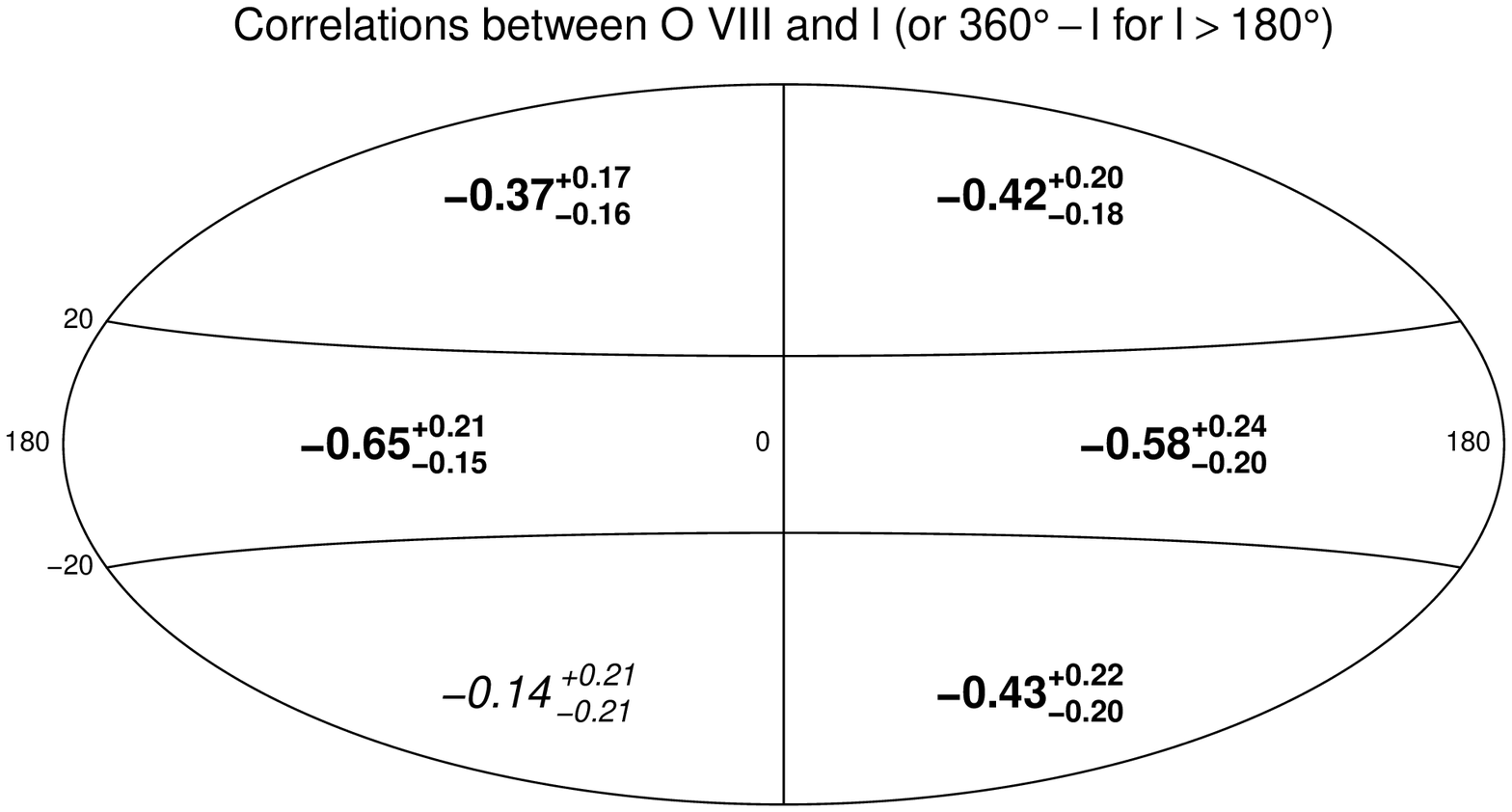}
\caption{Kendall's $\tau$ correlation coefficients for \Iovii\ (left) and \Ioviii\ (right) against
  $l$ for directions with $0\degr \le l < 180\degr$ or against $360\degr-l$ for directions with
  $180\degr \le l < 360\degr$, for different regions of the sky. The upper pair of plots shows the
  results for intensities obtained with proton flux filtering (observations identified as
  SWCX-contaminated by \citet{carter11} are excluded, as are observations toward the Eridanus
  enhancement and the Magellanic Clouds). The lower pair of plots show the results for directions
  with multiple \xmm\ observations, using the minimum intensity measured in each direction (again,
  observations toward the Eridanus enhancement and the Magellanic Clouds are excluded). The
  uncertainties indicate the 90\%\ confidence interval on $\tau$ obtained by bootstrap
  resampling. The results are generally shown in boldface; the one case for which the confidence
  interval encompasses zero is shown in italics.
  \label{fig:Intensity-vs-l}}
\end{figure*}

The variation of the oxygen intensities with Galactic longitude and latitude can provide information
on the general geometry of the halo. This, in turn, provides information on the mechanism
responsible for heating the hot halo gas. For example, if the heating of the halo is dominated by
SN-driven outflows from the disk \citep[e.g.,][]{shapiro76,bregman80,joung06,henley10b}, and the SN
rate per unit area of the disk is fairly uniform in the vicinity of the Sun, then we would expect an
approximately plane-parallel geometry for the halo, with the same intrinsic halo brightness for all
longitudes. If the SN rate increases toward the Galactic Center, we would expect the halo brightness
to increase toward the inner Galaxy. An increase in the halo brightness toward the inner Galaxy may
also indicate that a halo of accreted extragalactic material centered on the Galactic Center
\citep[e.g.,][]{crain10} is contributing to the halo emission.

In practice it is difficult to determine the true halo geometry from our intensity measurements,
because of the unknown amount of SWCX emission contributing to each measurement. However, we can
gain some insight by looking for correlations between the observed intensities and Galactic
longitude and latitude. As we do not expect the SWCX intensity to be correlated with Galactic
longitude and latitude, any observed correlations most likely reflect variations in the observed halo
intensity, which may be due to variations in the intrinsic intensity or in \NH.

In Sections~\ref{subsubsec:LongitudeVariation} and \ref{subsubsec:LatitudeVariation} we will examine
the variation of the observed intensities with Galactic longitude and latitude, respectively. Then,
in Section~\ref{subsubsec:ModelingLatitudeVariation}, we will compare the observed variation with
Galactic latitude with the variation expected from simple models of the intrinsic halo emission, in
an attempt to constrain the variation of the intrinsic halo intensity with latitude.

\subsubsection{Variation with Galactic Longitude}
\label{subsubsec:LongitudeVariation}

Figure~\ref{fig:Intensity-vs-l} show the correlation coefficients (Kendall's $\tau$; e.g.,
\citealt{press92}) for the observed intensities against $l$ for directions with $0\degr \le l <
180\degr$ or against $360\degr-l$ for directions with $180\degr \le l < 360\degr$, for different
regions of the sky. In the top panels, we use the intensities obtained with proton flux filtering,
and exclude observations identified as SWCX-contaminated by \citet{carter11}, as well as
observations toward the Eridanus enhancement \citep[e.g.,][]{burrows93,snowden95b} and the Magellanic
Clouds. We calculate the 90\%\ confidence intervals on the correlation coefficients by bootstrap
resampling. In the upper pair of plots, all of the correlation coefficients are significantly
different from zero; they are shown in boldface print.

There is a clear tendency for the observed \OVII\ and \OVIII\ intensities to increase from the outer
Galaxy ($l = 180\degr$) to the inner Galaxy ($l = 0\degr$ or 360\degr); see also
Figure~\ref{fig:Maps}. This trend is seen at all latitudes. However, we discount the trend seen at
high latitudes in the northern Galactic hemisphere, as it is likely due to the presence of the
Scorpius-Centaurus (Sco-Cen) superbubble \citep{egger95} toward the inner Galaxy, and so does not
tell us anything about the halo. Similarly, the trends at low latitudes reflect variations in the
disk, rather than the halo. However, the trends seen toward high southern latitudes most likely
reflect real variation in the observed Galactic halo intensity, a conclusion that is supported by
the fact that the trends are generally still apparent when we use only $\min(\Iovii)$ or
$\min(\Ioviii)$ (the lower pair of panels in Figure~\ref{fig:Intensity-vs-l}; in only one case is
the correlation coefficient consistent with zero, and this is shown in italics in the figure). As
\NH\ is not correlated with $l$ in this region of the sky, these results imply that the intrinsic
halo intensity increases from the outer to the inner Galaxy. As noted above, this increase may
indicate an increase in the SN rate toward the inner Galaxy or the presence of a halo of accreted
extragalactic material centered on the Galactic Center.

\begin{figure*}[t]
\centering
\raisebox{0.25\height}{\includegraphics[width=0.24\linewidth]{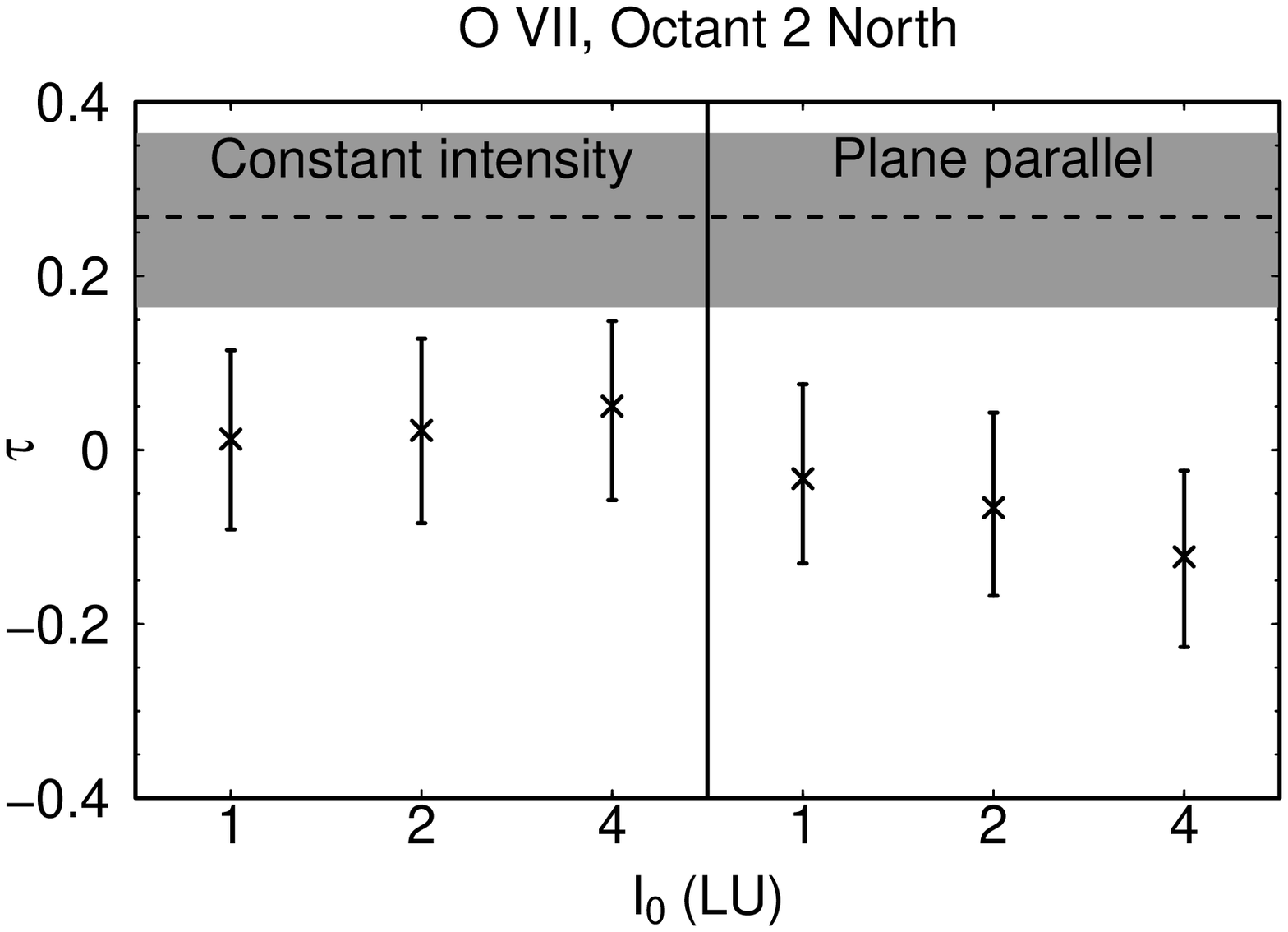}}
\hspace{0.04\linewidth}
\includegraphics[width=0.4\linewidth]{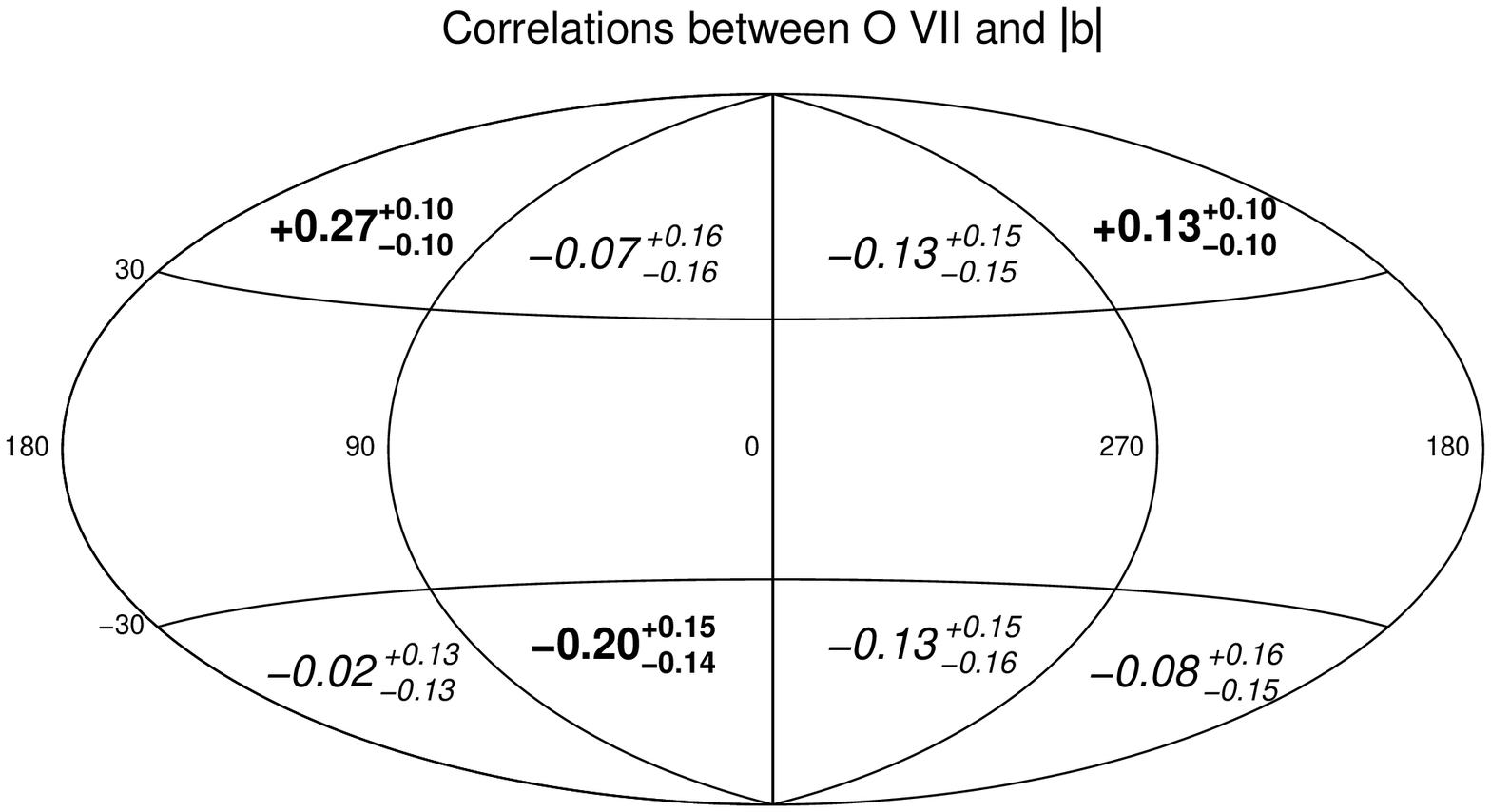}
\hspace{0.04\linewidth}
\raisebox{0.25\height}{\includegraphics[width=0.24\linewidth]{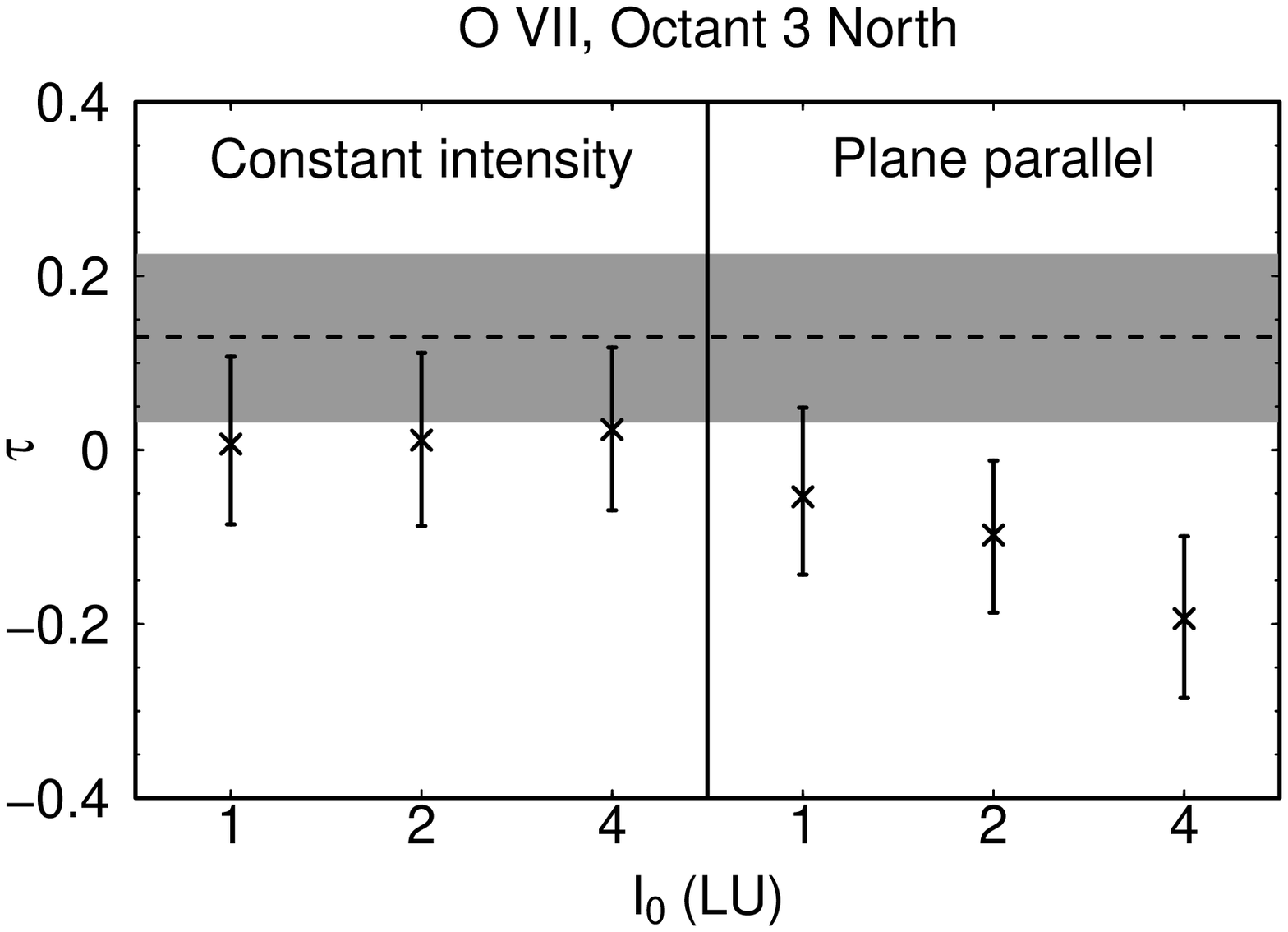}} \\
\includegraphics[width=0.24\linewidth]{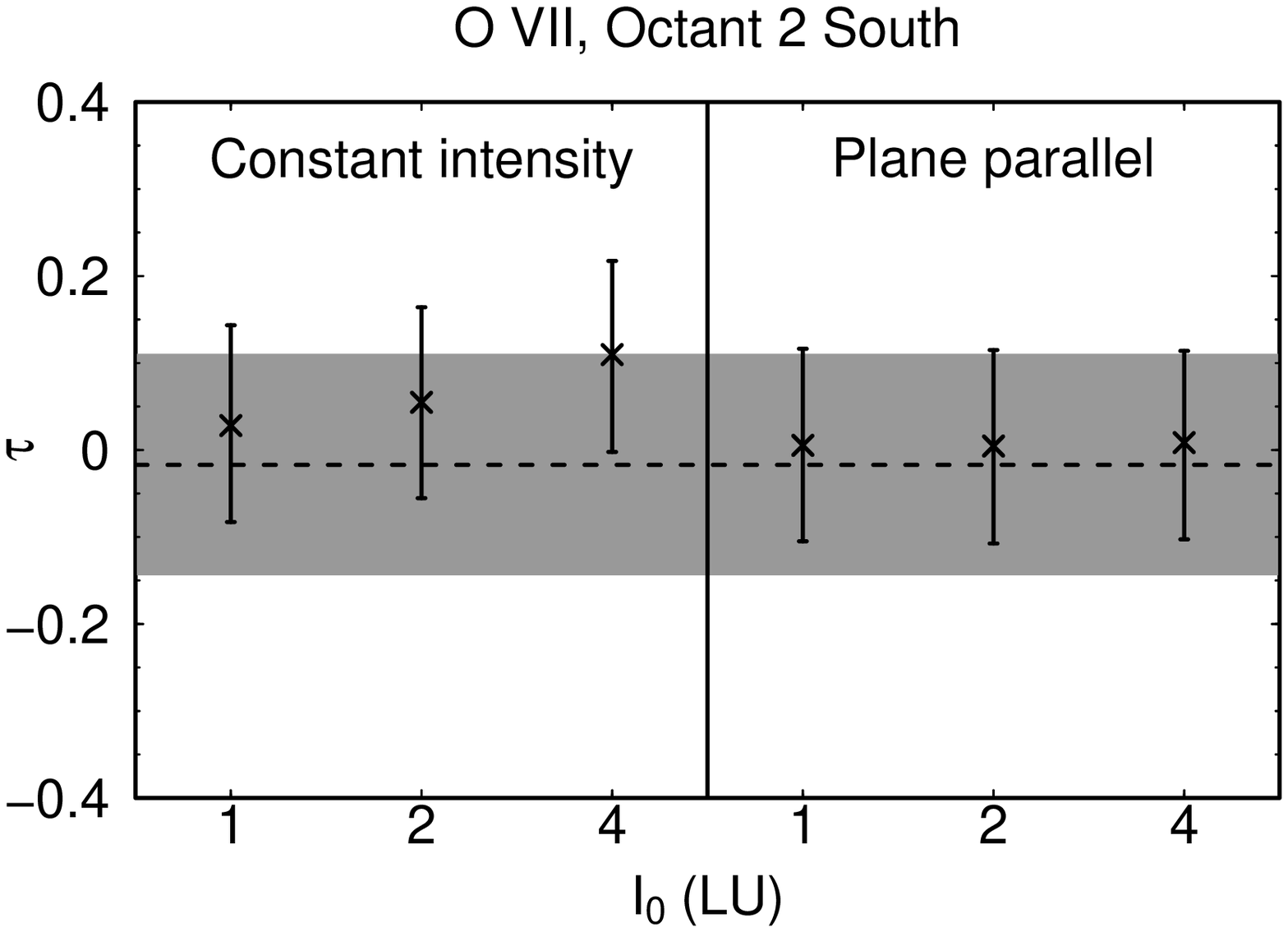}
\includegraphics[width=0.24\linewidth]{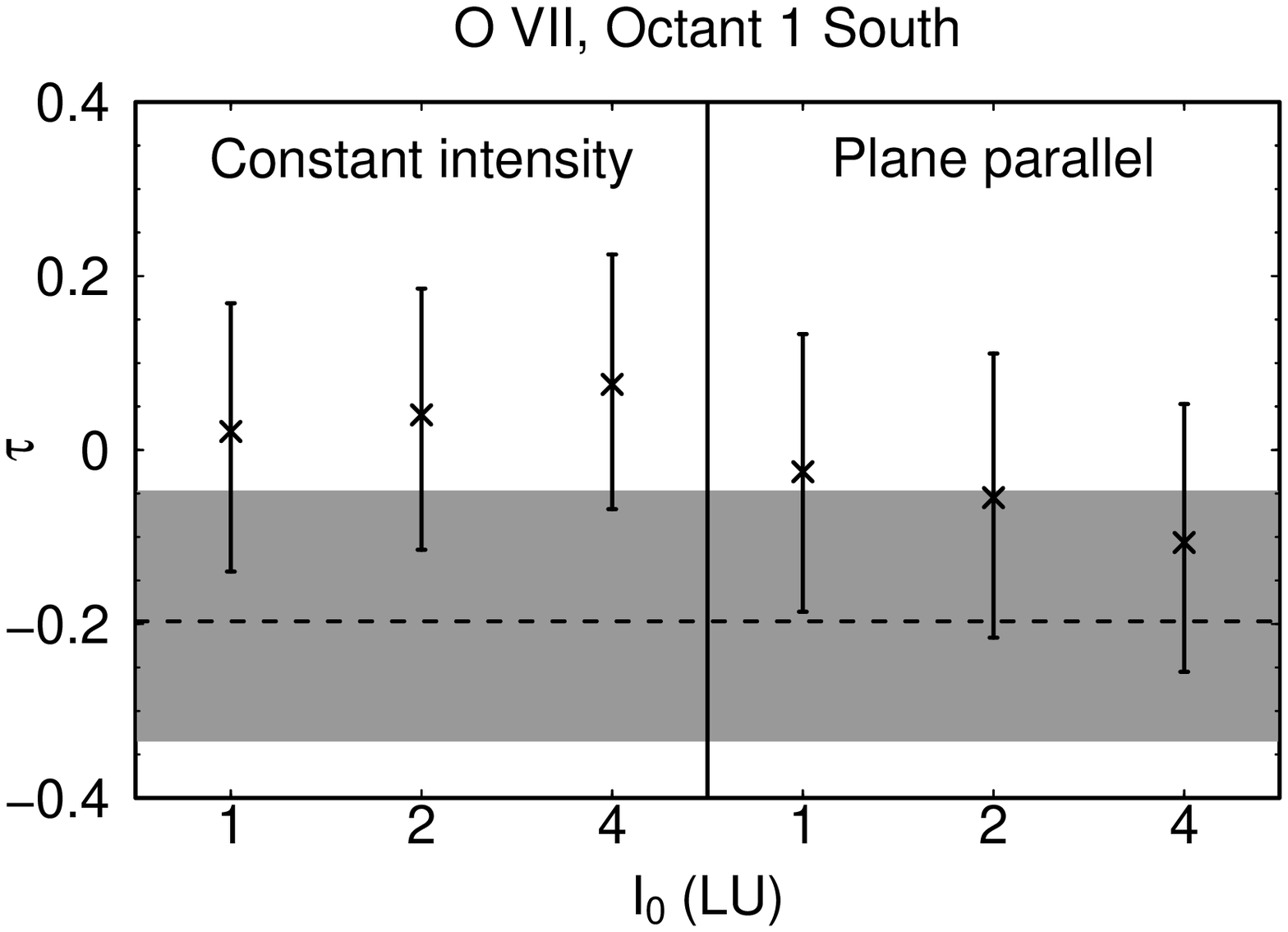}
\includegraphics[width=0.24\linewidth]{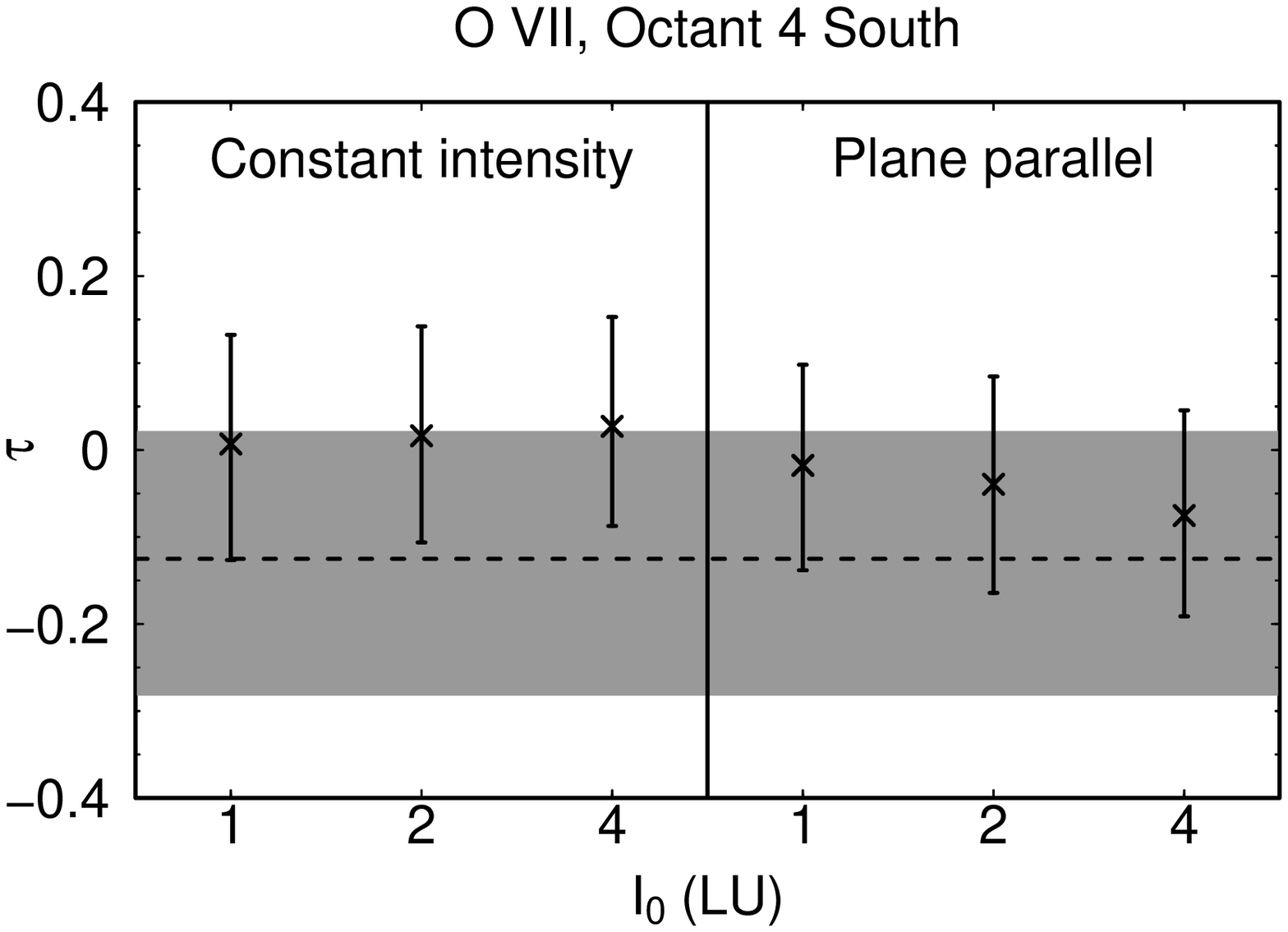}
\includegraphics[width=0.24\linewidth]{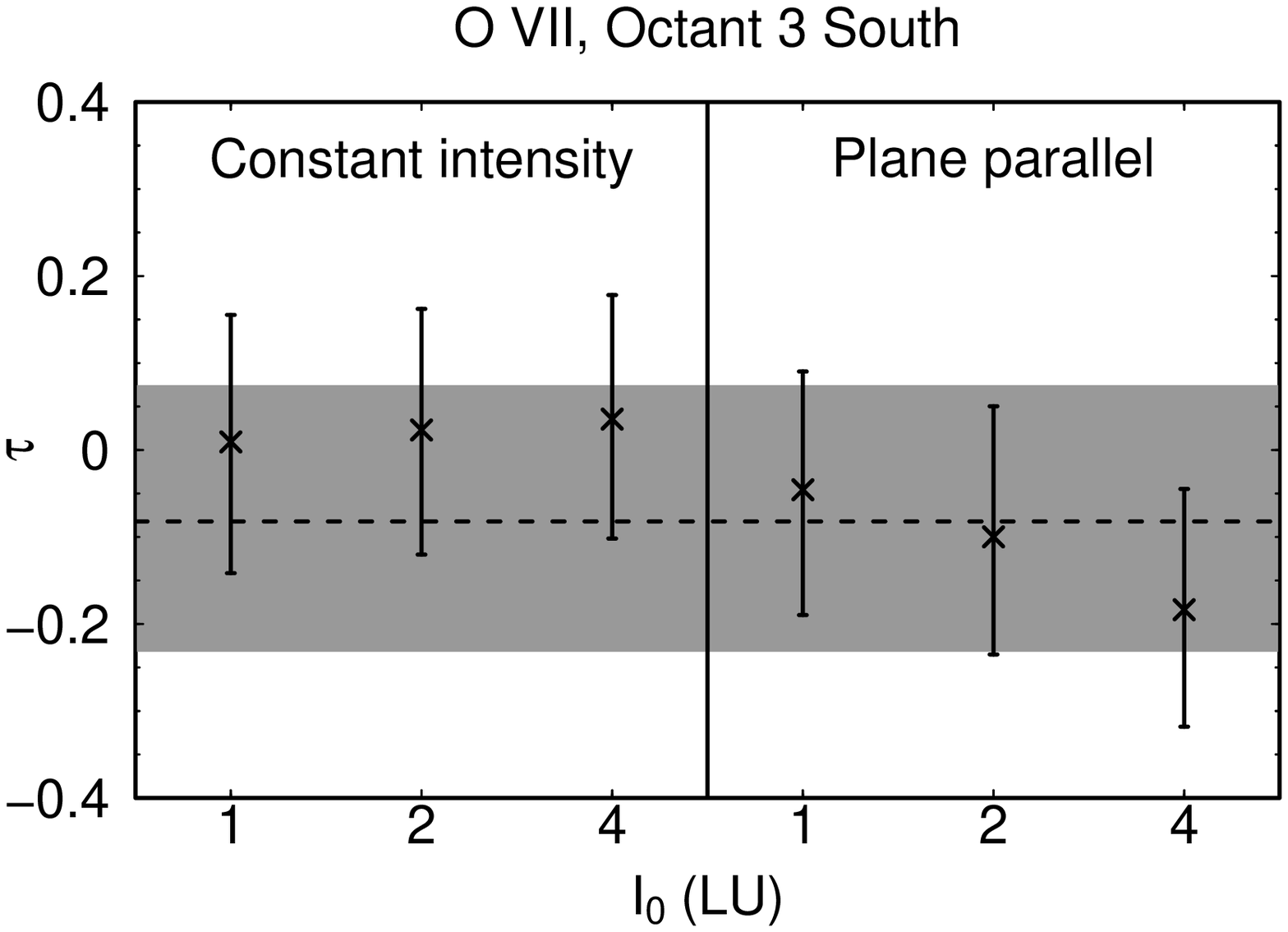}
\caption{\textit{Central upper panel:} Kendall's $\tau$ correlation coefficients for
  \Iovii\ (obtained with proton flux filtering) against $|b|$, for different regions of the sky. We
  consider only sightlines with $|b| \ge 30\degr$, and exclude observations identified as
  SWCX-contaminated by \citet{carter11}, or that are toward the Eridanus enhancement or the
  Magellanic Clouds. The uncertainties indicate the 90\%\ confidence interval on $\tau$ obtained by
  bootstrap resampling; if this interval does not (does) encompass zero, the results are shown in
  boldface (italics).  \textit{Surrounding panels:} Comparisons of the observed correlation
  coefficients with those expected from different halo models, for six different regions of the
  sky. In each panel, the horizontal dashed line shows the observed value of $\tau$, with the gray
  band indicating the 90\%\ bootstrap confidence interval. The errorbars indicate the
  90\%\ confidence intervals on the values of $\tau$ predicted by different halo models. The left
  half of each panel shows the predictions for models in which the intrinsic halo intensity is the
  same in all directions, while the right half of each panel shows the predictions for plane
  parallel halo models, in which the intrinsic halo intensity varies as $\cosec |b|$. For each
  model, we show predictions for different model normalizations, $I_0$, indicated on the $x$-axis.
  See the text for the details of how these model predictions were generated. Note that we do not
  show results for the 1st and 4th northern octants ($0\degr \le l < 90\degr$ and $270\degr \le l <
  360\degr$, respectively), because of the presence of the Sco-Cen superbubble in this region of the
  sky.
  \label{fig:o7-vs-b}}
\end{figure*}

\begin{figure*}
\centering
\raisebox{0.25\height}{\includegraphics[width=0.24\linewidth]{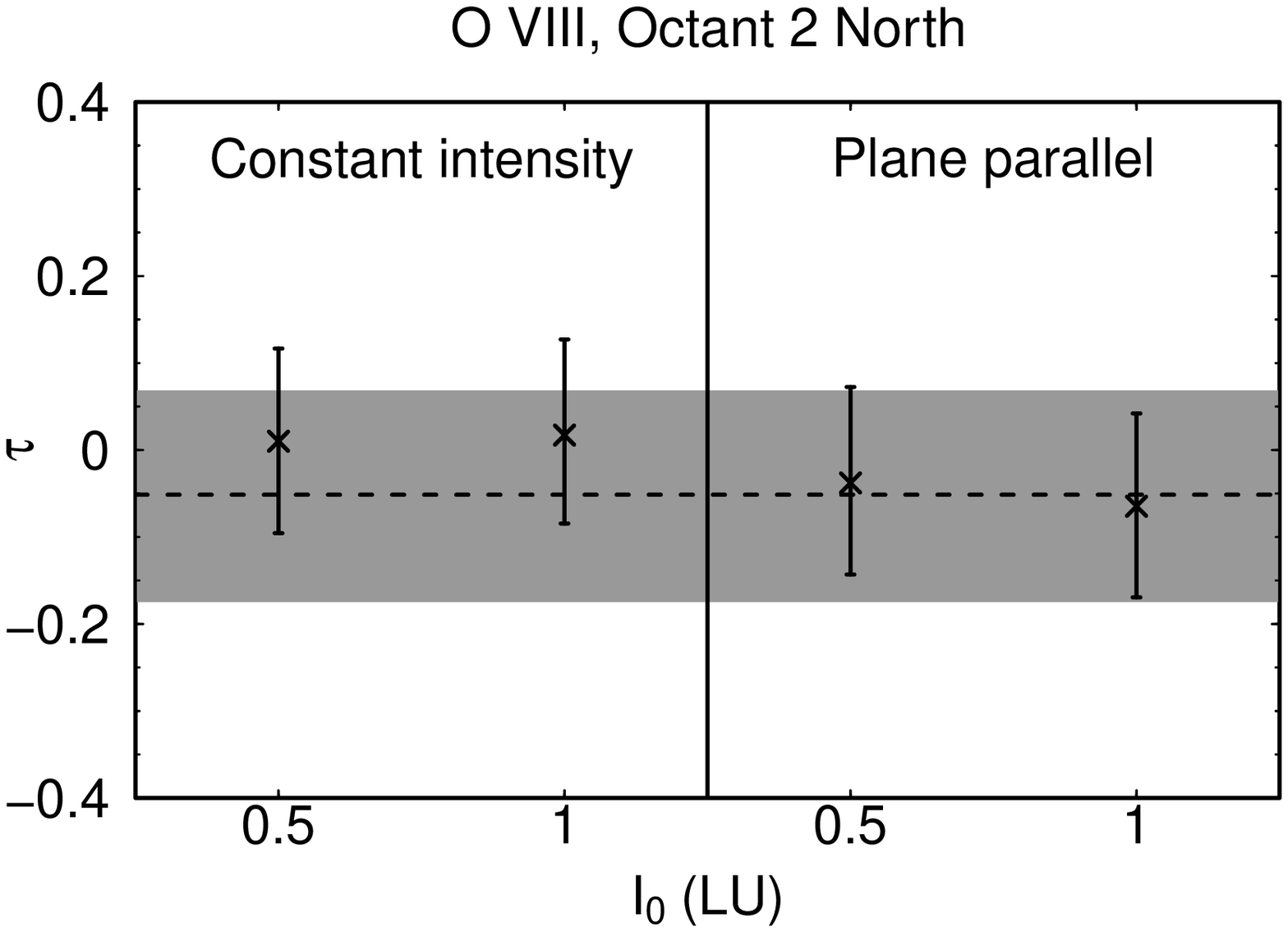}}
\hspace{0.04\linewidth}
\includegraphics[width=0.4\linewidth]{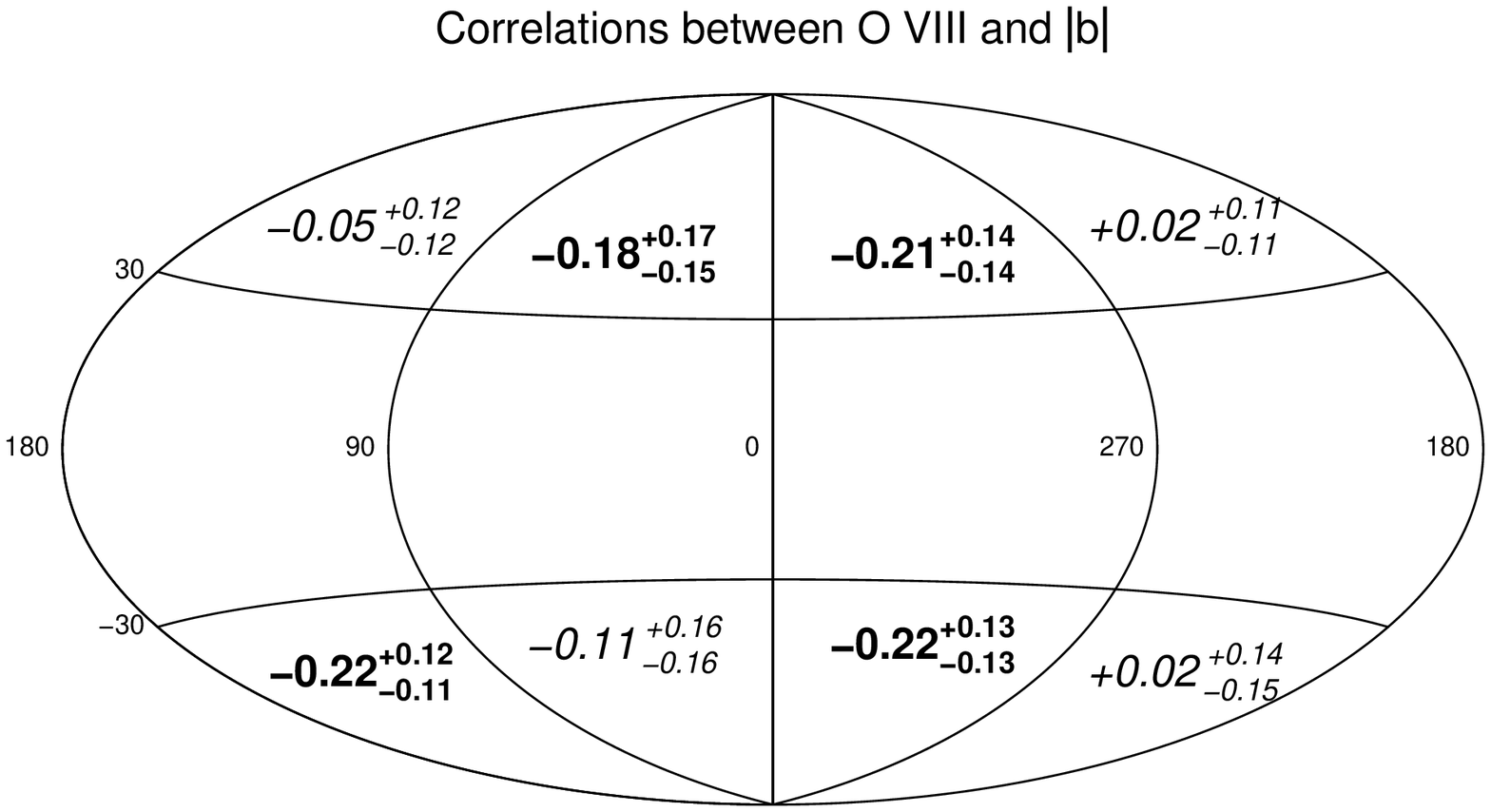}
\hspace{0.04\linewidth}
\raisebox{0.25\height}{\includegraphics[width=0.24\linewidth]{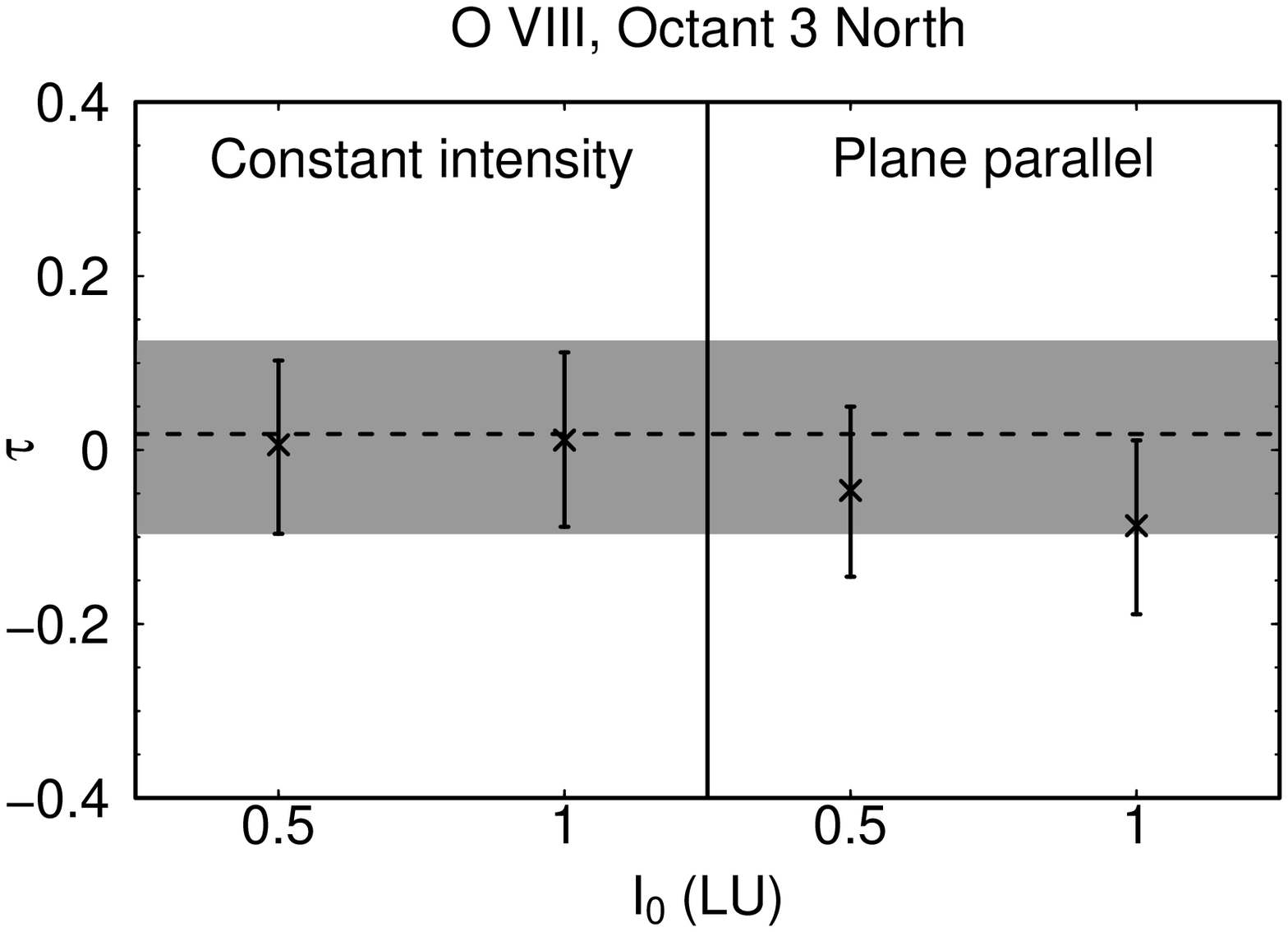}} \\
\includegraphics[width=0.24\linewidth]{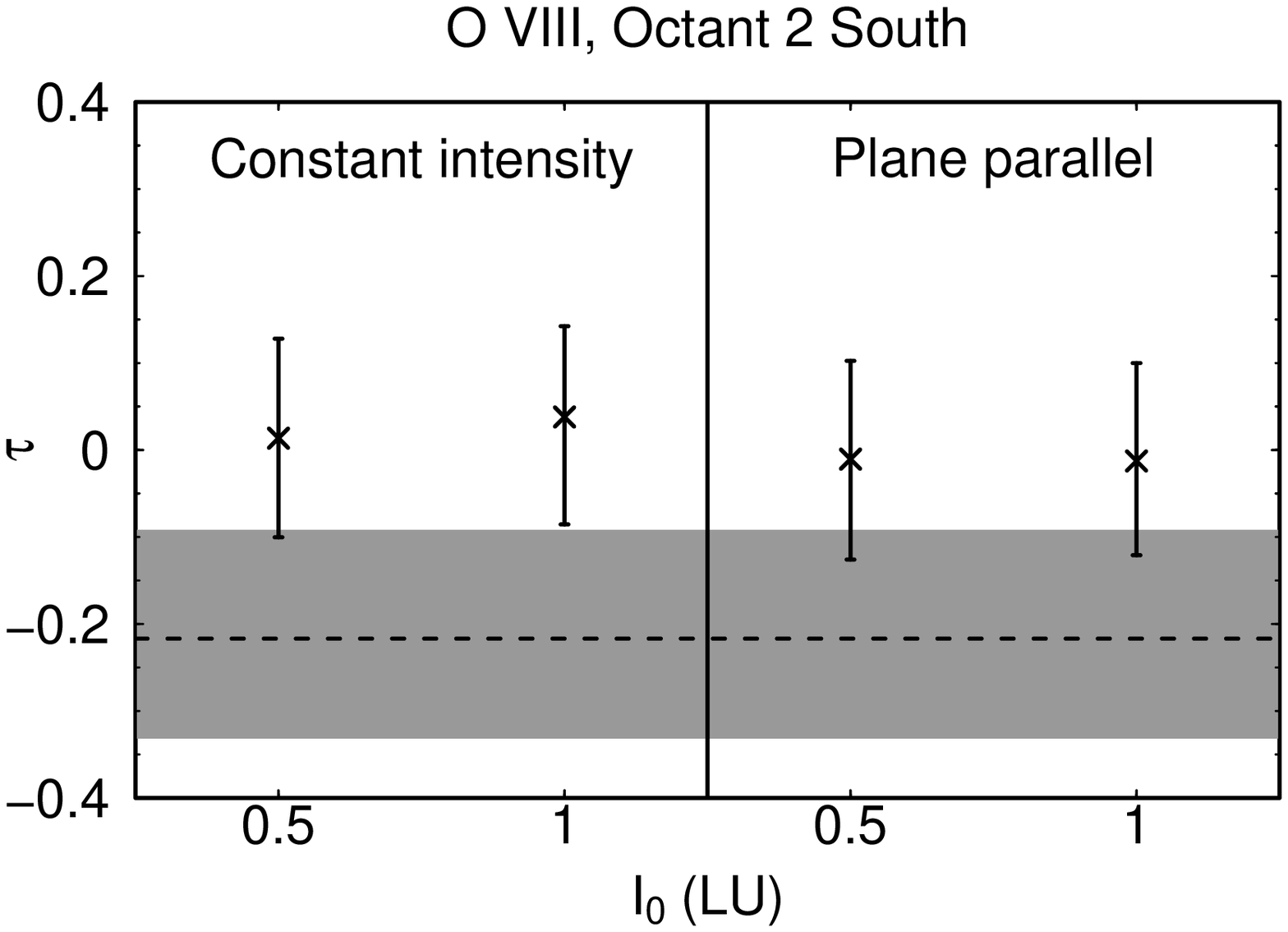}
\includegraphics[width=0.24\linewidth]{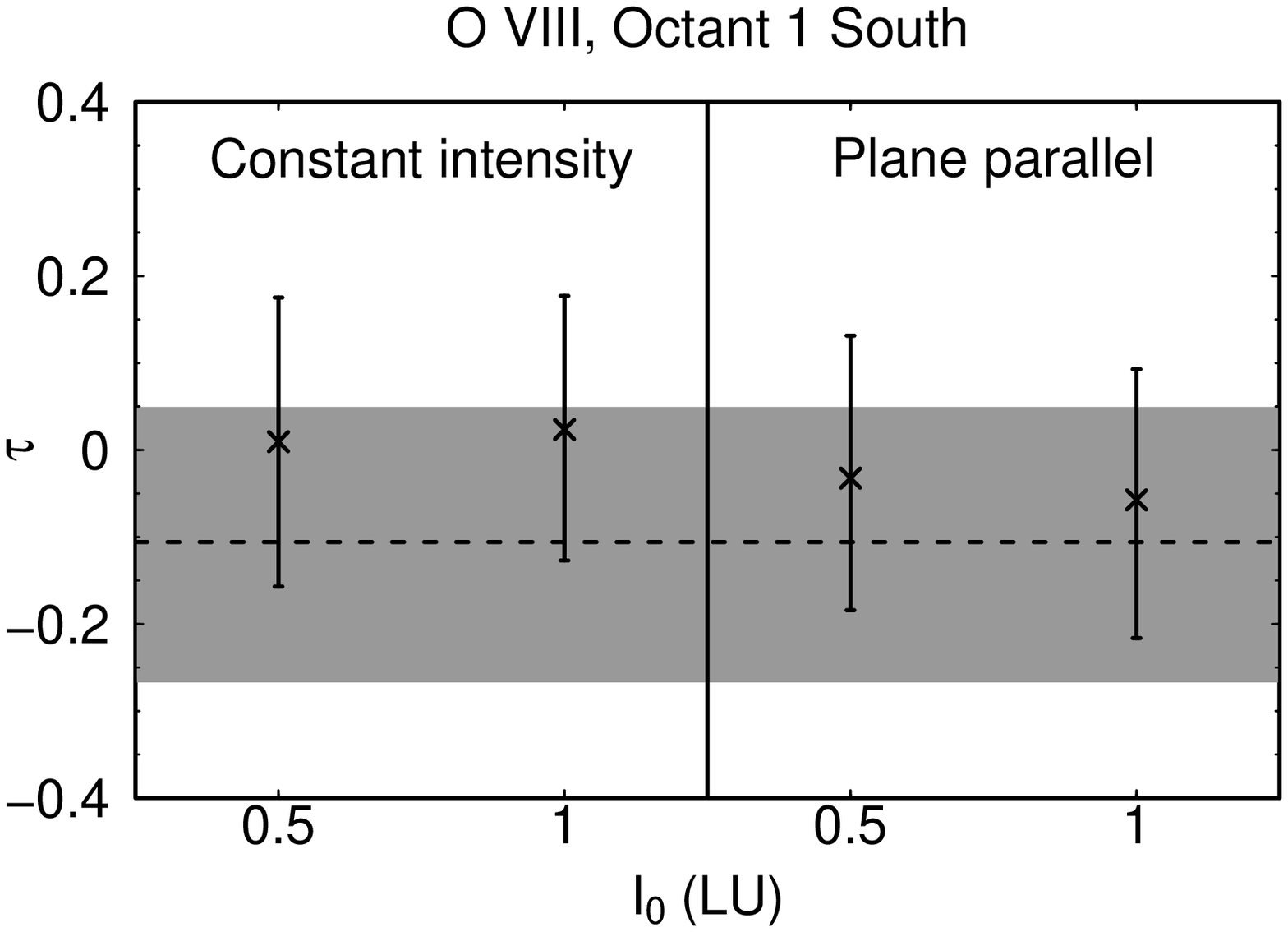}
\includegraphics[width=0.24\linewidth]{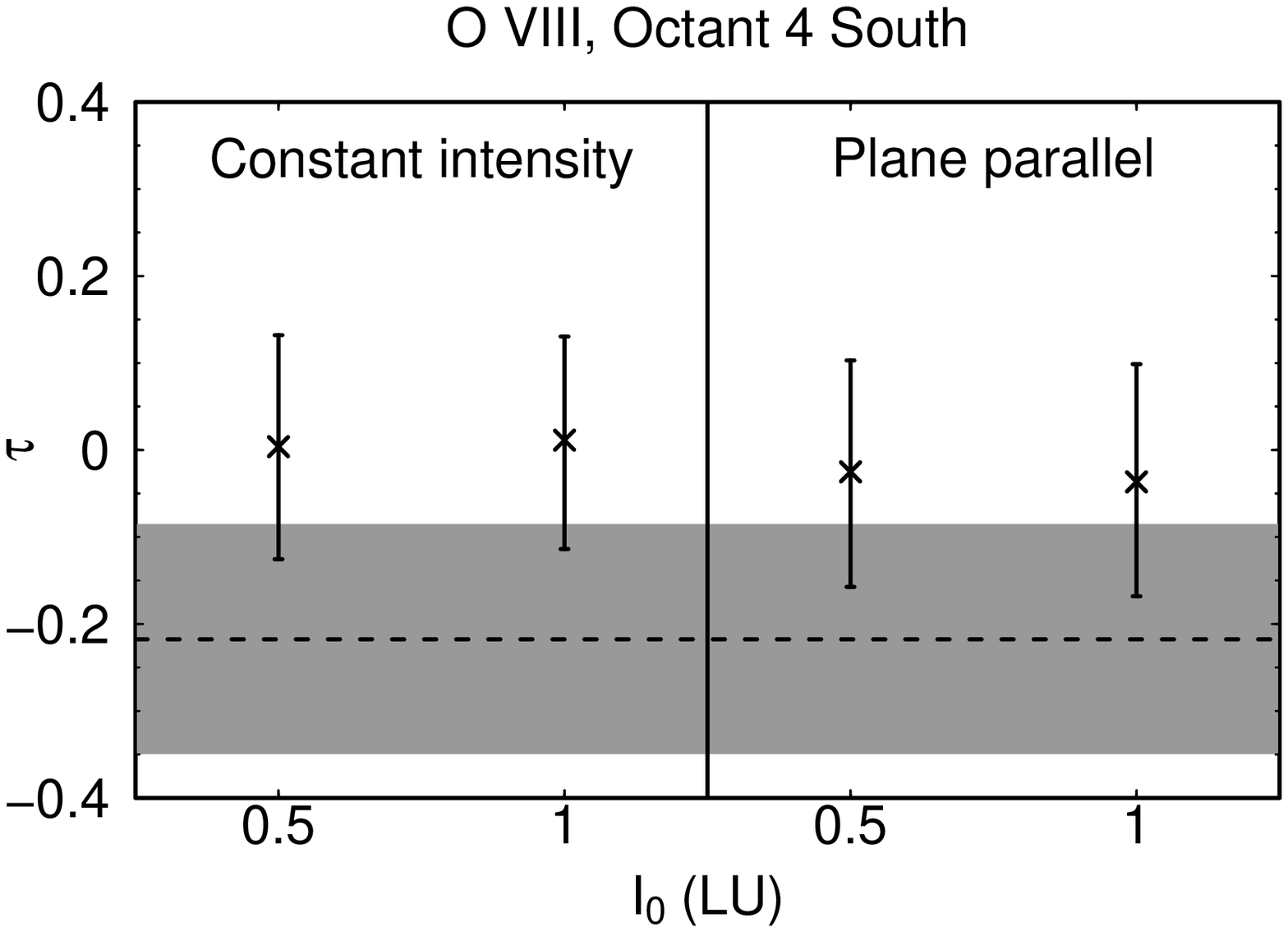}
\includegraphics[width=0.24\linewidth]{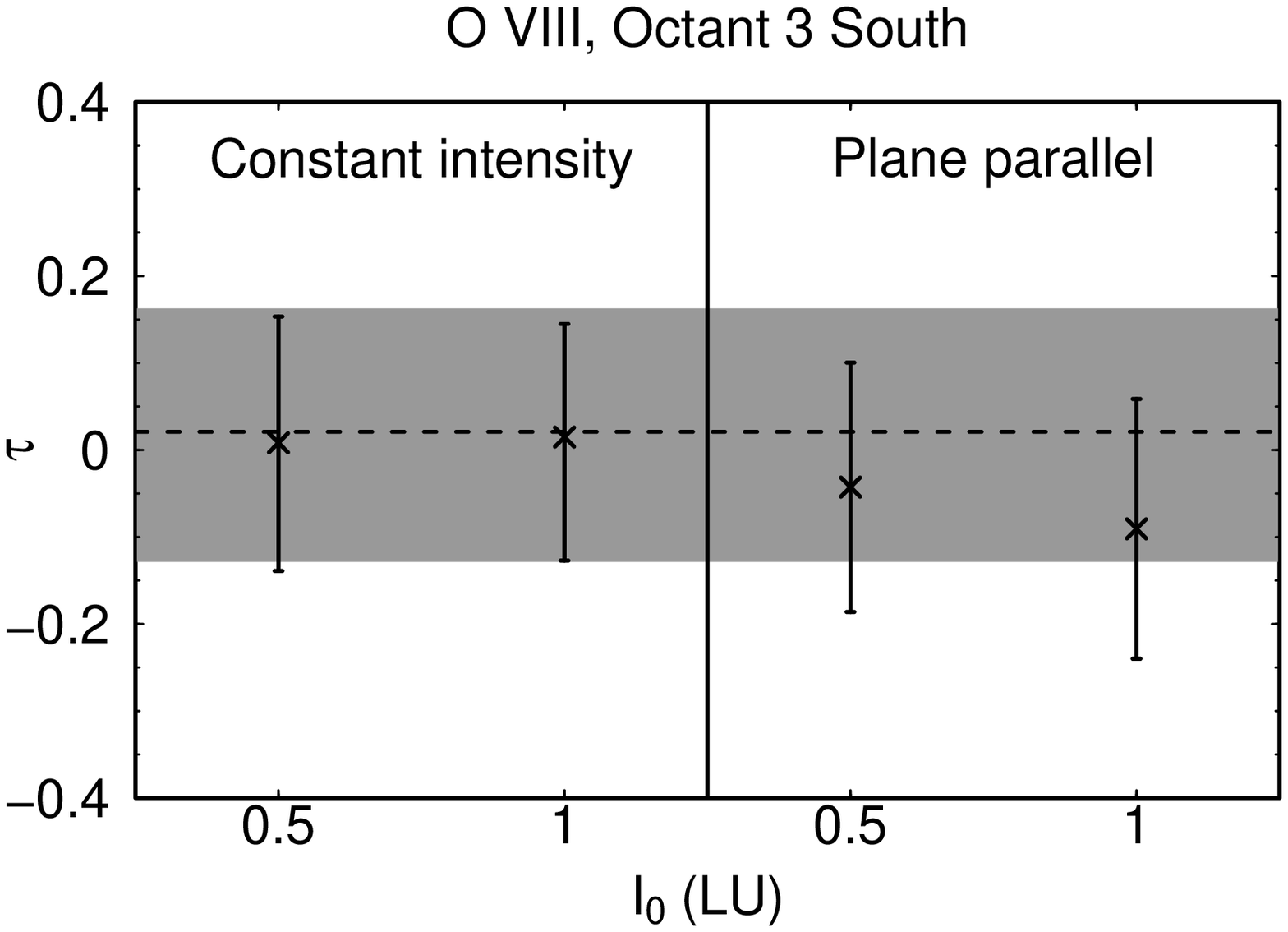}
\caption{As Figure~\ref{fig:o7-vs-b}, but for \OVIII. Note that the values of $I_0$ used to model
  the correlation coefficients here are different from those used in Figure~\ref{fig:o7-vs-b}.
  \label{fig:o8-vs-b}}
\end{figure*}

\subsubsection{Variation with Galactic Latitude}
\label{subsubsec:LatitudeVariation}

The central upper panels in Figure~\ref{fig:o7-vs-b} (\OVII) and Figure~\ref{fig:o8-vs-b} (\OVIII)
show the correlation coefficients for the observed intensities against $|b|$, for different regions
of the sky with $|b| \ge 30\degr$. As in the upper panels of Figure~\ref{fig:Intensity-vs-l}, we use
the intensities obtained with proton flux filtering, and exclude observations identified as
SWCX-contaminated by \citet{carter11}, as well as observations toward the Eridanus enhancement and
the Magellanic Clouds. As in Figure~\ref{fig:Intensity-vs-l}, the correlations that are
significantly different from zero are in boldface print, and those that are consistent with zero are
in italics.

The correlations between intensity and $|b|$ vary significantly across the sky. However, some
general trends do emerge. In the 1st and 4th northern octants ($0\degr \le l < 90\degr$ and
$270\degr \le l < 360\degr$, respectively), the \OVIII\ intensities are generally anticorrelated
with $|b|$ (i.e., the observed intensities tend to decrease toward the northern Galactic pole),
while the \OVII\ intensities are not significantly correlated with $|b|$. However, as this region of
the sky is dominated by emission from the Sco-Cen superbubble, these trends do not tell us anything
about the halo. In the 2nd and 3rd northern octants, the \OVII\ intensities are positively
correlated with $|b|$, while the \OVIII\ intensities are not significantly correlated with $|b|$. In
the southern hemisphere, the intensities are either anticorrelated with $|b|$ (i.e., they tend to
decrease toward the southern Galactic pole), or are not significantly correlated with $|b|$. In
general, the \OVII\ and \OVIII\ lines behave differently (i.e., in most southern octants, only one
of \OVII\ or \OVIII\ is significantly anticorrelated with $|b|$).

\subsubsection{Modeling the Variation with Galactic Latitude}
\label{subsubsec:ModelingLatitudeVariation}

In an attempt to constrain the geometry of the intrinsic halo emission, here we compare the observed
correlation coefficients with the correlation coefficients that we would expect to observe from each
of two different halo models. We use the correlation coefficients, instead of fitting halo models
directly to the observed intensities, because the uncertainty in the contribution from SWCX to each
intensity measurement prevents us from using standard least-squares fitting techniques.

In brief, we simulated many datasets in which each simulated intensity is the sum of the absorbed
model halo emission intensity, a SWCX intensity (which is randomly chosen from the observed $I -
\min(I)$ distribution in order to mimic the unpredictability of time-variable SWCX emission), and a
measurement error. See below for details.  We then calculate the correlation coefficient between the
model intensities and Galactic latitude for each simulated dataset, expecting that the random
time-variable SWCX emission will ``blur'' the correlation coefficient expected from the underlying
halo emission. Having done this procedure for many simulated datasets, we obtain a distribution of
simulated correlation coefficients, which we then compare with the observed correlation coefficient
(in particular, the distribution obtained by bootstrap resampling the observations, which indicates
the uncertainty on the observed correlation coefficient). Essentially, we are asking which halo
model(s) can reproduce the observed correlation coefficients, in the presence of the ``blurring''
effect of the time-variable SWCX emission.  Our method is described in more detail below.

\paragraph{Modeling the Correlation Coefficients for Different Halo Models}
For a given region of the sky, and for a given halo model, we generated 1000 sets of simulated
intensity measurements for each line (\OVII\ and \OVIII). Within each of these sets, the latitudes
and column densities of the simulated observations matched those of the actual observations in that
region of sky (we restricted ourselves to the observations that yielded oxygen measurements after
being subjected to the proton flux filtering; Table~\ref{tab:OxygenIntensities2}).  As noted above,
each simulated intensity measurement was the sum of the absorbed model halo emission, a randomly
chosen intensity due to the time-variable SWCX emission, and a randomly chosen positive or negative
intensity that mimics the measurement error. We examined two halo models: a plane-parallel model, in
which the intrinsic halo intensity is given by $I_0 \cosec |b|$, and a constant-intensity model, in
which the intrinsic halo intensity, $I_0$, is independent of $b$. For each model we tested different
values of $I_0$ -- the smaller the value of $I_0$, the more likely it is that any potentially
observable correlation due to the halo is obscured by the SWCX emission or the measurement error. We
assumed that the intensity of the time-variable SWCX emission followed the distribution given by our
$I - \min(I)$ measurements from the proton-flux-filtered data (gray histograms in
Figures~\ref{fig:IntensityHistogram}(c) and \ref{fig:IntensityHistogram}(d)). Note that our model
considers only the time-variable SWCX emission. We are implicitly assuming that any non-variable
SWCX emission does not vary strongly with Galactic latitude, and hence does not affect the
correlation coefficient between intensity and latitude.

Therefore, for the plane-parallel model, the $j$th simulated \OVII\ or \OVIII\ intensity (for $j=1$
to 1000) corresponding to the $i$th actual observation (where $i=1$ to $N$, $N$ being the number of
observations in the region of the sky being studied) is given by
\begin{equation}
  I_{\mathrm{sim},ij} = I_0 \cosec |b_i| \e^{-\sigma{\NH}_i} + I_{\mathrm{SWCX},ij} + \epsilon_{ij},
  \label{eq:Isim1}
\end{equation}
and for the constant-intensity model it is given by
\begin{equation}
  I_{\mathrm{sim},ij} = I_0 \e^{-\sigma{\NH}_i} + I_{\mathrm{SWCX},ij} + \epsilon_{ij}.
  \label{eq:Isim2}
\end{equation}
Here, $b_i$ is the Galactic latitude of the $i$th actual observation, ${\NH}_i$ is the column
density in the direction of this observation \citep{kalberla05}, $\sigma$ is the photoelectric
absorption cross-section appropriate for the emission line energy in question
\citep{balucinska92,yan98}, $I_{\mathrm{SWCX},ij}$ is the simulated SWCX intensity for this
simulated observation, and $\epsilon_{ij}$ represents the measurement error. As noted above, the
values of $I_{\mathrm{SWCX},ij}$ were drawn randomly from a distribution that follows the observed
$I - \min(I)$ distribution for the line in question (gray histograms in Figures~\ref{fig:I-Imin}(c)
and \ref{fig:I-Imin}(d) for \OVII\ and \OVIII, respectively). The values of $\epsilon_{ij}$ were
drawn randomly from a zero-mean normal distribution with standard deviations of 1.7 and 1.2~\LU\ for
\OVII\ and \OVIII, respectively (these are the 75th percentiles of the observed statistical +
systematic measurement errors).

We created such datasets for each octant of the sky, except for the 1st and 4th northern octants. We
omitted these regions because of the presence of the Sco-Cen superbubble. As we are interested in
the halo, we considered only high Galactic latitudes ($|b| \ge 30\degr$).

For each simulated dataset, we calculated the correlation coefficient for the intensity against
$|b|$. For each choice of oxygen emission line, halo model, and sky octant, the set of calculated
correlation coefficients forms a distribution. These distributions of simulated correlation
coefficients for \OVII\ are summarized by the errorbars in Figure~\ref{fig:o7-vs-b} -- the results
for the constant-intensity and plane-parallel halo models are shown in the left and right halves of
each panel, respectively, for different values of model $I_0$, as indicated on the $x$-axes. Each
errorbar indicates the 5th and 95th percentile of a given distribution of simulated correlation
coefficients, while the crosses indicate the medians. The corresponding results for \OVIII\ are
shown in Figure~\ref{fig:o8-vs-b}. Note that the values of $I_0$ used to model the \OVII\ and
\OVIII\ correlation coefficients differ -- larger values of $I_0$ tend to overpredict the
\OVIII\ intensities. For \OVIII, we find that we generally cannot distinguish between the
models. This is likely because the differences in the intensities predicted by the various models
are small compared to the uncertainties due to the SWCX emission and the measurement error.
Finally, note that, because the variation of \NH\ with $b$ is not the same in all octants, the same
model may produce different correlation coefficients in different octants.

\paragraph{Comparison with Observations}
For each octant of the sky in Figures~\ref{fig:o7-vs-b} and \ref{fig:o8-vs-b} for which we have
calculated model predictions, we have overplotted the observed correlation coefficient as a
horizontal dashed line, with a horizontal gray band indicating the 90\%\ confidence interval. Here,
we visually compare the above-described model predictions with the observed correlation
coefficients.

No single model compares well with the observed correlation coefficients in all octants for both
\OVII\ and \OVIII. Both halo models are generally in agreement with the observed \OVII\ correlation
coefficients in the four southern octants, although in the 1st southern quadrant, the plane-parallel
model may be favored over the constant-intensity model, and in the 2nd southern octant, the
plane-parallel model with $I_0 = 4~\LU$ tends to systematically overpredict the intensities at lower
latitudes (not plotted).  However, only the constant-intensity model is in good agreement with the
observed correlation coefficient in the 3rd northern octant, and neither model agrees with the
observed correlation coefficient in the 2nd northern octant.

For \OVIII, both halo models are generally in agreement with the observed correlation coefficients.
The exception is the 2nd southern quadrant, in which neither model agrees well with the observed
correlation coefficient.

The most important result from this analysis is that no single halo model seems to be able to
account for the observed variation in intensity with Galactic latitude in all directions. Some
octants seem to favor the plane-parallel geometry, others seem to favor a constant-intensity model,
while in others neither model is in good agreement with the data. Determining the true halo
intensity, and hence the true halo geometry, however, will require a better method for disentangling
the halo and SWCX contributions to the observed emission.

\section{SUMMARY}
\label{sec:Summary}

We have presented a new catalog of SXRB \OVII\ and \OVIII\ intensities, extracted from 1880 archival
\xmm\ observations. This catalog covers the full range of Galactic longitudes, and supersedes our
previous catalog (Paper~I), which covered $l = 120\degr$--240\degr. Our new analysis has some
improvements over our previous analysis, including the use of more recent software and calibration
data, the use of data from the \xmm\ Serendipitous Source Catalogue for point source removal, and
the estimation of the systematic errors on the oxygen intensities (see
Section~\ref{subsec:DifferencesFrom1} for more details on the differences from Paper~I).

As in Paper~I, we attempted to reduce the affects of SWCX contamination by excluding the portions of
the \xmm\ observations for which the contemporaneous solar wind proton flux exceeded $2 \times
10^8~\pcmsq\ \ps$. Of the 1880 \xmm\ observations for which we have results, 1868 yielded oxygen
intensities without the proton flux filtering, and 1003 yielded results with this filtering. In
practice, the proton flux filtering tended to render unusable those observations that yielded
particularly bright oxygen intensities (i.e., the solar wind proton flux tended to exceed our
threshold of $2 \times 10^8~\pcmsq\ \ps$ during such observations). As a result, the set of filtered
observations generally yields lower oxygen intensities than the set of unfiltered observations.
However, among the set of observations for which the proton flux filtering did remove at least some
\xmm\ data, but not so much that the observation became unusable, we found that the filtered
observations did not yield systematically lower oxygen intensities than the unfiltered observations
(see Section~\ref{subsec:ProtonFluxFilteringEffects}).

Our main results were presented in Section~\ref{subsec:OxygenResults}. We found a large range of
intensities: 0.0--50.8 and 0.0--36.2~\LU\ for \OVII\ and \OVIII, respectively, for intensities
obtained without proton flux filtering (0.0--48.8 and 0.0--24.5~\LU, respectively, for intensities
obtained with this filtering). However, the typical \OVII\ and \OVIII\ intensities are $\sim$2--11
and $\la$3~\LU, respectively. Note that we excluded bright diffuse objects such as SN
remnants or galaxy clusters during our processing, so the brightest intensities that we measured
should not be contaminated by emission from such objects. The medians and quartiles of the oxygen
intensities are generally significantly higher than those in Paper~I, as the current paper includes
observations toward the inner Galaxy, which typically yield higher intensities. We also found that
the observations identified by \citet{carter11} as being contaminated by SWCX emission tended to
yield larger-than-typical oxygen intensities.

Our dataset includes measurements for 217 sightlines that have been observed more than once with
\xmm\ (see Section~\ref{subsec:MultipleObs}). For a given direction, the differences in the measured
oxygen intensities between observations can be attributed to the time-variable SWCX
emission. Typically, the variations in the \OVII\ and \OVIII\ intensities are $\la$5 and $\la$2~\LU,
respectively. However, some observations exhibit bright intensity enhancements, the brightest being
an \OVII\ enhancement of 24~\LU\ over another observation in the same direction (note that this
bright enhancement was also reported in Paper~I).

We discussed various aspects of our results in Section~\ref{sec:Discussion}. Some important results
are as follows:
\begin{enumerate}
\item The oxygen intensities generally decrease from solar maximum to solar minimum, at high and low
  ecliptic latitudes. While this is as expected for high ecliptic latitudes from heliospheric SWCX
  models, it is not as expected for low ecliptic latitudes \citep{koutroumpa06}. We also found that
  the \OVII\ intensities during solar maximum are higher toward high ecliptic latitudes, while those
  during solar minimum are higher toward low ecliptic latitudes (the differences are small but
  statistically significant). The \OVIII\ intensities, however, do not exhibit clear variation with
  ecliptic latitude (see Section~\ref{subsubsec:HeliosphericSWCX}).
\item The observed variations in oxygen intensity are in poor agreement with the predictions of a
  geocoronal SWCX model similar to that described in \citet{robertson03b}. This model takes into
  account the time-variable density of the solar wind in and beyond the magnetosheath, using the
  \citet{spreiter66} magnetosphere model and solar wind data from OMNIWeb. For each
  \xmm\ observation, the expected geocoronal SWCX emissivity is integrated along the line of sight,
  taking into account the position and orientation of \xmm. In
  Section~\ref{subsubsec:GeocoronalSWCX} we discussed various possible reasons for the discrepancy
  between the model predictions and the observations.
\item Approximately $\sim$40--50\%\ of the 3/4~\kev\ SXRB (specifically, the \rosat\ R4 count-rate)
  that is not due to the extragalactic background is due to the oxygen \Kalpha\ lines (see
  Section~\ref{subsubsec:FractionDueToOxygen}). This value is similar to the fraction derived from a
  high-resolution microcalorimeter spectrum of the SXRB \citep{mccammon02}. The fraction is
  independent of the oxygen \Kalpha\ intensity and the non-EPL 3/4~\kev\ surface brightness,
  implying that the spectrum of the non-EPL 3/4~\kev\ emission does not vary systematically with its
  brightness.
\item The above fraction is less than the fraction expected if the non-EPL 3/4~\kev\ SXRB were
  dominated by a combination of unabsorbed SWCX emission and absorbed emission from a
  $\sim$$(\mbox{2--3}) \times 10^6~\K$ CIE plasma in the halo. We ruled out the possibilities that
  the SWCX emission was brighter during the RASS than during the
  \xmm\ observations, or that the halo is much hotter than $3 \times 10^6~\K$. Instead, these
  results may indicate that the true SWCX emission is harder than expected, or that the halo is out
  of equilibrium and overionized, or that oxygen is depleted in the halo (see
  Section~\ref{subsubsec:ImplicationsForSources}). With our current measurements, we are unable to
  distinguish between these possibilities. However, we noted that it is possible that
  solar-abundance CIE models of the halo with $T \sim \mbox{(2--3)} \times 10^6~\K$ underestimate
  the halo emission at energies greater than those of the oxygen \Kalpha\ lines.
\item The oxygen intensities tend to increase from the outer to the inner Galaxy, which may indicate
  either that the SN rate increases toward the inner Galaxy (if the hot halo gas is mainly due to
  Galactic fountains) or the presence of a halo of accreted material centered on the Galactic
  Center (see Section~\ref{subsubsec:LongitudeVariation}).
\item No clear overall picture emerges when we look at the variation of the intensities with
  Galactic latitude: some regions of the sky exhibit a statistically significant positive
  correlation between intensity and latitude, some regions exhibit a statistically significant
  anticorrelation, and some regions exhibit no significant correlation or anticorrelation (see
  Section~\ref{subsubsec:LatitudeVariation}). We compared the observed correlation coefficients with
  those expected from two models of the intrinsic halo emission (a constant-intensity model
  (intrinsic intensity independent of $b$), and a plane-parallel model) and found that no one halo
  model appears to be able to explain the observations in all octants. Indeed, in some octants,
  neither model is in good agreement with the observations (see
  Section~\ref{subsubsec:ModelingLatitudeVariation}).
\end{enumerate}

Our catalog of oxygen intensities will provide useful constraints on models of the heliospheric and
geocoronal SWCX emission -- as noted above, the variation in the oxygen intensities measured from
different observations of the same direction can be attributed to the time-variable SWCX
emission. Then, equipped with suitable models of the SWCX emission, one can subtract the estimated
SWCX intensity for each observation, revealing the \OVII\ and \OVIII\ intensities due to the hot gas
in our Galaxy. From these intensities, the temperature and distribution on the sky of this hot
gas can be inferred.

An alternative method for constraining the halo oxygen emission, which we are currently
investigating, is to use Bayesian analysis. The observed $I - \min(I)$ distributions
(Figure~\ref{fig:IntensityHistogram}) can be used to specify the prior probability distribution for
the SWCX intensity in an arbitrary \xmm\ spectrum. Using this prior, one can then, in principle,
calculate the posterior probability distribution for the halo emission intensity.

Finally, it should be noted that the \xmm\ SXRB spectra that we have extracted contain more
information than just the diffuse oxygen line intensities. For example, by fitting appropriate
models to the spectra one can constrain the temperature and emission measure of the halo, and use
these results to test physical models for the hot halo gas (as in \citealt{henley10b})

\acknowledgements

This research is based on observations obtained with \xmm, an ESA science mission with instruments
and contributions directly funded by ESA Member States and NASA.  We acknowledge use of the R
software package \citep{R}. This research was funded by NASA grant NNX08AJ47G, awarded through the
Astrophysics Data Analysis Program.

\bibliography{references}

\clearpage

\end{document}